\documentclass[final,1p,times]{elsarticle}

\usepackage{amssymb}

\usepackage{amssymb}
\usepackage{stmaryrd}

\usepackage{bm}

\usepackage{url}
\usepackage{booktabs}
\usepackage{longtable}
\usepackage[usenames,dvipsnames]{color}
\usepackage[table]{xcolor}
\usepackage[hypertex]{hyperref}

\usepackage[final]{changes}
\definecolor{bkgnd}{cmyk}{0,0,0,0.10}
\definecolor{bkgnd}{cmyk}{0.06,0.06,0,0}
\rowcolors{1}{White}{White}

\newcommand{\appref}[1]{Appendix \ref{#1}}

\newcommand{\half}{\ensuremath{\frac{1}{2}}}
\newcommand{\halfl}{\ensuremath{{\scriptstyle \frac{1}{2}}}}
\newcommand{\real}{\text{Re}}
\newcommand{\imag}{\text{Im}}
\newcommand{\expect}{\mathbb{E}}
\newcommand{\conv}{\otimes}
\newcommand{\arccosh}{\ensuremath{\mathrm{arccosh}}}
\newcommand{\rHz}{\ensuremath{\sqrt{\mathrm{Hz}}}}

\newcommand{\un}   [1]{\ensuremath{\,\mathrm{#1}}}
\newcommand{\unit} [1]{\un{#1}}
\newcommand{\unitt}[1]{\ensuremath { \mathrm {#1} }}
\newcommand{\intd}    {\ensuremath{\,\mathrm{d }}}

\newcommand{\text} [1]{\ensuremath { \mathrm {#1} }}

\newcommand{\vc}   [1]{\ensuremath{\mathbf{#1}}}
\newcommand{\ten}  [1]{\ensuremath{\mathbf{#1}}}
\newcommand{\pder} [2]{\ensuremath{\frac{\partial #1}{\partial #2}}}
\newcommand{\pderl}[2]{\ensuremath{\partial #1/\partial #2}}
\newcommand{\tder} [2]{\ensuremath{\frac{\text{d} #1}{\text{d} #2}}}
\newcommand{\tderl}[2]{\ensuremath{\text{d} #1/\text{d} #2}}

\newcommand{\ket}  [1]{\ensuremath{\left | #1 \right \rangle }}
\newcommand{\bra}  [1]{\ensuremath{\left \langle #1 \right | }}

\newcommand{\strain}  {\ensuremath{\gamma}}

\newcommand{\kb}      {\ensuremath{k_\mathrm{B} }}

\newcommand{\hg}      {\ensuremath{h_{\text{g}}}}

\newcommand{\vg}      {\ensuremath{V_{\mathrm{g}}}}
\newcommand{\vgdc}    {\ensuremath{V_{\mathrm{g}}^{\mathrm{dc}}}}
\newcommand{\vgac}    {\ensuremath{V_{\mathrm{g}}^{\mathrm{ac}}}}

\newcommand{\vgstar}  {\ensuremath{V_{\mathrm{g}}^{*}}}

\newcommand{\cg}      {\ensuremath{C_{\mathrm{g}}}}

\newcommand{\udc}     {\ensuremath{u_{\mathrm{dc}}}}
\newcommand{\uac}     {\ensuremath{u_{\mathrm{ac}}}}
\newcommand{\Tdc}     {\ensuremath{T_{\mathrm{dc}}}}
\newcommand{\Tac}     {\ensuremath{T_{\mathrm{ac}}}}

\journal{Physics Reports}

\begin{document}

\begin{frontmatter}



\title{Mechanical systems in the quantum regime}


\author{Menno Poot\footnote{Present address: Department of Electrical Engineering, Yale University, New Haven, CT 06520,
USA}}
\ead{menno.poot@yale.edu}
\author{Herre S. J. van der Zant\corref{}}
\ead{h.s.j.vanderzant@tudelft.nl}

\address{Kavli Institute of Nanoscience, Delft University of Technology, P.~O.~B.~5046, 2600GA Delft, The Netherlands}

\begin{abstract}
Mechanical systems are ideal candidates for studying quantum
behavior of macroscopic objects. To this end, a mechanical
resonator has to be cooled to its ground state and its position
has to be measured with great accuracy. Currently, various
routes to reach these goals are being explored. In this review,
we discuss different techniques for sensitive position
detection and we give an overview of the cooling techniques
that are being employed. \replaced{The latter}{These} include
sideband cooling and active feedback cooling. The basic
concepts that are important when measuring on mechanical
systems with high accuracy and/or at very low temperatures,
such as thermal and quantum noise, linear response
theory\added{,} and backaction, are explained. From this, the
quantum limit on linear position detection is obtained and the
sensitivities that have been achieved in recent \added{opto and
nanoelectromechanical} experiments are compared to this limit.
The mechanical resonators that are used in the experiments
range from meter-sized gravitational wave detectors to
nanomechanical systems that can only be read out using
mesoscopic devices such as single-electron transistors or
superconducting quantum interference devices. A special class
of nanomechanical systems are bottom-up fabricated carbon-based
devices, which have very high frequencies and yet a large
zero-point motion, making them ideal for reaching the quantum
regime. The mechanics of some of the different mechanical
systems at the nanoscale is studied. We conclude this review
with an outlook of how state-of-the-art mechanical resonators
can be improved to study quantum {\it mechanics}
\end{abstract}


\end{frontmatter}

\tableofcontents


\section{Introduction} \label{sec:intro}
Mechanics is probably the most well-known branch of physics as
everyone encounters it in every-day life. It describes a wide
range of effects: from the motion of galaxies and planets on a
large scale, the vibrations of a bridge induced by traffic or
wind, the stability of a riding bicycle, to the trajectories of
electrons in an old-fashioned television on a microscopic scale.
In the early days of physics, mainly objects that could be seen or
touched were studied. Until the beginning of the twentieth century
it was thought that the three laws of motion obtained by Newton
described the dynamics of mechanical systems completely. However,
the development of better telescopes and microscopes enabled the
study of mechanical systems on both much larger and smaller length
scales. In the early 1900s, the rapid developments that led to the
theory of special and general relativity and quantum mechanics
showed that the laws of classical mechanics were not the whole
truth.

Relativistic corrections turn out to be important for objects
with large masses or with velocities approaching the speed of
light. It is therefore an important factor in astrophysics,
where one studies the dynamics of heavy objects like galaxies
and black holes or the bending of light by the curvature of
space-time. When the masses and velocities of the objects
involved are made smaller and smaller, the relativistic
corrections eventually vanish and one obtains the classical
laws of motion \cite{carroll_spacetime}.

Quantum mechanics, on the other hand, is particularly well
suited to describe the mechanics of objects at the other end of
the length-scale range, i.e., (sub)atomic objects. In the
beginning of the twentieth century, quantum theory successfully
explained the photo-electric effect, black-body radiation, and
the atomic emission spectra. Quantum mechanics is different
from classical and relativistic mechanics in the sense that
objects are no longer described by a definite position, but by
a wavefunction. This wavefunction evolves deterministically
according to the Schr\"odinger equation and its absolute value
squared should be interpreted as the position
probability-density function, the so-called Born rule
\cite{griffiths_qm}. To find the object at a particular
location one has to {\it measure} its position. This process,
however, inevitably disturbs the evolution of the wavefunction
\cite{caves_RMP, braginsky_science_QND}.

Quantum mechanics does not only describe processes at the
(sub)atomic scale successfully, but it also explains the
microscopic origin of many macroscopic effects such as the
electronic properties of solids, superfluidity and so on.
Unlike in relativity where one can simply take the limit $m, v
\rightarrow 0$, in quantum mechanics it is still not entirely
clear how the transition from quantum mechanics to classical
mechanics exactly happens
\cite{adler_science_quantum_theory_exact,
dunningham_science_cat_states}. Although Ehrenfest's theorem
implies that the expectation values of the momentum and
potential energy obey Newton's second law \cite{griffiths_qm},
the classical laws of motion cannot in all cases be recovered
by simply taking $\hbar \rightarrow 0$ as has recently been
shown in calculating the quantum dynamics of nonlinear
resonators \cite{katz_NJP_transition}.

Another issue that is still debated is how quantum mechanics
should be interpreted \cite{griffiths_qm,
leggett_science_quantum_measurement}: as the truth, as a tool
to calculate outcomes of an experiment, or as an incomplete
theory? This question has been asked since the early days of
quantum theory, resulting in the famous Einstein-Podolsky-Rosen
paper \cite{einstein_PRL_EPR}, but has still not been answered.
The interpretation of quantum mechanics and its transition to
the macroscopic world are related and can be reformulated into
the question ``Can a macroscopic object be put in a quantum
superposition?'' This was first illustrated by Schr\"odinger in
1935 with the famous dead-or-alive cat {\it gedanken}
experiment. Superpositions of small objects are readily
observed, a good example are the singlet and triplet spin
states in a molecule, but this becomes increasingly difficult
for larger and larger systems mainly due to decoherence
\cite{zurek_RMP_decoherence,
bose_PRA_decoherence_superposition}. Superpositions have for
example been created using the circulating current in
superconducting quantum interference devices (SQUIDs)
\cite{wal_science_superposition} and with fullerenes
\cite{arndt_nature_C60_superposition,
hackermueller_nature_C70_superposition}.

\begin{figure}[tb]
\centering
\includegraphics{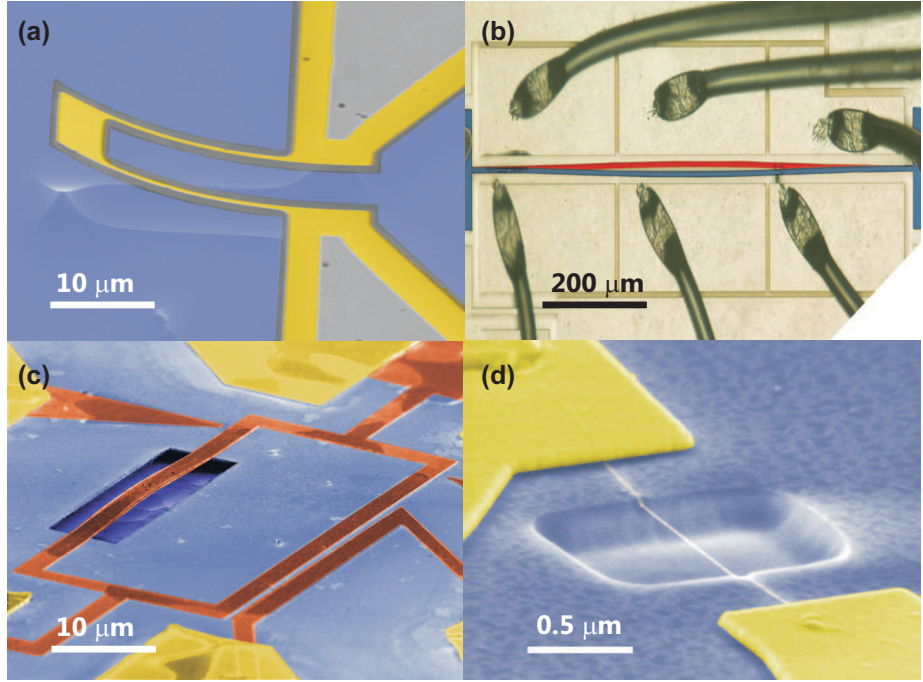}
\caption{Different types of MEMS/NEMS used at Delft University of Technology.
(a) Piezoresistive cantilever. A deflection of the cantilever changes the electrical resistance between the two sides. This resonator geometry can be used for mass detection.
(b) Bistable buckled beam. Blue and red indicates the two stable positions of the beam, which can be used to encode digital information \cite{roodenburg_APL_buckled}.
(c) A beam resonator that is embedded in the loop of a dc SQUID (red) \cite{etaki_natphys_squid}. A magnetic field couples the position of the beam to the magnetic flux through the loop.
(d) Suspended carbon nanotube (white) as a flexural resonator \cite{witkamp_NL_bendingmode}.
\label{fig:intro:resonator_types}}
\end{figure}

Nanomechanical devices \cite{cleland_nanomechanics,
ekinci_RSI_overview} are interesting candidates to further
increase the size of systems that can be put in a superposition
\cite{bose_PRA_decoherence_superposition,
marshall_PRL_superposition_mirror}. These systems are the
logical continuation of micromechanical devices which are made
using integrated-circuit technology, but then on a much
smaller, nanometer scale. Micro-electro\-mecha\-ni\-cal systems
(MEMS) are currently widely used as, for example,
accelerometers in airbags, pressure sensors, and in projectors.
When scaling these devices down to the nanometer scale, their
resonance frequencies increase and at the same time their mass
decreases. From an application point of view
nano-electromechanical systems (NEMS) may enable single-atom
mass-sensing, mechanical computing and efficient signal
processing in the radio-frequency and microwave bands. From a
scientific point of view these devices are interesting as they
can be cooled to temperatures so low that the resonator is
nearly always in its quantum-mechanical ground state. Figure
\ref{fig:intro:resonator_types} shows some examples of
miniature mechanical resonators.

In the last decade, many groups have pursued demonstration the
quantum limit of motion. The rapid progress in the development
of sensitive optical techniques and mesoscopic electronics have
led to detectors that have sensitivities that are approaching
the quantum limit on position detection \cite{caves_RMP}.
Moreover, very recently the long-sought quantum mechanical
behavior of mechanical resonators has become reality in two
different experiments indicating that the quantum regime of
mechanical motion has now been entered: Selected by Science
magazine as research breakthrough~\cite{science_breakthrough}
of the year 2010, the groups of Cleland and Martinis
demonstrated quantum mechanical behavior of a 6~GHz mechanical
resonator by coupling it to a superconducting flux
qubit~\cite{oconnell_nature_quantum_piezo_resonator}. At
dilution refrigerator temperatures the first ten energy states
of the harmonic mechanical oscillator, including the ground
state, could be probed. The measurement itself is performed on
the superconducting qubit which acts as a two-level quantum
system whose response changes when the occupation of the
mechanical resonator state changes. A large coupling between
the two quantum systems is achieved by using the piezoelectric
properties of the mechanical resonator material. In a different
approach Teufel {\it et al.}~\cite{teufel_arXiv_groundstate}
use a superconducting microwave cavity to \replaced{actively
cool}{show that} a mechanical drum resonator \deleted{can be
actively cooled} to such low temperatures that it is in the
quantum mechanical ground state for most of the time. The drum
resonator is integrated in the superconducting resonant circuit
to provide strong phonon-photon coupling and the measurements
use concepts developed for optical cavities to achieve
efficient cooling.

These two breakthrough experiments not only show that
non-classical behavior can be encoded in the motion of a
mechanical resonator but also open a new exciting research
field involving quantum states of motion. In this review we
will summarize the main theoretical and experimental
discoveries that have led to the demonstration of quantum
motion. Our main focus will be on the mechanics, i.e., on the
mechanical properties of resonators and on the different
optical and electronic detection schemes that have been
developed to measure their displacements. Since measuring
always means that the detector has to be coupled to the
resonator, we will also discuss the coupling and the
consequences it may have on the measurement itself.
Furthermore, the advantages and disadvantages of the various
approaches will be discussed.

\subsection{Ground-state cooling and quantum-limited position detection}
The concept of the quantum limit on mechanical motion detection
and its implications became relevant in the 1970s when more and
more sensitive gravitational wave detectors
\cite{thorne_RMP_gravitational_waves} were designed (see for
example Refs. \cite{billing_JPE_GW} and \cite{weber_GRG_GW}),
raising questions on the (im)possibility to violate Heisenberg
uncertainty principle \cite{griffiths_qm, caves_RMP,
braginsky_quantum_measurement}. Nowadays, these issues are
important when measuring micro- and nanomechanical devices with
very sensitive detectors or at very low temperatures.

There are two important considerations when approaching the
quantum limit. First, the thermal occupation is important,
which is defined in Sec. \ref{ssec:basic:noise} as $\overline
n=(k_B T_R/h f_R) -1/2$ where $f_R$ is the resonance frequency
and $T_R$ is the resonator temperature. A value for $\overline
n$ that is below 1 indicates that the resonator is in its
ground state most of the time (see Sec. \ref{ssec:basic:noise}
for a more detailed description of the thermal occupation).
Second, the zero-point motion sets the ultimate limit on the
resonator displacement. At high temperatures and in the absence
of actuation, Brownian motion determines the resonator
position. As temperature decreases, the displacement decreases
\deleted{as well} to the point that the quantum regime is
reached ($\overline n \lesssim 1$). Zero-point motion of the
undriven beam remains with an amplitude of $u_0=
\sqrt{\hbar/4\pi m f_R}$ with $m$ the resonator mass. To
observe the zero-point motion, one has to detect it. This
sounds trivial, but it turns out that a measurement on a
quantum system inevitably disturbs it. Quantum mechanics sets a
limit to the precision of continuously measuring the position
of the resonator, the standard quantum limit. A detector at
that limit is therefore called quantum-limited. Detection of
the zero-point motion thus requires resonant frequencies higher
than about 1~GHz (where $\overline n = 1$ corresponds to a
temperature of 50~mK, which can be reached in a dilution
refrigerator) in combination with a position detection scheme
that meets the quantum limit. Alternatively, a low-frequency
resonator can be actively cooled to its quantum mechanical
ground state. In experiments, both approaches are pursued and
we discuss the status of both of them.
%
\rowcolors{1}{white}{bkgnd}

\begin{table}[htbp]
\centering \caption{Overview of recent key experiments with
micro- and nanomechanical resonators in chronological order.
Several types of resonators and detection methods are used by
different groups in the field \added{and are measured} at
different temperatures $T$. The table shows the resonance
frequency $f_R$, quality factor $Q$ and the mass $m$ of the
resonator. From this, the zero-point motion $u_0$ is
calculated.}
\begin{tabular}{rllcrcccr}
\addlinespace \toprule
{\bf }                 & {\bf Group}            & {\bf Resonator} & $\mathbf{f_R \un{(MHz)}}$ & {\bf $\mathbf{Q}$}     & {\bf $\mathbf{T \un{(K)}}$} & {\bf $\mathbf{m \un{(kg)}}$}     & {\bf $\mathbf{u_0 \un{(fm)}}$}   & {\bf Ref.} \\
\midrule
1                         & Roukes                    & SiC beam                  & 1029                      & 500                       & 4.2                       &  $3.4 \cdot 10^{-17} $    &           15              & \cite{huang_nature_GHz} \\
2                         & UCSB                      & GaAs beam                 & 117                       & 1700                      & 0.030                     &  $2.8 \cdot 10^{-15} $    &          5.1              & \cite{knobel_nature_set} \\
3                         & Schwab                    & SiN/Au beam               & 19.7                      & 35000                     & 0.035                     &  $9.7 \cdot 10^{-16} $    &           21              & \cite{lahaye_science} \\
4                         & LMU                       & Si/Au cantilever          & $7.30 \cdot 10^{-3} $     & 2000                      & 295                       &  $8.6 \cdot 10^{-12} $    &           12              & \cite{metzger_nature_cooling} \\
5                         & Schwab                    & SiN/Al beam               & 21.9                      & 120000                    & 0.030                     &  $6.8 \cdot 10^{-16} $    &           24              & \cite{naik_nature} \\
6                         & LKB Paris                 & Si micromirror            & 0.815                     & 10000                     & 295                       &  $1.9 \cdot 10^{-7} $     &  $7.3 \cdot 10^{-3} $     & \cite{arcizet_nature_cavity} \\
7                         & UCSB                      & AFM cantilever            & $1.25 \cdot 10^{-2} $     & 137000                    & 295                       &  $2.4 \cdot 10^{-11} $    &          5.3              & \cite{kleckner_nature_feedback} \\
8                         & Vienna                    & SiO$_2$/TiO$_2$ beam      & 0.278                     & 9000                      & 295                       &  $9.0 \cdot 10^{-12} $    &          1.8              & \cite{gigan_nature_cavity} \\
9                         & LKB Paris                 & Si micromirror            & 0.814                     & 10000                     & 295                       &  $1.9 \cdot 10^{-7} $     &  $7.4 \cdot 10^{-3} $     & \cite{arcizet_PRL_sensitive_monitoring} \\
10                        & MPI-QO                    & Silica toroid             & 57.8                      & 2890                      & 300                       &  $1.5 \cdot 10^{-11} $    &        0.10               & \cite{schliesser_PRL_cavity_cooling} \\
11                        & Roukes                    & SiC/Au cantilever         & 127                       & 900                       & 295                       &  $5.0 \cdot 10^{-17} $    &           36              & \cite{li_natnano_piezo} \\
12                        & LIGO                      & Micromirror               & $1.72 \cdot 10^{-4} $     & 3200                      & 295                       &  $1.0 \cdot 10^{-3} $     &  $7.0 \cdot 10^{-3} $     & \cite{corbitt_PRL_opticalspring} \\
13                        & LIGO                      & Micromirror               & $1.27 \cdot 10^{-5} $     & 19950                     & 295                       &  $1.0 \cdot 10^{-3} $     &      0.026                & \cite{corbitt_PRL_cooling_mK} \\
14                        & LMU                       & Si micromirror            & 0.547                     & 1059                      & 300                       &  $1.1 \cdot 10^{-14} $    &           37              & \cite{favero_APL_cooling_micromirror} \\
15                        & JILA                      & Gold beam                 & 43.1                      & 5000                      & 0.25                      &  $2.3 \cdot 10^{-15} $    &          9.2              & \cite{flowers_PRL_APC} \\
16                        & IBM                       & Si cantilever             & $2.60 \cdot 10^{-3} $     & 55600                     & 2.2                       &  $3.2 \cdot 10^{-13} $    &         100               & \cite{poggio_PRL_feedback} \\
17                        & NIST                      & Si cantilever             & $7.00 \cdot 10^{-3} $     & 20000                     & 295                       &  $1.0 \cdot 10^{-10} $    &          3.5              & \cite{brown_PRL_circuit_cooling} \\
18                        & LKB Paris                 & Micromirror               & 0.711                     & 16000                     & 295                       &  $7.4 \cdot 10^{-4} $     &  $1.3 \cdot 10^{-4} $     & \cite{caniard_PRL_backaction_cancelation} \\
19                        & Harris                    & SiN membrane              & 0.134                     & 1100000                   & 294                       &  $3.9 \cdot 10^{-11} $    &          1.3              & \cite{thompson_nature_cavity_membrane} \\
20                        & Vienna                    & Si cantilever             & 0.557                     & 2000                      & 35                        &  $4.0 \cdot 10^{-11} $    &        0.61               & \cite{groeblacher_EPL_sideband} \\
21                        & ANU                       & Mirror on beam            & $8.48 \cdot 10^{-5} $     & 44500                     & 300                       &  $6.9 \cdot 10^{-4} $     &        0.10               & \cite{mow_PRL_feedback_cooling} \\
22                        & MPI-QO                    & Silica toroid             & 74.0                      & 57000                     & 295                       &  $1.0 \cdot 10^{-11} $    &        0.11               & \cite{schliesser_natphys_sideband} \\
23                        & JILA                      & Al beam                   & 0.237                     & 2300                      & 0.040                     &  $2.0 \cdot 10^{-15} $    &         133               & \cite{regal_natphys_cavity} \\
24                        & Roukes                    & SiC/Au beam               & 428                       & 2500                      & 22                        &  $5.1 \cdot 10^{-17} $    &           20              & \cite{feng_natnano_selfsustaining} \\
25                        & IBM                       & Si cantilever             & $4.95 \cdot 10^{-3} $     & 22500                     & 4.2                       &  $2.0 \cdot 10^{-12} $    &           29              & \cite{poggio_natphys_QPC} \\
26                        & Delft                     & AlGaSb beam               & 2.00                      & 18000                     & 0.020                     &  $6.1 \cdot 10^{-13} $    &          2.6              & \cite{etaki_natphys_squid} \\
27                        & AURIGA                    & Al bar                    & $8.65 \cdot 10^{-4} $     & 1200000                   & 4.2                       &  $1.1 \cdot 10^{3} $      &  $3.0 \cdot 10^{-6} $     & \cite{vinante_PRL_feedback} \\
28                        & AURIGA                    & Al bar                    & $9.14 \cdot 10^{-4} $     & 880000                    & 4.2                       &  $1.1 \cdot 10^{3} $      &  $2.9 \cdot 10^{-6} $     & \cite{vinante_PRL_feedback} \\
29                        & Alberta                   & Si cantilever             & 1040                      & 18                        & 295                       &  $2.0 \cdot 10^{-17} $    &           20              & \cite{liu_natnano_timedomain} \\
30                        & Tang                      & Si beam                   & 8.87                      & 1850                      & 295                       &  $1.3 \cdot 10^{-15} $    &           27              & \cite{li_nature_harnessing_forces} \\
31                        & JILA                      & Al beam                   & 1.53                      & 300000                    & 0.050                     &  $6.2 \cdot 10^{-15} $    &           30              & \cite{teufel_PRL_stripline_cooling} \\
32                        & JILA                      & Al beam                   & 1.53                      & 10000                     & 0.050                     &  $6.2 \cdot 10^{-15} $    &           30              & \cite{teufel_PRL_stripline_cooling} \\
33                        & LMU                       & SiN beam                  & 8.90                      & 150000                    & 295                       &  $1.8 \cdot 10^{-15} $    &           23              & \cite{unterreithmeier_nature_dielectric} \\
34                        & Queensland                & Silica toroid             & 6.272                     & 545                       & 300                       &  $3.0 \cdot 10^{-8} $     &  $6.7 \cdot 10^{-3} $     & \cite{lee_PRL_feedback_toroid} \\
35                        & Tang                      & Si cantilever             & 13.86                     & 4500                      & 295                       &  $4.5 \cdot 10^{-16} $    &           37              & \cite{li_natnano_broadband_transmission} \\
36                        & Vienna                    & Si cantilever             & 0.945                     & 30000                     & 5.3                       &  $4.3 \cdot 10^{-11} $    &        0.45               & \cite{groeblacher_natphys_cooling} \\
37                        & MPI-QO                    & Silica toroid             & 65.0                      & 2000                      & 1.65                      &  $7.0 \cdot 10^{-11} $    &      0.043                & \cite{schliesser_natphys_lowoccupation} \\
38                        & Painter                   & Si ph. crystal            & 8.2                       & 150                       & 360                       &  $4.3 \cdot 10^{-14} $    &          4.9              & \cite{eichenfield_nature_photonic_crystal} \\
39                        & Oregon                    & Silica sphere             & 118.6                     & 3400                      & 1.4                       &  $2.8 \cdot 10^{-11} $    &      0.050                & \cite{park_natphys_sideband} \\
40                        & LIGO                      & Susp. mirror              & $1.23 \cdot 10^{-4} $     &                           & 300                       &  $2.7 \cdot 10^{0} $      &  $1.6 \cdot 10^{-4} $     & \cite{abbott_NJP_kg_groundstate} \\
41                        & Painter                   & Silica double tor.        & 8.53                      & 4070                      & 300                       &  $1.5 \cdot 10^{-13} $    &          2.6              & \cite{lin_PRL_disks} \\
42                        & JILA                      & Al beam                   & 1.04                      & 160000                    & 0.015                     &  $1.1 \cdot 10^{-14} $    &           27              & \cite{teufel_natnano_beyond_SQL} \\
43                        & MPI-QO / LMU              & SiN beam                  & 8.07                      & 10000                     & 300                       &  $4.9 \cdot 10^{-15} $    &           15              & \cite{anetsberger_natphys_nearfield} \\
44                        & Schwab                    & SiN/Al beam               & 6.30                      & 1000000                   & 0.020                     &  $2.1 \cdot 10^{-15} $    &           25              & \cite{rocheleau_nature_lown} \\
45                        & MPI-QO / LMU              & SiN beam                  & 8.30                      & 30000                     & 300                       &  $3.7 \cdot 10^{-15} $    &           16              & \cite{anetsberger_PRA_far_below_SQL} \\
46                        & UCSB                      & AlN FBAR                  & 6170                      & 260                       & 0.025                     &  $2.8 \cdot 10^{-12} $    &      0.022                & \cite{oconnell_nature_quantum_piezo_resonator} \\
47                        & Delft                     & AlGaSb beam               & 2.14                      & 24000                     & 0.015                     &  $6.1 \cdot 10^{-13} $    &          2.5              & \cite{poot_APL_feedback} \\
48                        & NIST                      & Al drum                   & 10.69                     & 360000                    & 0.020                     &  $4.8 \cdot 10^{-14} $    &          4.0              & \cite{teufel_arXiv_groundstate} \\
\bottomrule
\end{tabular}
\label{tab:intro:short}
\end{table}

\rowcolors{1}{white}{white}

\subsection{Bottom-up and top-down nanomechanical devices}
Table \ref{tab:intro:short} provides an overview of the
different groups that have performed experiments with
mechanical resonators that approach the quantum limit in
position-detection sensitivity or that have been cooled to a
low resonator temperature. The micro- and nanomechanical
devices listed in this table are made using so-called top-down
fabrication techniques, which are also employed in the
semiconductor industry. Different groups use different types of
resonators: doubly clamped beams, singly clamped cantilevers,
radial breathing modes of silica microtoroids, membranes,
micromirrors and macroscopic bars. We have indicated the
resonator mass, which ranges from 20~ag to 1000~kg, and the
resonator frequency, which ranges from about 10~Hz to a few
GHz. From these numbers the zero-point motion $u_0$ has been
calculated; for top-down devices it is generally on the order
of femtometers.

These top-down nanoscale structures listed in Table
\ref{tab:intro:short} (e.g. beams, cantilevers and microtoroids)
are made by etching parts of a larger structure, for example a
thin film on a substrate, or by depositing material (evaporating,
sputtering) on a resist mask that is subsequently removed in a
lift-off process. In both cases, patterning of resist is needed,
which is done using optical or electron-beam lithography.
State-of-the-art top-down fabricated devices have thicknesses and
widths of less than 100 nanometer.

A major drawback of making smaller resonators to increase their
frequency, is that the quality factor decreases
\cite{ekinci_RSI_overview}. The quality factor is a measure for
the dissipation in the system. A low \added{quality factor or
``}Q-factor'' means a large dissipation, and this is an
unwanted property for resonators in the quantum regime. The
associated decoherence of quantum states then limits the time
for performing operations with these states. For example, in
the experiments of
Ref.~\cite{oconnell_nature_quantum_piezo_resonator} where $Q$
is on the order of a few hundred, the time for the manipulation
of quantum states is limited to only 6~ns. \deleted{In the
table, the measured Q-factor has been listed. A general trend
is that the smaller the device dimensions, the lower the
Q-factor and therefore the larger the dissipation.} The
decrease in Q-factor with device dimensions is often attributed
to the increase in surface-to-volume
ratio~\cite{ekinci_RSI_overview, imboden_APL_diamond,
unterreithmeier_PRL_damping}. An explanation comes from the
fabrication  which may introduce defects at the surface during
the micro-machining processes that are involved. These defects
provide channels for dissipation, resulting in the low
Q-factor.

Having this in mind, a different approach is to use the small
structures that nature gives us, to build or assemble
mechanical resonators. Bottom-up devices are expected not to
suffer from excessive damping, as their surface can be
defect-free at the atomic scale. Examples are inorganic
nanowires\footnote{Sometimes ``nanowire'' is also used for
top-down fabricated devices. Here, this term is used
exclusively for grown wires.}, carbon nanotubes and few-layer
graphene. The last two are examples of carbon-based materials.
Using these bottom-up materials, mechanical devices with true
nanometer dimensions can be made with the hope that surface
defects can be eliminated. High Q-values are therefore expected
for these devices, which, combined with their low mass, make
them ideal building blocks for nano-electromechanical systems
\deleted{(NEMS)}.
%
%
%
%
%
%
\rowcolors{1}{white}{bkgnd}

\begin{table}[tbp]
\caption{Overview of recent experiments with bottom-up resonators.
Several types of resonators are used: carbon nanotubes (CNT),
nanowires (NW), single-layer graphene (SLG), and few-layer
graphene (FLG) or graphene oxide (FLGO) sheets. The table shows
the resonance frequency $f_R$ and quality factors at room and
cryogenic temperature ($Q_{\mathrm{RT}}$ and $Q_{\mathrm{cryo}}$
resp.). $T_{\mathrm{min}}$ is the lowest temperature at which the
resonator is measured. $m$ is the mass of the resonator and $\ell$
is its length. From these data, the zero-point motion $u_0$ is
calculated. \label{tab:intro:bottomup}}
{\small \begin{centering}

\begin{tabular}{ccrrrrrrrrl}
\addlinespace \toprule {\bf Group} & {\bf Type} &
\parbox[c]{8mm}{\centering $\mathbf{f_R}$ \\ $(\unitt{MHz})$}
 & $\mathbf{Q}_{\mathrm{RT}}$ & $\mathbf{Q}_{\mathrm{cryo}}$
 &
\parbox[c]{5mm}{\centering $\mathbf{T_{\mathrm{min}}}$ \\ $(\unitt{K})$}
 &
\parbox[c]{8mm}{\centering $\mathbf{\ell}$ \\ $(\unitt{\mu m})$}
 &
\parbox[c]{15mm}{\centering $\mathbf{m}$ \\ $(\unitt{kg})$}
 &
\parbox[c]{8mm}{\centering $\mathbf{u_0}$ \\ $(\unitt{pm})$}
 & {\bf Ref.} \\

\midrule
Cornell               & CNT                   & 55                    & 80                    &                       & 300                   & 1.75                  & $7.4 \cdot 10^{-21} $ & 4.52                  & \cite{sazonova_nature} \\

Delft                 & CNT                   & 60                    & 100                   &                       & 300                   & 1.25                  & $1.0 \cdot 10^{-20} $ & 3.66                  & \cite{witkamp_NL_bendingmode} \\
Berkeley              & CNT                   & 350                   & 440                   &                       & 300                   & 0.5                   & $5.3 \cdot 10^{-20} $ & 0.67                  & \cite{jensen_NL_CNT_radio} \\
ICN                   & CNT                   & 3120                  & 8                     &                       & 300                   & 0.77                  & $5.8 \cdot 10^{-20} $ & 0.22                  & \cite{garcia_PRL_dfm_nanotube} \\
ICN                   & CNT                   & 154                   & 20                    &                       & 300                   & 0.265                 & $1.1 \cdot 10^{-19} $ & 0.69                  & \cite{garcia_PRL_dfm_nanotube} \\
ICN                   & CNT                   & 573                   & 20                    &                       & 300                   & 0.193                 & $4.0 \cdot 10^{-22} $ & 6.03                  & \cite{garcia_PRL_dfm_nanotube} \\
ICN                   & CNT                   & 167                   & 200                   & 2000                  & 5                     & 0.9                   & $1.4 \cdot 10^{-21} $ & 6.03                  & \cite{lassagne_NL_masssensing} \\
Berkeley              & CNT                   & 328.5                 & 1000                  &                       & 300                   & 0.205                 & $1.6 \cdot 10^{-21} $ & 4.01                  & \cite{jensen_natnano_masssensing} \\
CalTech               & CNT                   & 230                   &                       & 200                   & 6                     & 0.45                  & $4.8 \cdot 10^{-22} $ & 8.73                  & \cite{chiu_NL_masssensing} \\
Purdue                & CNT                   & 0.9                   & 3                     &                       & 300                   & 12.6                  & $4.1 \cdot 10^{-16} $ & 0.15                  & \cite{biedermann_NT_doppler} \\
Chalmers              & CNT                   & 62                    & 50                    &                       & 300                   & 2.05                  & $2.2 \cdot 10^{-17} $ & 0.08                  & \cite{eriksson_NL_CNT_relay} \\
Delft                 & CNT                   & 360                   &                       & 120000                & 0.02                  & 0.8                   & $5.3 \cdot 10^{-21} $ & 2.09                  & \cite{huettel_NL_highQ} \\
ICN                   & CNT                   & 50                    & 40                    & 600                   & 4                     & 1                     & $1.3 \cdot 10^{-21} $ & 11.41                 & \cite{lassagne_science_coupled_cnt} \\
U. Wash.              & CNT                   & 410                   &                       &                       & 64                    &                       &                       &                       & \cite{wang_science_nanotube_phase_transitions} \\
U. Illinois           & CNT                   & 4                     & 240                   &                       & 300                   & 6.2                   & $5.9 \cdot 10^{-18} $ & 0.59                  & \cite{cho_NL_NL_MWNT} \\
LPMCN                 & CNT                   & 73.6                  & 160                   &                       & 300                   & 2                     & $8.5 \cdot 10^{-21} $ & 3.66                  & \cite{gouttenoire_SM_FM_nanotube} \\
\midrule
CalTech               & Pt NW                 & 105.3                 &                       & 8500                  & 4                     & 1.3                   & $4.0 \cdot 10^{-17} $ & 0.045                 & \cite{husain_APL_nanowire} \\
CalTech               & Si NW                 & 215                   &                       & 5750                  & 25                    & 1.6                   & $1.9 \cdot 10^{-17} $ & 0.045                 & \cite{feng_NL_NW_magnetomotive} \\
U. Penn.              & GaN NW                & 2.235                 & 2800                  &                       & 300                   & 5.5                   & $1.9 \cdot 10^{-16} $ & 0.141                 & \cite{nam_NL_GaN_nanowire} \\
Yale                  & GaN NW                & 8.584                 & 1008                  &                       & 300                   & 2.04                  & $1.6 \cdot 10^{-17} $ & 0.245                 & \cite{henry_NL_AlN_NW} \\
Lyon                  & SiC NW                & 0.043                 & 159000                &                       & 300                   & 93                    & $9.8 \cdot 10^{-15} $ & 0.141                 & \cite{perisanu_APL_highQ_SiC} \\
Alberta               & Si NW                 & 1.842                 & 4200                  &                       & 300                   & 5.2                   & $2.4 \cdot 10^{-17} $ & 0.437                 & \cite{belov_JAP_si_nanowire} \\
Alberta               & Si NW                 & 1.842                 & 2000                  & 10000                 & 77                    & 11.2                  & $4.6 \cdot 10^{-16} $ & 0.099                 & \cite{belov_JAP_si_nanowire} \\
Penn. State           & Si NW                 & 1.928                 & 4830                  &                       & 300                   & 11.8                  & $2.4 \cdot 10^{-15} $ & 0.043                 & \cite{li_natnano_largearea_nanowire} \\
Penn. State           & Rh NW                 & 7.186                 & 1080                  &                       & 0.3                   & 5.8                   & $4.4 \cdot 10^{-15} $ & 0.016                 & \cite{li_natnano_largearea_nanowire} \\
CalTech               & Si NW                 & 96                    & 550                   &                       & 300                   & 1.8                   & $5.3 \cdot 10^{-18} $ & 0.129                 & \cite{he_NL_NW_piezo_mixer} \\
Tsinghua U.           & ZnO NW                & 26                    & 5                     &                       & 300                   & 12                    & $1.9 \cdot 10^{-14} $ & 0.004                 & \cite{zhu_nanotech_ZnO_nanowire} \\
U. Illinois           & Si NW                 & 0.208                 & 10000                 &                       & 300                   & 14.4                  & $5.1 \cdot 10^{-17} $ & 0.887                 & \cite{nichol_APL_nanowire_interferometer} \\
U. Mich.              & SnO$_2$ NW            & 59                    & 2200                  &                       & 300                   & 2.5                   & $2.2 \cdot 10^{-16} $ & 0.025                 & \cite{fung_APL_nanowire} \\
Madrid                & Si NW                 & 2.2                   & 2000                  &                       & 300                   & 7.5                   & $5.5 \cdot 10^{-16} $ & 0.083                 & \cite{gil_natnano_nanowire} \\
\midrule
Cornell               & SLG                   & 70.5                  & 78                    &                       & 300                   & 1.1                   & $1.4 \cdot 10^{-18} $ & 0.287                 & \cite{bunch_science_grapheneresonators} \\
ICN                   & FLG                   & 32                    & 64                    &                       & 300                   & 2.8                   & $2.7 \cdot 10^{-17} $ & 0.098                 & \cite{garcia_NL_imaging_graphene} \\
Cornell               & FLG                   & 160                   & 25                    &                       & 300                   & 4.75                  & $4.5 \cdot 10^{-16} $ & 0.011                 & \cite{bunch_NL_impermeable} \\
NRL                   & FLGO                  & 57.6                  & 3000                  &                       & 300                   & 2.75                  & $7.8 \cdot 10^{-17} $ & 0.043                 & \cite{robinson_NL_graphene_oxide} \\
Cornell               & FLG                   & 8.36                  & 97                    &                       & 300                   & 8                     & $3.5 \cdot 10^{-17} $ & 0.169                 & \cite{shivaraman_NL_expitaxial_graphene} \\
Columbia              & SLG                   & 130                   & 125                   & 14000                 & 5                     & 3                     & $2.2 \cdot 10^{-18} $ & 0.169                 & \cite{chen_natnano_graphene_mixing} \\
TIFR                  & SLG                   & 64                    & 250                   & 2100                  & 7                     & 2.5                   & $1.9 \cdot 10^{-17} $ & 0.083                 & \cite{singh_NT_graphene_mixing} \\
\midrule
TIFR                  & NbSe$_2$              & 24                    &                       & 215                   & 10                    & 3.3                   & $1.8 \cdot 10^{-15} $ & 0.014                 & \cite{sengupta_PRB_CDW_NEMS} \\
\bottomrule
\end{tabular}

\end{centering}}
\end{table}

\rowcolors{1}{white}{white}


%
\begin{figure}[tb]
\centering
\includegraphics{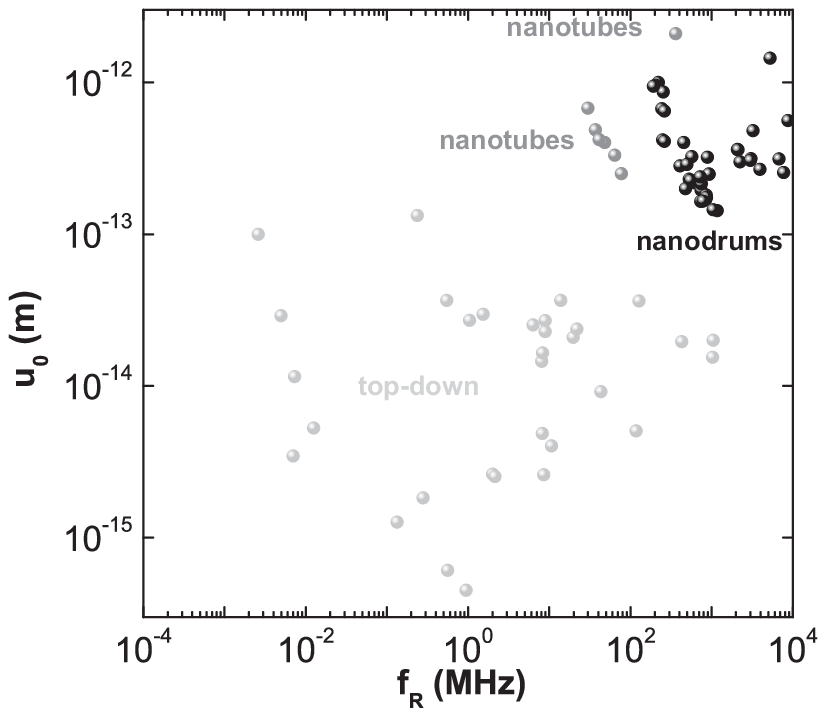}
\caption{Comparison of the resonance frequency $f_R$ and
zero-point motion $u_0$ of top-down (Table \ref{tab:intro:short})
and two kinds of bottom-up devices: Suspended carbon nanotubes
\cite{witkamp_NL_bendingmode, huettel_NL_highQ,
steele_science_strong_coupling} and graphene nanodrums
\cite{poot_APL_nanodrums}. The resonance frequencies and
zero-point motion are much larger for the bottom-up devices.
\label{fig:intro:topdown_bottomup}}
\end{figure}

Table \ref{tab:intro:bottomup} shows the properties of
\deleted{the} mechanical resonators that have been made so far
using bottom-up fabricated devices. Their frequencies are high:
by choosing the right device geometry resonances in the UHF
band (300 MHz - 3 GHz) are readily made, as Table
\ref{tab:intro:bottomup} shows. When comparing the quality
factors and zero-point motion of these devices, it is clear
that the nanowire performance is more or less comparable to
top-down fabricated devices since their thickness is of the
order of 10-100~nm, about the size of the smallest top-down
fabricated devices.

Due to their low mass $m$ and high strength (see the mechanical
properties listed in Table \ref{tab:nems:materials} in Sec.
\ref{sec:nems}) the frequencies of carbon-based resonators are
high and their zero-point motion $u_0$ large. Note, that in
Table \ref{tab:intro:short} $u_0$ was given in femtometer,
whereas in Table \ref{tab:intro:bottomup} it is given in
picometer. Figure \ref{fig:intro:topdown_bottomup} illustrates
this point more clearly. The quantum regime with a large $u_0$
and small $\overline n$ is positioned in the upper right
corner. When cooled to dilution refrigerator temperatures,
carbon-based resonators would therefore be in the ground-state
while exhibiting relatively large amplitude zero-point
fluctuations.

Position detectors for bottom-up NEMS, however, are not yet as
sophisticated as those for the larger top-down counterparts.
Consequently, non-driven motion at cryogenic temperatures
(\replaced{i.e.,}{either} Brownian or zero-point motion) nor
active cooling has been reported for carbon-based NEMS. The
devices always need to be actuated to yield a measurable
response. This is at least in part due to their small size,
which makes the coupling to the detector small as well.
Nevertheless impressive progress in understanding the
electromechanical properties of bottom-up resonators has been
made in recent years using so-called self-detecting schemes. In
these schemes, the nanotube both acts as the actuator and
detector of its own motion. Ultra-high quality factors have now
been demonstrated for carbon nanotube resonators at low
temperatures \cite{huettel_NL_highQ} as well as a strong
coupling between electron transport and mechanical motion
\cite{steele_science_strong_coupling,
lassagne_science_coupled_cnt}.

\subsection{Carbon-based materials}
\label{ssec:nems:carbonbased} In this Report we pay special
attention to the mechanics of bottom-up materials as this has not
been reviewed in such detail as the mechanics of the silicon-based
top-down devices. In particular we will focus on the carbon-based
materials as they have extraordinary mechanical and electrical
properties.

\begin{figure}[tb]{
\centering
\includegraphics{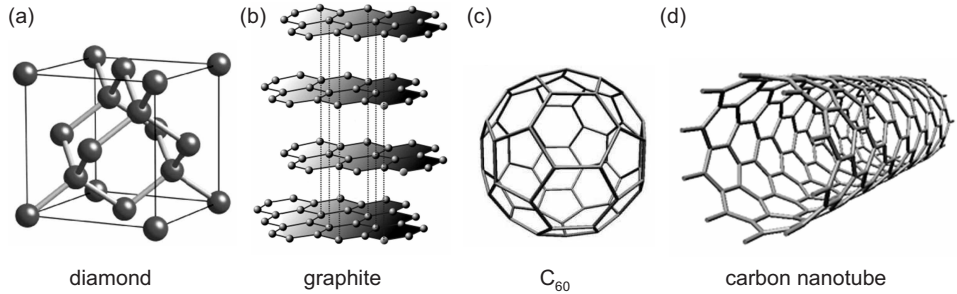}
\caption{The structure of the
different allotropes of carbon. (a) Diamond has two intertwined
face-centered cubic lattices. (b) Graphite consists of stacked
planes of hexagonally ordered carbon atoms. A single plane is
called a ``graphene sheet''. (c) A $C_{60}$ buckyball molecule.
(d) A single-walled carbon nanotube, which can be viewed as a
graphene sheet that has been rolled up and sewn together. (c) and
(d)
Reprinted from Nanomedicine: Nanotechnology, Biology and
Medicine, {\bf 4} 3, M. Foldvari, M. Bagonluri, Carbon
nanotubes as functional excipients for nanomedicines: I.
pharmaceutical properties, 173--182, Copyright (2008), with
permission from Elsevier.}\label{fig:intro:allotropes}}
\end{figure}

Carbon exists in many different forms, ranging from amorphous
coal to crystalline graphite and diamond. Diamond has a
face-centered cubic structure as shown in Fig.
\ref{fig:intro:allotropes}a and is one of the hardest materials
known. Its Young's modulus (Table \ref{tab:nems:materials}) is
extremely high: about $1 \un{TPa}$. Graphite has a very
different crystal structure: it consists of stacked planes of
carbon atoms in a hexagonal arrangement (Fig.
\ref{fig:intro:allotropes}b). Its Young's modulus for in-plane
stress is nearly as high as that of diamond, but it is much
lower for out-of-plane stress, as Table
\ref{tab:nems:materials} will indicate. This difference is
caused by the nature of the bonds holding the carbon atoms
together. Atoms in one of the planes are covalently bonded to
each other, whereas different planes are held together by the
much weaker van der Waals force.

Graphite and diamond were already known for millennia, but in
the last decades novel allotropes of carbon were discovered.
First, in 1985 $C_{60}$ molecules, called
\added{Buckminsterfullerenes or ``}buckyballs'', were
synthesized \cite{kroto_nature_C60}. Then in the early 1990s
carbon nanotubes were discovered \cite{iijima_nature_cnt}.
These consist of cylinders of hexagonally ordered carbon atoms;
similar to what one would get if one were to take a single
layer of graphite and role it up into a cylinder. In 2004
another allotrope called ``graphene'' was identified
\cite{novoselov_science_atomicthin,
novoselov_PNAS_2d_crystals}. This is a single layer of
graphite, which is, unlike a nanotube, flat. Graphene is
usually deposited onto a substrate using mechanical
exfoliation, and high-quality sheets of $\un{mm}$-size have
been made using this technique \cite{geim_science_status}.
Another way of making graphene devices is to grow it directly
on a substrate, and in a semi-industrial process meter-sized
sheets have been reported
\cite{bae_natnano_graphene_touch_screen}. Although truly
two-dimensional structures are not energetically stable
\cite{LL_statistical_pysics}, graphene can exist due to fact
that it contains ripples that stabilize its atomically thin
structure \cite{meyer_nature_TEM, fasolino_NM_ripples,
liu_APL_bending_rigidity, gazit_PRB_buckling_graphene}. Figure
\ref{fig:intro:allotropes} shows the structure of the four
different carbon allotropes.

Although nanomechanical devices have been made out of diamond
using top-down fabrication techniques
\cite{sekaric_APL_diamond_NEMS, wang_IEEE_GHZ_diamond,
imboden_APL_diamond, gaidarzhy_APL_diamond_GHz}, in this Review
we focus on bottom-up fabricated carbon-based NEMS made from
few-layer graphene or suspended carbon nanotubes.

\subsection{Outline of this review}
This Review consists of this introduction, three main Sections,
one on nanomechanics, one on backaction and cooling, and the
last one focusses on different types of detectors. These are
supplemented with a final Section summarizing some prospects
and future directions in the field of quantum electromechanical
systems or QEMS in short. In Section \ref{sec:nems} we will
discuss mechanics at the nanoscale. From continuum mechanics,
the general equations of motion are obtained. These are
illustrated by studying the dynamics of a number of
nanomechanical devices, such as beam and string-like
resonators, buckled beams and suspended carbon nanotubes. We
will demonstrate that the dynamics of a particular vibrational
mode of the resonator is that of a harmonic oscillator. Another
important point is that in nanoscale devices, tension is a
crucial property that must be included in the analysis. As we
will show it also provides a unique tool for the tuning of
resonators when their thickness is on the atomic to nanometer
scale.

In Section \ref{sec:basic} we discuss the properties of the
(quantum) harmonic oscillator and we study the effects of
backaction. A measurement always influences the measured object
itself and this is called backaction. It has important
consequences for linear detection schemes, and we will show
that backaction ultimately limits their position resolution to
what is known as the standard quantum limit. This limit will be
derived in several ways. Backaction, however, can also be used
to one's advantage as it provides a way to cool resonator
modes. This can be done by backaction alone (self-cooling).
Alternatively, the resonator temperature can be lowered by
adopting active cooling protocols. An overview of the two
\replaced{most popular}{main} protocols will be given,
including a summary of the main achievements.

In general, two distinct \replaced{approaches}{methods} for
position detection of mechanical oscillators are used: optical
and electrical. In Sec. \ref{sec:detectors}, we give an
overview of the different detection schemes. We will discuss
the use of optical cavities in various forms, optical
waveguides and their analogues in solid state devices in which
electrons play the role of photons. In addition, we will
summarize the concepts behind capacitive and inductive
actuation and detection as well as the self-detection schemes
that are used in bottom-up NEMS devices. Special attention will
be paid to the achieved position resolution and to the
limitations that prevent the standard quantum limit to be
reached. Furthermore, we explain the mechanisms behind the
backaction and, when possible, quantify the coupling between
resonator and detector.

\section{Mechanics at the nanoscale} \label{sec:nems}
\replaced{Nearly, a}{A}ll objects, including macroscopic and
nanometre-sized systems, have particular frequencies at which
they can resonate when actuated. These frequencies are called
the eigenfrequencies or normal frequencies; when actuated at an
eigenfrequency, i.e. on resonance, the amplitude of vibration
can become very large and this can have catastrophic
consequences. Examples include the breaking of a glass by sound
waves and the collapse of the Tacoma bridge which was set in
motion by the wind flowing around it.

Continuum mechanics provides the tools to calculate these
frequencies. For simple geometries such as cantilevers,
doubly-clamped beams or thin plates, the frequencies of
flexural or torsional modes can be calculated analytically.
However, in many NEMS experiments more complicated geometries
are used, such as
microtoroids~\cite{schliesser_AP_toroid_overview}, suspended
photonic crystal
structures~\cite{eichenfield_nature_photonic_crystal,
eichenfield_nature_optomechanical_crystals}, or film bulk
acoustic resonators (FBARs)
\cite{oconnell_nature_quantum_piezo_resonator,
dunn_APL_modecoupling} and the vibrational mode may have more a
complicated shape than the simple flexural or torsional motion.
To calculate the eigenfrequencies and mode shapes in these
cases one has to rely on numerical calculations. Popular
implementations include the software packages ANSYS and COMSOL
which are based on the finite-element method. These packages
have additional advantages as they provide a means to model
other properties of the nanomechanical system: This includes
electrostatic interactions (see e.g. Refs.
\cite{erbe_PRL_shuttle, witkamp_APL_paddle,
unterreithmeier_nature_dielectric}), thermal effects
\cite{wiederhecker_nature_spoke_disks,
schliesser_AP_toroid_overview}, and electromagnetic (optical)
properties~\cite{li_nature_harnessing_forces,
eichenfield_nature_photonic_crystal,
pernice_OE_modeling_waveguides}.

In this Section, we will summarize some main results of
continuum mechanics needed to describe the experiments
discussed in this Report. We will review analytical expressions
for the eigenfrequencies of simple structures such as beams,
buckled beams and strings. This material has been described in
\replaced{several}{many} textbooks (see e.g. the book by A.
Cleland \cite{cleland_nanomechanics}). Less attention has been
devoted to thin beams or plates with nanometer-sized cross
sections made from, for example, carbon nanotubes or graphene.
These resonators are in a different regime than top-down
devices that generally have larger sizes. In particular, the
deflection of these nano-resonators can exceed their thickness
or radius so that tension-induced nonlinear effects start to
play a role. We derive the equations for describing these
nonlinear effects in nanobeams and show that the induced
tension can be used to tune the frequencies over a large range.
We end this section with a discussion on the mechanics of
(layered) graphene resonators that can be viewed as miniature
drums.

\subsection{Continuum mechanics}
\label{ssec:nems:continuum} To describe the motion of mechanical
objects, the dynamics of all particles (i.e., atoms and electrons)
which make up the oscillator should, in principle, be taken into
account. It is, however, known that for large, macroscopic objects
this is unnecessary and that materials can be accurately described
as a continuum with the mechanical behavior captured by a few
parameters such as the elasticity tensor. Molecular dynamics
simulations \cite{dequesnes_nanotech_MD_nanotube,
pattrick_PRB_buckling_MD, reddy_nanotech_MD_graphene,
samadikhah_MRS_graphene_MD, scarpa_nanotech_MD_graphene,
neek_arXiv_nanoindentation_MD} and experiments
\cite{witkamp_NL_bendingmode, lefevre_PRL_scaling,
poot_APL_nanodrums} demonstrate that even for nanometer-sized
objects continuum mechanics is, with some modifications, still
applicable. This means that the dynamics of the individual
particles is irrelevant when one talks about deflections and
deformations; the microscopic details do, however, determine the
material properties and therefore also the values of macroscopic
quantities like the Young's modulus or the Poisson ratio.

The basis of continuum mechanics lies in the relations between
strain and stress in a material. The strain tells how the
material is deformed with respect to its relaxed state. After
the deformation of the material, the part that was originally
at position $\vc{x}$ is displaced by $\vc{u}$ to its new
location $\vc{x}+\vc{u}$. The strain describes how much an
infinitesimal line segment is elongated by the deformation
$\vc{u}(x,y,z)$ and is given by\footnote{In this Report, the
so-called Einstein notation \cite{chung_continuum} for the
elements of vectors and tensors is employed. When indices
appear on one side of an equal sign only, one sums over them,
without explicitly writing the summation sign. For example,
$x_i = R_{ij}x_j$ reads as $x_i = \sum_{j=1}^3 R_{ij}x_j$. The
index runs over the three cartesian coordinates $(x,y,z)$,
where $x_1 = x$, $x_2 = y$ and $x_3 = z$. Finally, the symbols
$\vc{\hat x}_i$ ($\vc{\hat x}$, $\vc{\hat y}$ and $\vc{\hat
z}$) denote the unit vectors in the fixed rectangular
coordinate system (which form a basis) so that a vector
$\vc{r}$ can be expressed as $\vc{r} = r_i \vc{\hat x_i}$.}
\cite{chung_continuum}:
\begin{equation}
\strain_{ij} = \half\left(\pder{u_i}{x_j} + \pder{u_j}{x_i} +
\pder{u_m}{x_i}\pder{u_m}{x_j} \right).\label{eq:nems:strain}
\end{equation}
This definition shows that strain is symmetric under a reversal
of the indices, i.e., $\strain_{ij} = \strain_{ji}$. The
diagonal elements (i.e., $i = j$) of the first two terms are
the normal strains, whereas the off-diagonal elements ($i \ne
j$) are the shear strains. Eq. \ref{eq:nems:strain} is exact,
but the last, non-linear term is only relevant when the
deformations are large \cite{cleland_nanomechanics} and will
not be considered in this work.

\begin{figure}[tb]{
\centering
\includegraphics{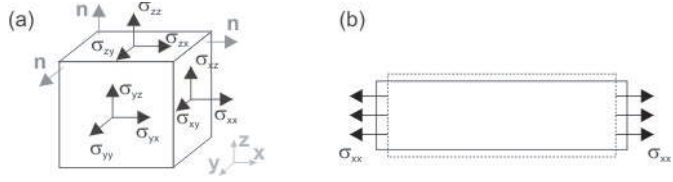}
\caption{(a) Visualization of the stress tensor on a cubic element
$\Delta V$. The force per unit area is the inner product of
the stress tensor and the normal vector of the surface $\vc{n}$.
(b) Deformation of a plate under plane stress. The original plate
(dotted) is deformed by the stress $\sigma_{xx}$.
\label{fig:nems:continuum}}}
\end{figure}

To deform a material external forces have to be applied, which
in turn give rise to forces inside the material. When the
material is thought of as composed of small elements, each
element feels the force exerted on its faces by the neighboring
elements. The magnitude and direction of the force depend not
only on the location of the element in the material but also on
the {\it orientation} of its faces (see Fig.
\ref{fig:nems:continuum}a). The force $\vc{\delta F}$ on a
small area $\delta A$ of the element is given by:
\begin{equation}
\delta F_i = \sigma_{ij} n_j \delta A,
\end{equation}
where $\vc{n}$ is the vector perpendicular to the surface and
$\bm{\sigma}$ is the stress tensor. Now consider an element of
the material with mass $\Delta m$ and volume $\Delta V$. When
it is moving with a speed $\vc{v} = \vc{\dot u}$, its momentum
$\vc{\Delta p}$ is:
\begin{equation}
\vc{\Delta p} \equiv \int_{\Delta m}
\vc{v} \intd m = \int_{\Delta V} \rho \vc{v} \intd V,
\end{equation}
where $\rho$ is the mass density. The rate of change of
momentum equals the sum of the forces working on the element.
These forces include the stress $\bm{\sigma}$ at the surface
and body forces $\vc{F_b}$ that act on the volume element
$\Delta V$:
\begin{equation}\label{eq:nems:cauchy1int}
\tder{\vc{\Delta p}}{t} = \int_{\Delta V} \vc{F_b} \intd V
+ \int_{\delta (\Delta V)} \bm{\sigma} \intd A.
\end{equation}
Examples of body forces are gravity, with $\vc{F_b} = \rho g
\vc{\hat z} $ and the electrostatic force $q \bm{\mathcal{E}}$,
where $g$ is the gravitational acceleration, $q$ is the charge
density, and $\bm{\mathcal{E}}$ is the local electric field.
Using the Green-Gauss theorem, the integral over the boundary
of the element can be converted into an integral over the
volume: $\int_{\delta (\Delta V)} \bm{\sigma} \intd A =
\int_{\Delta V} \pderl{\sigma_{ij}}{x_i}
\replaced{\cdot}{\times} \vc{\hat x}_i \intd V$. Eq.
\ref{eq:nems:cauchy1int} should hold for {\it any} element
because so far nothing has been specified about the shape or
size of the element. This then yields Cauchy's first law of
motion \cite{chung_continuum}:
\begin{equation} \label{eq:nems:cauchy1}
\rho \ddot u_j = \pder{\bm{\sigma}_{ij}}{x_i} + F_{b,j}.
\end{equation}
A similar analysis for the angular momentum yields Cauchy's
second law of motion: $\sigma_{ij} = \sigma_{ji}$. With these
equations (and boundary conditions) the stress distribution can
be calculated for a given applied force profile
$\vc{F_b}(x,y,z)$.

\subsection{Elasticity}
\label{ssec:nems:elasticity} The stress tensor gives the forces
acting inside the material, whereas the strain tensor describes
the local deformation of the material. These two quantities
are, of course, related to each other. When the deformations
are not too large, the stress and strain tensor are related
linearly via the the fourth-rank elasticity tensor $\ten{E}$:
\begin{equation} \label{eq:nems:hooke_tensor}
\sigma_{ij} = E_{ijkl}\strain_{kl}
\end{equation}
The properties of the stress and strain tensor imply that
$E_{ijkl} = E_{jikl} = E_{ijlk} = E_{klij}$, so $\ten{E}$ has at
most 21 independent elements out of a total of $3\times 3\times
3\times 3=81$ elements. This makes it possible to express Eq.
\ref{eq:nems:hooke_tensor} in a convenient matrix representation:
\begin{equation} \label{eq:nems:hooke_matrix}
\left[\begin{array}{c}
\sigma_{xx} \\
\sigma_{yy} \\
\sigma_{zz} \\
\sigma_{xz} \\
\sigma_{yz} \\
\sigma_{xy} \\
\end{array}\right] =
\left[\begin{array}{cccccc}
E_{xxxx} & E_{xxyy} & E_{xxzz} & E_{xxxz} & E_{xxyz} & E_{xxxy} \\
E_{xxyy} & E_{yyyy} & E_{yyzz} & E_{yyxz} & E_{yyyz} & E_{yyyx} \\
E_{xxzz} & E_{yyzz} & E_{zzzz} & E_{zzzx} & E_{zzzy} & E_{zzxy} \\
E_{xxxz} & E_{yyxz} & E_{zzzx} & E_{xzzx} & E_{xzzy} & E_{yxxz} \\
E_{xxyz} & E_{yyyz} & E_{zzzy} & E_{xzzy} & E_{yxxz} & E_{xyyz} \\
E_{xxxy} & E_{yyyx} & E_{zzxy} & E_{yxxz} & E_{xyyz} & E_{xyyx} \\
\end{array}\right]
\left[\begin{array}{c}
\strain_{xx} \\
\strain_{yy} \\
\strain_{zz} \\
2\strain_{xz} \\
2\strain_{yz} \\
2\strain_{xy} \\
\end{array}\right],
\end{equation}
or in short hand notation\footnote{\added{Note that there are
three different notations for the elasticity tensor: $\ten{E}$
is the actual \emph{tensor} with \emph{elements} $E_{ijkl}$.
Finally, the elements can also be written in a \emph{matrix}
$[E]$.}}: $[\sigma] = [E][\strain]$. The inverse of the
elasticity tensor is called the compliance tensor $\ten{C}$,
which expresses the strain in terms of the stress:
\begin{equation}
\strain_{ij} = C_{ijkl}\sigma_{kl} \text{, or~} [\strain] = [C]
[\sigma].
\end{equation}
The number of independent elements of $\ten{E}$ is further
reduced when the crystal structure of the material has
symmetries \cite{cleland_nanomechanics, newnham_materials,
LL_elasticity}. The most drastic example is an isotropic
material, whose properties are the same in all directions. In
this case, only two independent parameters remain: the Young's
modulus $E$ and Poisson's ratio $\nu$. The compliance matrix is
in this case given by:
\begin{equation}
[C] = \left[\begin{array}{cccccc}
1/E     & -\nu/E    & -\nu/E& 0 & 0 & 0 \\
-\nu/E   & 1/E      & -\nu/E& 0 & 0 & 0 \\
-\nu/E  & -\nu/E    & 1/E   & 0 & 0 & 0 \\
0       & 0         & 0     & 1/G & 0 & 0 \\
0       & 0         & 0     & 0 & 1/G & 0 \\
0       & 0         & 0     & 0 & 0 & 1/G \\
\end{array}\right] \hspace{1cm} G = \frac{E}{2+2\nu},
\end{equation}
where $G$ is the shear modulus. By inverting $[C]$, the
elasticity matrix is obtained:
\begin{equation}
[E] = \frac{1}{(1+ \nu)(1 - 2\nu^2)}\left[\begin{array}{cccccc}
E(1-\nu)& E\nu      & E\nu      & 0 & 0 & 0 \\
E\nu    & E(1-\nu)  & E\nu      & 0 & 0 & 0 \\
E\nu    & E\nu      & E(1-\nu)  & 0 & 0 & 0 \\
0       & 0         & 0         & G & 0 & 0 \\
0       & 0         & 0         & 0 & G & 0 \\
0       & 0         & 0         & 0 & 0 & G \\
\end{array}\right].
\label{eq:nems:elasticity_isotropic}
\end{equation}

When a plane stress $\sigma_{xx}$ is applied, a material will
be stretched in the x-direction, as illustrated in Fig.
\ref{fig:nems:continuum}b. The resulting strain $\strain_{xx} =
\sigma_{xx}/E$ induces stress in the y- and z-directions, which
are nulled by a negative strain (i.e., contraction) in these
directions that is $\nu$ times smaller than the strain in the
x-direction. This follows directly from the structure of the
compliance matrix. In the opposite situation where a plane
strain is applied, the stress can directly be calculated using
the elasticity matrix, Eq. \ref{eq:nems:elasticity_isotropic}
\cite{chung_continuum}. The Young's modulus and Poisson's ratio
of materials that are frequently used for nanomechanical
devices are indicated in Table \ref{tab:nems:materials}. The
Young's modulus of most semiconductors (Si, GaAs, InAs),
insulators (Si$_3$N$_4$, SiC, SiO$_2$) and metals is of the
order of $100 \un{GPa}$. This is much larger than the values
for soft materials such as polymers (typically between 0.1 and
1 GPa) which have also been used for nanomechanical devices
\cite{yamazaki_JJAP_pmma_nems}, but still smaller than that of
diamond and graphite. Their Young's modulus of slightly less
than 1 TPa combined with a low mass density makes these
carbon-based materials ideal to build high-frequency
resonators. Also, the large spread in the mass density of the
metals should be noted.
%
\begin{table}[t]
\centering
\caption {Mechanical properties of materials that are used in
nanomechanical devices. Most materials have a density $\rho$
around $3\cdot 10^3\un{kg/m^3}$ and a Young's modulus of the order
of $10^2 \un{GPa}$. The carbon-based materials graphite and
diamond are slightly lighter, but much stiffer. Compiled from
Refs. \cite{cleland_nanomechanics} and \cite{lide_chemphys}.}
\begin{tabular}{lrrr}
\addlinespace \toprule
Material                  & $\rho~(10^3 \un{kg/m^3})$ & $E~(\unitt{GPa})$            & $\nu$~~~~ \\
\midrule
Silicon                   &        2.33               &      130.2                &        0.28~~ \\
Si$_3$N$_4$               &        3.10               & 357~~~                    &        0.25~~ \\
SiC                       &        3.17               &      166.4                &        0.40~~ \\
SiO$_2$ (crystaline)      &        2.65               &        85.0               &        0.09~~ \\
SiO$_2$ (amorphous)       &        2.20               &  $\sim 80~~~$             &        0.17~~ \\
Diamond                   &        3.51               &      992.2                &        0.14~~ \\
Graphite (in-plane)       &        2.20               & 920~~~                    &        0.052 \\
Graphite (out-of-plane)   &        2.20               & 33~~~                     &        0.076 \\
Aluminum                  &        2.70               &        63.1               &        0.36~~ \\
Gold                      &       19.30               &        43.0               &        0.46~~ \\
Platinum                  &       21.50               & 136.3                     &        0.42~~ \\
Niobium                   &        8.57               &      151.5                &        0.35~~ \\
GaAs                      &        5.32               &        85.3               &        0.31~~ \\
InAs                      &        5.68               &        51.4               &        0.35~~ \\
\bottomrule
    \end{tabular}
  \label{tab:nems:materials}
\end{table}

\begin{figure}[tb]{\centering
\includegraphics{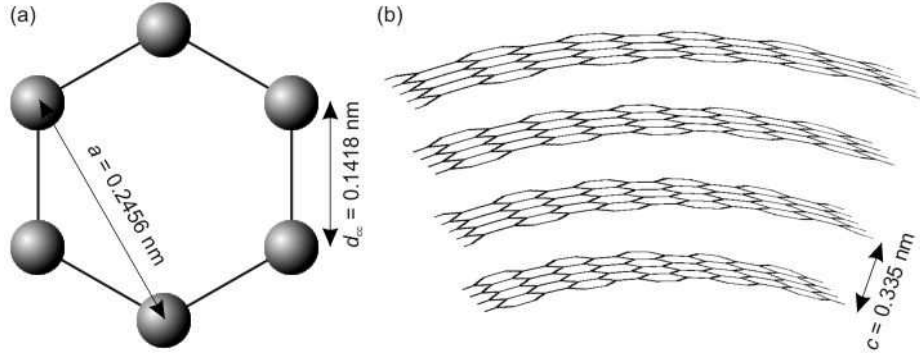}
\caption[The unit cell of graphene and bending of few-layer
graphene]{(a) The unit cell of graphene with the dimensions
indicated. $a$ is the length of the two translation vectors
$\vc{a}_{1,2} = \halfl a [\pm 1, \sqrt{3}, 0]$, and $d_{cc}$ is
the distance between two carbon atoms. The area of the unit cell
is $\halfl \sqrt{3} a^2 = 5.22 \cdot 10^{-20} \un{m^2}$. (b)
Bending of a few-layer graphene sheet. The equilibrium distance between the
graphene layers is $c = 0.335 \un{nm}$
\cite{slonczewsk_PR_grapite}.} \label{fig:nems:graphenesheet}}
\end{figure}

For non-isotropic materials, the Young's modulus and Poisson's
ratio depend on the direction of the applied stress and are
defined as: $E_i = 1/C_{iiii}$ and $\nu_{ij} = -
C_{iijj}/C_{iiii}~(i \ne j) \label{p:nems:poisson}$
\cite{turley_JPhysD_anisotropy}. Graphite is highly
anisotropic, and its mechanical properties are important for
studying many carbon-based nanomechanical devices. It consists
of layers of carbon atoms that are stacked on top of each other
(see Fig. \ref{fig:intro:allotropes}b and
\ref{fig:nems:graphenesheet}b) with an inter-layer spacing $c =
0.335 \un{nm}$ \label{p:nems:graphite}
\cite{slonczewsk_PR_grapite}. The individual layers are called
graphene sheets. The unit cell of graphene consists of a
hexagon with a carbon atom on each corner, as illustrated in
Fig. \ref{fig:nems:graphenesheet}a. Each of the six carbon
atoms lies in three different unit cells; a single unit cell
thus contains two carbon atoms. The sides of the hexagon have a
length $d_{cc} = 0.14 \un{nm}$, so that the unit cell has an
area of $5.22 \cdot 10^{-20} \un{m^2}$, and the two-dimensional
mass density is $\rho_{2d} = 6.8 \cdot 10^{-7} \un{kg/m^2}$
\label{p:nems:rho2d}.
The graphene planes are not located exactly above each other
but every other layer is shifted by half the unit cell, or
equivalently, it is rotated by $60^{\text{o}}$ around an axis
through one of the carbon atoms. Three of the six atoms are on
top of the atoms in the other layer and the other three are
located at the center of the hexagon below them. The six-fold
rotational symmetry ensures that the elastic properties are the
same when looking in any direction along the planes, i.e., they
are isotropic in those directions \cite{newnham_materials,
LL_elasticity}. On the other hand, the mechanical properties
for deformations perpendicular to the planes are quite
different. It is therefore convenient to introduce the in- and
out-of-plane Young's modulus, $E_\boxempty$ and $E_\perp$
respectively, and the corresponding Poisson's ratios
$\nu_\boxempty$ and $\nu_\perp$. They are defined such that the
compliance matrix is given by\footnote{Note that this
definition is slightly different from the conventional
definition of the Poisson's ratio in an anisotropic material
that was given on page \pageref{p:nems:poisson}.}:
\begin{equation}
[C] = \left[\begin{array}{cccccc}
1/E_\boxempty               & -\nu_\boxempty/E_\boxempty    & -\nu_\perp/E_\perp  & 0     & 0     & 0 \\
-\nu_\boxempty/E_\boxempty  & 1/E_\boxempty                 & -\nu_\perp/E_\perp  & 0     & 0     & 0 \\
-\nu_\perp/E_\perp          & -\nu_\perp/E_\perp            & 1/E_\perp           & 0     & 0     & 0 \\
0     & 0     & 0           & 1/G_\boxempty & 0 & 0 \\
0     & 0     & 0           & 0 & 1/G_\boxempty & 0 \\
0     & 0     & 0           & 0 & 0 & 1/G_\perp \\
\end{array}\right],
\label{eq:nems:compliance_graphite}
\end{equation}
where\footnote{The values of the elastic constants depend on
the quality of the graphite samples. Therefore, slightly
different values can be found in the literature. Compare, for
example, the data in Refs. \cite{blakslee_JAP_elasticgrapite}
and \cite{michel_PSSB_elastic_constants} with the values in
Ref. \cite{newnham_materials}} $E_\boxempty = 0.92 \un{TPa}$,
$E_\perp = 33 \un{GPa}$, $G_\boxempty = 1.8 \un{GPa}$, $G_\perp
= 0.44 \un{TPa}$, $\nu_\boxempty = 0.052$ and $\nu_\perp =
0.076$ \cite{newnham_materials}. The large in-plane stiffness
is the reason that carbon nanotubes and graphene have very high
Young's moduli of about 1 TPa, which makes them one of the
strongest materials known. Note, that the\added{se six} elastic
constants are not independent as the in-plane shear modulus is
given by $G_\boxempty = (E_{1111} - E_{1122})/2$ for a material
with hexagonal symmetry.

\subsection{Energy, bending rigidity and tension}
\label{ssec:nems:eulerbernoulli} In the previous Section the
relation between the stress and strain in a material was given.
Here, we focus on the energy needed to deform the material.
From this, the equations of motion are derived. For small
deformations, the potential energy $U$ depends quadratically on
the strain. It should be invariant under coordinate
transformations \cite{chung_continuum}, leading to
\begin{equation}
U = \int_V U' \intd V\ \hspace{5mm} \text{with} \hspace{5mm} U' =
\half E_{ijkl}\strain_{ij} \strain_{kl} \label{eq:nems:energy},
\end{equation}
where $U'$ is the potential energy density. From this
definition it follows that the stress is given by $\sigma_{ij}
=
\partial U' / \partial \strain_{ij}$. For an isotropic material
Eq. \ref{eq:nems:energy} reduces to \cite{LL_elasticity}:
\begin{equation}
U' = \half \frac{E}{1+\nu}\left(\strain_{ij}^2 +
\frac{\nu}{1-2\nu} \strain_{kk}^2 \right).
\label{eq:nems:energy_isotropic}
\end{equation}
Although Eq. \ref{eq:nems:energy} is valid for any mechanical
system in the linear regime, it is not straightforward to analyze
a system this way. Therefore, we focus on two simple and
frequently used geometries where the equation of motion can be
obtained without too much effort, namely plates and beams.

\subsubsection{Plates}
\begin{figure}[tb]{
\centering
\includegraphics{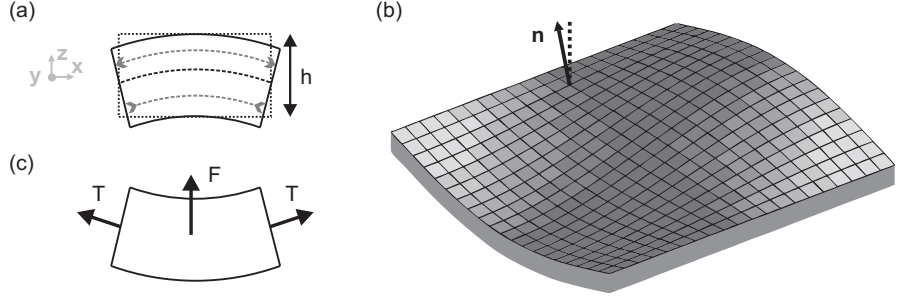}
\caption{(a) Bending of a plate with thickness $h$. The top part
of the plate is extended, whereas the bottom is compressed. The
black dashed line indicates the neutral plane. (b) The normal
vector of the top surface of a slightly deflected plate makes a
small angle with the unit vector $\vc{\hat z}$ (dotted). (c) If
tension $T$ is present in a plate, a net vertical force $\vc{F}$
results when the displacement profile has a finite curvature.
 \label{fig:nems:bending}}}
\end{figure}

A plate is a thin object that is long and wide, i.e., $h \ll
\ell \lesssim w$. When a torque is applied to it, it bends, as
illustrated in Fig. \ref{fig:nems:bending}a. The top part,
which was initially at $z = h/2$, is extended whereas the
bottom part of the beam, originally at $z = -h/2$, is
compressed. There is a plane through the plate where the
longitudinal strain is zero: the so-called neutral plane. The
vertical displacement of this plane is indicated by $u(x,y)$
and for small deflections it lies midway through the plate
\cite{LL_elasticity}, which we take at $z = 0$. Because of the
small deflection and small thickness this is called the
thin-plate approximation.

Consider the top (or bottom) face of the plate that is shown in
Fig. \ref{fig:nems:bending}b: because there is no material
above (below) that surface, there cannot be a normal force
$\vc{F_n}$ at this surface (except at the clamping points). In
other words, the perpendicular stress components vanishes:
$F_{n,i} = \sigma_{ij}n_j = 0$. For a thin plate, the normal
vector $\vc{n}$ at the top and bottom face points in the
z-direction i.e., $\vc{n} = \pm \vc{\hat z}$ to first order in
the displacement $u$ or, equivalently, in the radius of
curvature $R_c^{-1}$. The condition for vanishing stress thus
becomes: $\sigma_{xz} = \sigma_{yz} = \sigma_{zz} = 0$ at the
faces. This  not only holds at the faces but also inside the
material because the plate is thin and the stress cannot build
up. For an isotropic material the displacement and strain
fields that satisfy these requirements are
\cite{LL_elasticity}:
\begin{eqnarray}
u_x = &-z \cdot \pderl{u}{x},   & \strain_{xx} = -z  \cdot \pderl{^2 u}{x^2} \nonumber \\
u_y = &-z \cdot \pderl{u}{y},   & \strain_{yy} = -z  \cdot \pderl{^2 u}{y^2} \nonumber\\
u_z = & u,  & \strain_{zz} =  z\nu/(1-\nu) \cdot \nabla^2 u \nonumber \\
\strain_{xz} = & \strain_{yz} = 0, & \strain_{xy} = -z  \cdot
 \pderl{^2 u}{x \partial y}. \label{eq:nems:fields}
\end{eqnarray}
The vertical displacement of the material $u_z$ is thus equal
to the deflection $u$ for every value of $z$. Besides
\replaced{this}{a} vertical displacement there is also a
horizontal displacement ($u_x$ and $u_y$) induced when $u$
changes. In that case, the material displaces in different
directions above and below the neutral plane (Fig.
\ref{fig:nems:bending}a) as expressed by the proportionality
with $z$. Moreover, the strain components averaged over the
thickness $h$, $\overline \strain_{ij} (x,y)$, are all zero, as
the contributions above and below the neutral plane cancel each
other.

To proceed, Eq. \ref{eq:nems:fields} is inserted into Eq.
\ref{eq:nems:energy_isotropic} and the integration over $z$ in Eq.
\ref{eq:nems:energy} is carried out. This yields the energy needed
to bend the plate:
\begin{equation}
U_B = \frac{Eh^3}{24(1-\nu^2)}\int \!\!\! \int \left[
\left(\pder{^2 u}{x^2}+\pder{^2 u}{y^2}\right)^2 +
2(1-\nu)\left(\left\{\pder{^2 u}{x \partial y}\right\}^2 -
\pder{^2 u}{x^2}\pder{^2 u}{y^2}\right) \right] \intd x \intd y
\label{eq:nems:energy_bending}
\end{equation}

It is possible that the plate is not only bent, but that it is
also under a longitudinal tension $\ten{T} = \int \bm{\sigma}
\intd z$ (positive for tensile tension, negative for
compressive tension). The tension is tangential to the surface
and from Fig. \ref{fig:nems:bending}c it is clear that the
longitudinal tension results in a restoring force in the
z-direction when the plate is bent, i.e., when $\pderl{^2
u}{x^2} \neq 0$. The tension deforms the plate, as indicated in
Fig. \ref{fig:nems:continuum}b. The displacement results in a
strain field $\overline \strain_{\alpha \beta} =
\halfl(\pderl{\overline u_\beta}{x_\alpha}+ \pderl{\overline
u_\alpha}{x_\beta} + \pderl{\overline u}{x_\alpha} \cdot
\pderl{\overline u}{x_\beta})$, where Greek indices run over
the x and y coordinate only. Using Eq. \ref{eq:nems:energy} the
work done by applying the tension \replaced{is}{can be}
calculated. The resulting stretching energy is:
\begin{equation}
U_T = \half \int \!\!\! \int \strain_{\alpha\beta} T_{\alpha\beta}
\intd x \intd y, \text{~where~} T_{\alpha\beta} = \int_{-h/2}^{h/2}
\sigma_{\alpha\beta} \intd z \equiv h \overline \sigma_{\alpha\beta}.
\end{equation}
The equation of motion for the vertical deflection of the plate is
obtained when the variation of the total potential energy $U = U_B
+ U_T + U_F$ is considered ($U_F = - \int \!\!\! \int F u \intd
x\intd y$ includes the effect of an external force per unit area
$F$ in the z-direction) for an arbitrary variation in the
displacement $u \rightarrow u + \delta u$. This yields the
equation of motion for the plate \cite{LL_elasticity}:
\begin{equation}
\rho h \pder{^2 u}{t^2} + \left(D\nabla^4 - \pder{}{x_\alpha}
T_{\alpha\beta} \pder{}{x_\beta}\right) u(x, y) = F(x,y),
\label{eq:nems:plate}
\end{equation}
where $\rho$ is the mass density of the material\footnote{In
principle, the first term should read $\rho h \pderl{^2 u}{t^2}
+ \rho h^3\pderl{}{t^2}(\pderl{^2 u}{x^2} + \pderl{^2
u}{y^2})/12$ as the material is also moving in the x and y
direction. These corrections are, however, negligible when
$h/\ell \ll 1 $.}. The first term in Eq. \ref{eq:nems:plate} is
the inertial term and the last term is the external force
acting on the plate. The term in brackets is the force
resulting from the deformation of the plate. Here, $D =
Eh^3/12(1-\nu^2)$ is the so-called bending rigidity of the
plate, the prefactor in Eq. \ref{eq:nems:energy_bending} that
quantifies how much energy it costs to bend a unit area of the
plate. The tension makes Eq. \ref{eq:nems:plate} nonlinear as a
displacement of the plate elongates it and thereby induces
tension, i.e. $T_{\alpha\beta} = T_{\alpha\beta}[u(x',
y')](x,y)$. However, for small deformations the
displacement-induced tension is small and will be overwhelmed
by the bending rigidity or by tension induced by the clamping.
Then the tension is independent of $u$ and Eq.
\ref{eq:nems:plate} is linear.

\subsubsection{Beams}
\begin{figure}[tb]{
\centering
\includegraphics{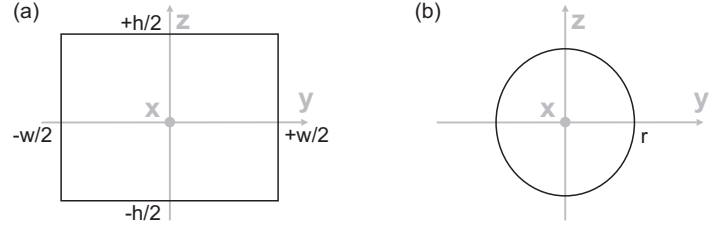}
\caption{Cross-sections of a rectangular beam with thickness $h$
and width $w$ (a) and a cylindrical beam with radius $r$ (b).
 \label{fig:nems:beam}}}
\end{figure}

Table \ref{tab:intro:short}  shows that many nanomechanical
devices are doubly-clamped beams or cantilevers instead of
plates i.e., they have a width that is much smaller then their
length. For a beam, the normal components of the stress on the
sides should also vanish, i.e., $\sigma_{yy} = \sigma_{xy} = 0$
and the integration of Eq. \ref{eq:nems:energy} over $y$ can be
done directly. This implies that the tension $T_{\alpha\beta}$
is only in the x-direction. Moreover, the material displaces in
the x-direction so that $T = T_{xx}$ is independent of $x$.
Combining all of this yields the equation of motion in the
``thin-beam'' approximation: the Euler-Bernoulli equation with
tension:
\begin{equation}
\rho A \frac{\partial^2 u}{\partial t^2} + D\frac{\partial^4
u}{\partial x^4} - T \frac{\partial^2 u}{\partial x^2} = F,
\label{eq:nems:beam}
\end{equation}
where the crosssection area $A$ equals $wh$ for a rectangular
beam and for a cylindrical beam $A = \pi r^2$ (see Fig.
\ref{fig:nems:beam}). The structure of Eq. \ref{eq:nems:beam}
is similar to that of the the equation of motion for a plate,
Eq. \ref{eq:nems:plate}. The first term is the acceleration and
the term on the right-hand side is the external force. The
restoring force due to the bending stiffness depends on the
fourth order derivative of the displacement w.r.t. $x$. For the
tension, this is a second-order derivative. The bending
rigidity\footnote{Note that the units of the bending rigidity,
tension and external force are different from the case of a
plate due to the integration over the y-coordinate. $D$ is
given in $\unit{N m^2}$ instead of $\unit{N m} (=\unit{J})$ and
$T$ is now in $\un{N}$ instead of $\un{N/m}$. The external
force \added{$F$} is given per unit length instead of per unit
area. From the context it should be clear what the meaning of
the different symbols is.} $D = Eh^3w/12(1-\nu^2)$ can be
written as the product of the Young's modulus and the second
moment of inertia $D=EI/(1-\nu^2)$, where the small correction
$(1-\nu^2)$ is often omitted \cite{cleland_nanomechanics}. For
a rectangular beam the second moment of inertia is $I =
h^3w/12$; for a (solid) cylinder with radius $r$ it is $I = \pi
r^4 /4$ \cite{cleland_nanomechanics}. The bending rigidity
depends strongly on the dimensions of the device. It can be as
small as $10^{-24} \unit{N m^2}$ for a carbon nanotube (Sec.
\ref{sssec:nems:nanotubes}) or as large as $10^{-5} \unit{N
m^2}$ for a millimetre-sized mirror
\cite{arcizet_nature_cavity}.
\subsection{Examples}
\subsubsection{Cantilevers and doubly-clamped beams}
\label{sssec:nems:beams} 
\begin{figure}[tb]{
\centering
\includegraphics{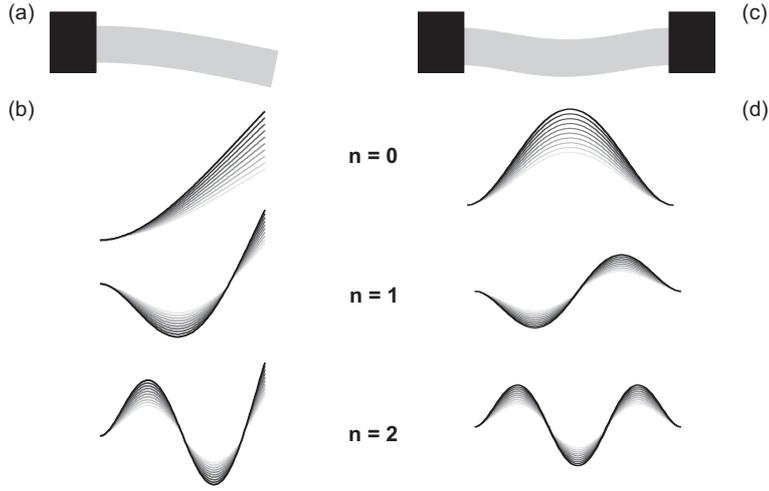}
\caption{Schematic of a singly-clamped cantilever (a) and a
doubly-clamped beam (c). (b) and (d) show the shape of the first
three ($n = 0,1,2$) flexural modes of these flexural resonators.
This shape is calculated using Eqs. \ref{eq:nems:cantilevershape}
and \ref{eq:nems:beamshape} respectively.
\label{fig:nems:cantileverbeam}}}
\end{figure}

A cantilever is a beam that is clamped
on one side ($x = 0$) and free on the other side ($x = \ell$).
Because a cantilever is not fixed on the\added{, say,} right
side, the tension is zero apart from more exotic cases where
electrostatic forces \cite{jensen_NL_CNT_radio,
eriksson_NL_CNT_relay} or surface tension
\cite{gurtin_PMA_surface, lachut_PRL_surface} acts on the free
end. With $T = 0$, the eigenmodes $u_n$ and (angular)
eigenfrequencies $\omega_n$ satisfy:
\begin{equation}
\omega_n^2 \rho A  u_n = D\frac{\partial^4 u_n}{\partial x^4}.
\label{eq:nems:eigenequation}
\end{equation}
The solution to this equation is a linear combination of the
regular and hyperbolic sine and cosine functions ($\sin(kx)$,
$\cos(kx)$, $\sinh(kx)$, and $\cosh(kx)$ resp.), where $k^4 =
\omega^2 D/\rho A$. Their coefficients are determined by the
boundary conditions. At the fixed end, the displacement is zero
and the beam is horizontal: $u(0) = 0$ and $u'(0) = 0$, where
$'$ denotes differentiation with respect to $x$. At the free
end, the force in the z-direction and the torque vanish, so
$u''(\ell) = 0$ and $u'''(\ell) = 0$. There are thus four
boundary conditions and four unknown coefficients. This system
always has a trivial solution $u_n = 0$ where all four
coefficients are zero. There are, however, certain values $k =
k_n$ where one of the four boundary conditions is automatically
satisfied\replaced{. These values}{and that} correspond to the
eigenmodes\added{ of the flexural resonator}. Using the other
three boundary conditions, three of the coefficients are
expressed in the fourth one (which we call $c_4$). For a
cantilever we define $\alpha_n \equiv k_n \ell$ and the n-th
eigenmode is:
\begin{equation}
u_n(x) = c_4 \left(\sin(\alpha_n \frac{x}{\ell}) - \sinh(\alpha_n
\frac{x}{\ell}) - \frac{\sin(\alpha_n) +
\sinh(\alpha_n)}{\cos(\alpha_n) + \cosh(\alpha_n) }\left[
\cos(\alpha_n \frac{x}{\ell}) - \cosh(\alpha_n \frac{x}{\ell})
\right] \right). \label{eq:nems:cantilevershape}
\end{equation}
Note, that if $u_n(x)$ is an eigenmode of the cantilever, then $c
\cdot u_n(x)$ is one as well, for every value of $c$. In the
following part we will therefore use the normalized eigenfunctions
$\xi_n$, which satisfy $\ell^{-1}\int_0^\ell \xi_n^2 \intd x = 1$.
The eigenfrequencies are given by:
\begin{equation}
\cos(\alpha_n) \cosh(\alpha_n) + 1 = 0, ~~~~ \omega_n = 2\pi f_n
\alpha_n^2 \ell^{-2} \sqrt{D/\rho A}. \label{eq:nems:cantilever}
\end{equation}
This equation can only be solved numerically. The first few
solutions $\alpha_n$ are indicated in Table
\ref{tab:nems:mode_freqs} and Fig.
\ref{fig:nems:cantileverbeam}b shows the corresponding mode
shapes. Unlike in a string under tension, a type of resonator
that will be discussed in the next section, \replaced{which
has}{where} $f_n/f_0 = (n+1)$, the eigenmodes do not have a
harmonic spectrum. Also, the average displacement of the mode,
defined as $\mu_n = \ell^{-1}\int_0^\ell \xi_n \intd x$,
decreases with increasing mode number $n$. This is because a
part of the cantilever is moving upwards and an other part is
moving downwards.
\begin{table}[t]
\centering \caption{Normalized eigenfrequencies and average
mode deflection $\mu_n$ for the first 5 flexural eigenmodes for
cantilevers and doubly-clamped beams. $\mu_n = \ell^{-1}
\int_0^\ell \xi_n(x) \intd x$ indicates the displacement of the
mode averaged along the length of the resonator per unit
deflection. This number is important to calculate the detection
efficiency for detectors that couple over the whole length of
the beam (p\pageref{pg:basic:average_displacement}). $\alpha_n$
and $\beta_n$ are the solutions of Eq. \ref{eq:nems:cantilever}
and \ref{eq:nems:beam} respectively which determine the
eigenfrequencies. For large $n$ the solutions approach
$\alpha_n \rightarrow (n + 1/2)\pi$ and $\beta_n \rightarrow (n
+ 3/2)\pi$. \label{tab:nems:mode_freqs}}
\begin{tabular}{c|rrr|rrr}
\addlinespace \multicolumn{1}{c}{\bf{Mode}}    &
\multicolumn{3}{c}{\bf{Cantilever}} &
\multicolumn{3}{c}{\bf{Beam}} \\
\midrule
    $n$ & $\alpha_n$ & $f_n/f_0$ & $\mu_n$ & $\beta_n$ & $f_n/f_0$ & $\mu_n$ \\
\midrule
    0          & 1.875      & 1.000      & 0.783      & 4.730      & 1.000      & 0.831 \\
    1          & 4.694      & 6.267      & 0.434      & 7.853      & 2.757      & 0 \\
    2          & 7.855      & 17.547     & 0.254      & 10.996     & 5.404      & 0.364 \\
    3          & 10.996     & 34.386     & 0.182      & 14.137     & 8.933      & 0 \\
    4          & 14.137     & 56.843     & 0.141      & 17.279     & 13.344     & 0.231 \\
\bottomrule
\end{tabular}
\end{table}

The analysis for the flexural eigenmodes of a doubly-clamped
beam, which does not have a free end but which is clamped at
both $x = 0$ and $x = \ell$, closely follows that for a
cantilever. The difference is that the boundary conditions at
$x = \ell$ are now $u(\ell) = 0$ and $u'(\ell) = 0$. With
$\beta_n \equiv k_n \ell$ this yields:
\begin{equation}
u_n(x) = c_4 \left(\sin(\beta_n \frac{x}{\ell}) - \sinh(\beta_n
\frac{x}{\ell}) - \frac{\sin(\beta_n) +
\sinh(\beta_n)}{\cos(\beta_n) + \cosh(\beta_n) }\left[
\cos(\beta_n \frac{x}{\ell}) - \cosh(\beta_n \frac{x}{\ell})
\right] \right), \label{eq:nems:beamshape}
\end{equation}
and
\begin{equation} \cos(\beta_n) \cosh(\beta_n) - 1 = 0, ~~~~
\omega_n = 2\pi f_n = \beta_n^2 \ell^{-2} \sqrt{D/\rho A}.
\label{eq:nems:beammode}
\end{equation}
Table \ref{tab:nems:mode_freqs} shows that \added{$\beta_n >
\alpha_n$, so that} the eigenfrequencies of a
\added{clamped-clamped} beam are higher than that of a
cantilever with the same dimensions. This is because the
additional clamping makes it stiffer. Moreover, due to symmetry
$\mu_n$ vanishes for the odd modes of a beam. Figure
\ref{fig:nems:cantileverbeam} shows the first three mode
shapes.

Equations \ref{eq:nems:cantilever} and \ref{eq:nems:beammode} show
that when the resonator is made shorter while its transverse
dimensions are kept the same, the eigenfrequencies increase due to
the $\ell^{-2}$ term in the characteristic frequency scale
$\omega_{ch} \equiv \ell^{-2} \sqrt{D/\rho A}$
\cite{cleland_APL_AlN, babei_gavan_JMM_undercut}. On the other
hand, the crosssectional area and the bending rigidity depend on
the size of the resonator. The exact scaling is $\omega_{ch}
\propto h/\ell^2$ and $\omega_{ch} \propto r/\ell^2$ for a
rectangular and cylindrical resonator respectively . When making
all dimensions of a resonator smaller \cite{li_natnano_piezo}, the
resonance frequencies thus increase inversely proportionally with
the size. This scaling is the reason that nanomechanical
resonators can have very high resonance frequencies of more than 1
GHz \cite{huang_nature_GHz}.

\subsubsection{String resonators} \label{sssec:nems:string}
So far we have not considered the effect of tension in the
examples of flexural resonators. In certain types of materials
there can be so much stress that the bending rigidity hardly
contributes to the restoring force. The resonator is then a string
under tension instead of a beam. Eq. \ref{eq:nems:beam} shows that
this is the case when $T \gg D / \ell^2$, or when this is
rewritten in the (longitudinal) strain $\strain$, $\strain \gg
(h/\ell)^2/12$. Tension is thus more important in resonators with
a large aspect ratio $\ell/h$. The eigenfrequencies of a string
under tension are $f_n = \sqrt{T/\rho A} \times (n+1)/2\ell$ and
the corresponding modeshapes are $\xi_n(x) = \sqrt{2} \sin(\pi n x
/\ell)$.

Tension in the resonator arises when materials with different
thermal expansion coefficients or different lattice constants
are used. An example of the former is given by Regal {\it et
al.} \cite{regal_natphys_cavity}, where a $50 \un{\mu m}$ long
aluminum beam resonator is annealed at $150$ to $350 \un{^o
C}$, thereby increasing the resonance frequency from $237
\un{kHz}$ to $2.3 \un{MHz}$. Stress can also be induced by
growing heterostructures with different lattice constants. If
the layer with the resonator has a smaller lattice constant
that the layer underneath it, then the resonator is strained.
By engineering the heterostucture, different amounts of tension
can be induced in the resonator
\cite{yamaguchi_APL_strained_GaAs,
watanabe_APEX_cooling_strained}. Another way of inducing
tension is by placing the resonator on a flexible substrate
that can be bent. As the top part of the substrate is
elongated, the resonator becomes strained. With this technique,
the resonance frequency of a resonator has been tuned by more
than a factor 5 \cite{verbridge_NL_bending}. The most commonly
used high-stress material is silicon nitride. Under the
appropriate growing conditions thin films with stresses of
$\sim 1 \un{GPa}$ can be obtained \cite{verbridge_JAP_damping}
and resonators made using this material have extremely high
quality factors \cite{verbridge_APL_highQ, fong_APL_SiN_highQ}.
Recently, high-stress SiN has therefore become a very popular
material for nanomechanical resonators
\cite{rocheleau_nature_lown, unterreithmeier_nature_dielectric,
anetsberger_PRA_far_below_SQL, verbridge_NL_bending,
unterreithmeier_PRL_damping}. Another important type of string
resonators are suspended carbon nanotubes. Since in this system
the displacement can readily be of the same order as the radius
of the tube, nonlinear effects are important\replaced{. T}{and
We will treat t}hese systems \added{will be treated} in
\deleted{a separate section,} Sec. \ref{sssec:nems:nanotubes}.

Although high $Q$-values seem to be a general observation for
strained beams, the mechanism behind the increase is not
completely clear. Recall that the Q-factor is proportional to
the ratio between the energy stored in the oscillator and the
energy dissipated per cycle. \replaced{Supported by
experiments, the authors in
Ref~\cite{unterreithmeier_PRL_damping} demonstrate that stress
does not substantially change the dissipation rate in strained
beam mechanical oscillators but rather significantly increases
the elastic energy stored in the resonator. They argue that the
microscopic origin of the damping lies in localized defect
states in the material.}{Supported by experiments, the authors
in Ref.~\cite{unterreithmeier_PRL_damping} argue that stress
does not change the dissipation in strained beam mechanical
oscillators but rather increases the energy stored in the
resonator. The microscopic origin of this energy storage is
unknown but speculations hint at localized defect states in the
material.} In the abovementioned experiments by Regal {\it et
al.} the quality factor of an aluminum beam resonator was
increased by a factor 50 by annealing it
\cite{regal_natphys_cavity}. Although the induced tension also
increased the resonance frequency by a factor 10, the quality
factor increased more. This means that in this case the
annealing did not just increase the energy in the resonator,
but it actually reduced the damping rate ($\gamma_R =
\omega_R/Q$), which is the ratio of the resonance frequency
$\omega_R$ and the quality factor. Also in this system the
physical mechanism behind this remarkable increase is still
unclear.

\subsubsection{Buckled beams} \label{sssec:nems:buckling}
In the case of a string resonator the tension is tensile, i.e.,
the resonator is elongated. It is also possible that the resonator
is compressed. In that case, the tension is negative. When the
tension exceeds a critical value, it is energetically favorable
for the beam to have a non-zero flexural displacement; the beam
buckles. Buckled beams form an important class of (nano)mechanical
resonators, and can be formed when the resonator consists of
layers of different materials \cite{carr_IEEE_buckled_beams,
lawrence_NJP_buckled, etaki_natphys_squid}, or using thermal
expansion \cite{roodenburg_APL_buckled}.
\begin{figure}[tb]{
\centering
\includegraphics{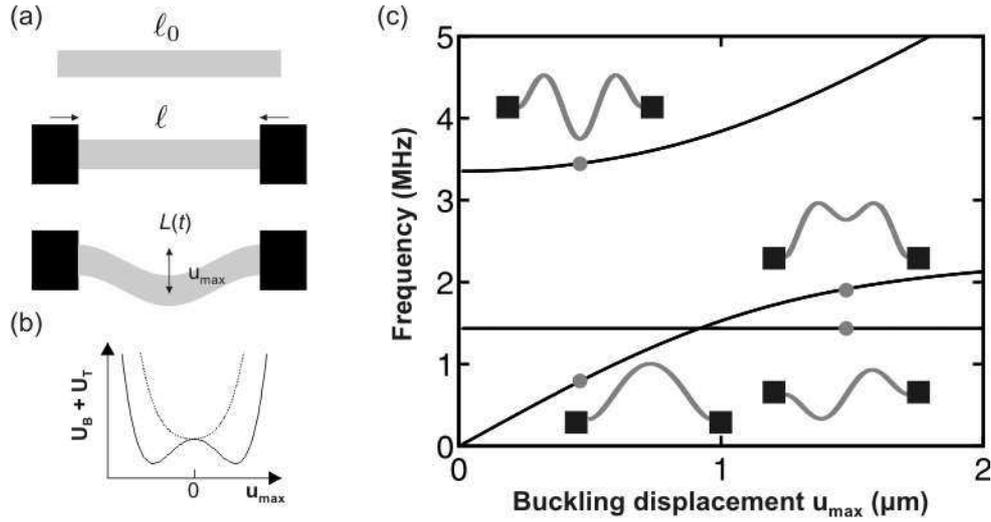}
\caption{(a) The distance between the clamping points $\ell$ of a
doubly clamped beam differs from the length of the free beam
$\ell_0$. If the induced tension is large enough, it is
energetically favorable for the beam to displace, releasing strain
energy at the cost of bending energy. This is called buckling. Due
to the displacement $u(x)$ the length of the beam is extended to
$L(t)$. (b) The total potential energy $U_T + U_B$ for a beam
below (dotted) and above (solid) the buckling threshold. In the
former case the potential energy \added{only} has a \added{single} minimum at
$u_{\mathrm{max}} = 0$, whereas the latter has two minima at
non-zero deflection. (c) An example of the calculated
eigenfrequencies of a $50 \un{\mu m}$-long beam with
$D=3.20\cdot10^{-15} \unit{J}$ \cite{etaki_natphys_squid}. The
mode shapes at the position of the dots are shown.
\label{fig:nems:buckled_beam}}}
\end{figure}

The starting point for the analysis of buckled beams is the
Euler-Bernoulli beam equation (Eq. \ref{eq:nems:beam}) with no
external force, i.e., $F = 0$. The boundary conditions are the
same as for a doubly-clamped beam: $u(0) = u(\ell) =0$ and $u'(0)
= u'(\ell) = 0$ \cite{etaki_natphys_squid, nayfeh_buckled}. The
negative tension has two contributions: first the beam is
compressed by the fact that it is clamped. Figure
\ref{fig:nems:buckled_beam}a shows that the length of the free
beam ($\ell_0$) and that of the clamped beam ($\ell$) are
different when the clamping point exert a longitudinal tension on
the beam. This is the so-called residual tension, $T_0 =
EA\strain_0$, with $\strain_0 = (\ell - \ell_0)/\ell_0$. The
second, positive, contribution comes from the stretching of the
beam when it flexures. The resulting length of the flexed beam is
denoted by $L = \int_0^\ell \{1 + (\pderl{u}{x})^2\}^{1/2} \intd
x$. Combining both effects gives for small ($\pderl{u}{x} \ll 1$)
deflections:
\begin{equation}
T \approx T_0 + \frac{EA}{2\ell}\int_0^\ell \left(\frac{\partial
u}{\partial x}\right)^2\;\mathrm{d}x. \label{eq:nems:tension}
\end{equation}
In the absence of driving, both the displacement and the tension
are time-independent and the static deflection $\udc$ satisfies:
\begin{equation}
 D\frac{\partial^4 \udc}{\partial x^4} - \Tdc \frac{\partial^2
\udc}{\partial x^2}  = 0, \text{~~with~~} \Tdc =  T_0 +
\frac{EA}{2\ell}\int_0^\ell \left(\frac{\partial \udc}{\partial
x}\right)^2\;\mathrm{d}x.\label{eq:nems:dcbuckled}
\end{equation}
This only has a non-trivial solution $\udc \neq 0$ when $\Tdc =
n^2T_c$, where $T_c = -4\pi^2D/L^2$ is the critical tension at
which the beam buckles and $n$ is an integer. The solution is
then $\udc(x) = u_{\mathrm{max}}[1-\cos(2\pi nx/\ell)]/2$. When
an initially unstrained beam is compressed slightly, work is
done and the energy stored in $U_T$ increases. The potential
energy has a single minimum around $u_{\mathrm{max}} = 0$ in
this case; see the dotted line in Fig.
\ref{fig:nems:buckled_beam}b. When the compressive residual
tension $T_0$ is made more negative than $T_c$, it becomes
energetically favorable for the beam to convert a part of $U_T$
into the bending energy $U_B$. The potential energy now has two
minima at non-zero static displacements, as illustrated by the
solid line in Fig. \ref{fig:nems:buckled_beam}b. The beam
buckles to a displacement that keeps the tension exactly at
$T_c$ (for $n$=1). The value of the displacement depends on the
residual tension that caused it: $u_{\mathrm{max}} = \pm
2\ell/\pi \added{\cdot}([T_c-T_0]/EA)^{1/2}$ for $T_0 \le T_c <
0$. Note, that in the absence of a static force, the beam does
not have a preferential direction of buckling.

To find the eigenmodes of the buckled beam, we do not only
focus on the static deflection, but we also include the
dynamics of the displacement. When driving the modes of the
beam, the total deflection $u$ is the sum of the static
($\udc$) and an oscillating part ($\uac$). The time-dependent
displacement satisfies:
\begin{equation} \rho A \frac{\partial \uac^2}{\partial t^2} +
D\frac{\partial^4 \uac}{\partial x^4} - \Tdc \frac{\partial^2
\uac}{\partial x^2}  = \Tac\frac{\partial^2 \udc}{\partial x^2},
\label{eq:nems:sepbeam}
\end{equation}
with:
\begin{equation}
\Tac = \frac{EA}{\ell} \int_0^\ell \frac{\partial \udc}{\partial
x}\frac{\partial \uac}{\partial x}\;\mathrm{d}x.
\label{eq:nems:Tdynamic}
\end{equation}
Note, that both sides of Eq. \ref{eq:nems:sepbeam} are linear
in $\uac$ and that the static displacement $\udc$ thus acts as
an effective spring constant for ac motion, as indicated by the
r.h.s. of Eq. \ref{eq:nems:sepbeam}. The eigenfrequencies of
buckled beams were calculated by Nayfeh {\it et al.}
\cite{nayfeh_buckled}. As an example the eigenfrequencies and
modes for the beam used in Ref. \cite{etaki_natphys_squid} are
shown in Fig. \ref{fig:nems:buckled_beam}b. At zero buckling
the frequency of the fundamental mode is zero as the potential
energy is quartic in the displacement: the quadratic terms in
$U_B$ and $U_T$ cancel each other exactly. The frequency of the
fundamental mode increases with increasing buckling due to the
contribution of $\Tac$. The first higher mode has an
eigenfrequency $\omega_1/2\pi = 1.44 \un{MHz}$ and is
independent of $u_\mathrm{max}$, as the mode is anti-symmetric
around the node, giving $\Tac = 0$. When $u_\mathrm{max}$ is
increased to $0.92 \un{\mu m}$, the two lowest modes cross and
the fundamental mode is higher in frequency than the first odd
mode.\footnote{The modes are classified by their shape and not
by the ordering of eigenfrequencies. The fundamental mode is
the mode without a node.} Thus, when the length of the beam,
bending rigidity and the buckling displacement are known, the
eigenfrequencies and modeshapes of a buckled beam can be
calculated.

\subsubsection{Nanobeams}
\label{sssec:nems:nanotubes} In Sec. \ref{sssec:nems:string} we
showed that tension can overwhelm the influence of bending
rigidity in nanomechanical devices with high aspect ratios. For
these thin wires the displacement can be of the order of the
resonator thickness and in this case nonlinear effects
connected with the deflection-induced tension become important.
Suspended carbon nanotubes resonators are prototypical examples
where the induced tension can be so large that the tubes are
tuned from bending (beam-like) to tension dominated
(string-like) \cite{witkamp_NL_bendingmode,
sapmaz_PRB_cnt_theory}.

The induced tension is a key feature of \deleted{the} thin
resonators as it can be used to electrically tune their
frequency over a large range. In this Subsection we derive the
equations for the frequency tuning and give some typical
numbers for single-wall CNTs. We will start with considering
the device geometry of a suspended CNT resonator and derive
some of the basic equations describing the electrostatics of
the problem. The analysis is, however, also applicable to other
thin string-like resonators such as multi-wall CNTs, suspended
graphene nanoribbons, or long suspended nanowires made from
inorganic materials (see Table~\ref{tab:intro:bottomup}).

\begin{figure}[tb]{
\centering
\includegraphics{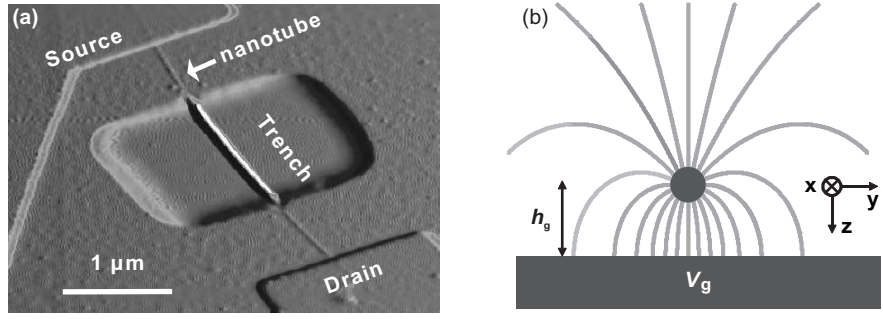}
\caption{(a) an AFM image of a suspended nanotube device connected
to the source and drain electrode. The tube is suspended above the
trench only. The suspended part of this device is $\ell = 1.25
\un{\mu m}$ long and the radius of the tube is $r = 1.4 \un{nm}$.
(b) Field lines of the electrostatic potential induced by the gate
electrode. When the distance between the tube and the gate $h_g$
changes, the gate capacitance $C_g$
changes.\label{fig:nems:nanotube}}}
\end{figure}

Fig. \ref{fig:nems:nanotube}a shows an atomic force microscope
image of a suspended single-wall carbon nanotube in a three
terminal geometry \cite{poot_PSSB_modelling_CNT}. The nanotube
is connected to source and drain electrodes, enabling
electrical transport measurements. The tube is suspended above
a gate electrode at a distance $\hg$, which cannot only be used
to change the electrostatic potential on the tube, but also to
drive the resonator
\cite{poncharal_science_electrostatic_driving} and to induce
tension. An electrostatic force can be applied to the tube by
applying a voltage $\vg$ between a gate electrode (Fig.
\ref{fig:nems:nanotube}b) and the nanotube. The potential
energy\footnote{This is the potential energy for the tube, in
contrast to the energy stored in the capacitor: $+\cg \vg^2 /
2$. The difference in the sign is because the voltage source
performs work when the capacitance changes \cite{griffiths_ed},
which should also be taken into account.} depends on the
capacitance between the tube and the gate $\cg$, and is $U_F =
-\cg \vg^2 / 2$. The potential energy depends on the distance
between the gate and the tube $\hg - u(x)$ via the gate
capacitance, which means that there is a force acting acting on
the nanotube. The electrostatic potential energy is written as:
$U_F = -\int_0^\ell c_g(x)\vg^2/2 \intd x$, where $c_g(x)$ is
the capacitance per unit length, and the potential energy
equals by definition $U_F = - \int_0^\ell F u \intd x$. The
electrostatic force per unit length is thus $F(t) = \halfl
\pderl{c_g}{u} \vg^2$.

To calculate the displacement dependence of the capacitance, we
first consider the spatial profile of the electrostatic
potential. Under the assumption that the screening effect of
the source and drain electrodes is negligible, the tube is
viewed as an infinitely long grounded solid cylinder, suspended
above a conducting plate at an electrostatic potential ${\phi(z
= 0) = \vg}$. The potential profile for $u=0$ is given by
\cite{kwok_complex}:
\begin{equation}
\phi(y,z) = V_g-V_g\frac{1}{\mathrm{arccosh}(\hg /r)}
\ln\left(\frac{\left[z+\sqrt{\hg^2-r^2}\right]^2+y^2}
{\left[z-\sqrt{\hg^2-r^2}\right]^2+y^2}\right).
\end{equation}
The field lines associated with this potential are shown in Fig.
\ref{fig:nems:nanotube}b. The deflection of the nanotube is
included by replacing $\hg$ with $\hg - u$. After dividing the
induced charge by the gate voltage, the capacitance per unit
length $c_g(x)$ is obtained\footnote{This expression might appear
different from those in Refs \cite{lefevre_PRL_scaling} and
\cite{sapmaz_PRB_cnt_theory}, but note that $\arccosh(x) = \ln(x +
\sqrt{x^2-1}) \approx \ln(2x)$ for $x \gg 1$.}:
\begin{equation}
c_g(x) =
\frac{2\pi\epsilon_0}{\mathrm{arccosh}{\left([\hg-u(x)]/r\right)}}
\approx
\frac{2\pi\epsilon_0}{\mathrm{arccosh}{\left(\hg/r\right)}}+
\frac{2\pi\epsilon_0}{\sqrt{\hg^2-r^2}\;
\mathrm{arccosh}^2{\left(\hg/r\right)}}u(x).
\label{eq:nems:capacity}
\end{equation}
The approximation in Eq. \ref{eq:nems:capacity} is allowed
because the displacement $u$ is typically much smaller than
$\hg$. This in contrast to top-down fabricated devices, where
higher order terms can be important and electrostatic softening
of the spring constant might occur
\cite{kozinsky_APL_dynamicrange_tuning,
unterreithmeier_nature_dielectric}. This softening is thus due
to nonlinear capacitance terms, which should be contrasted to
the nonlinearities due to the displacement-induced tension that
we will discuss further on in this Section. For thin resonators
under high built-in tension, however, the nonlinear capacitance
terms can become relevant again. This has been nicely
illustrated in graphene resonators, which show a small downward
shift in the frequency around zero-gate voltage
\cite{singh_NT_graphene_mixing}. We will not consider these
effects in the following, but we finally note that the
electrical softening is closely related to the optical spring
that will be discussed in Sec. \ref{ssec:basic:sideband}.

The electrostatic force per unit length is now:
\begin{equation}
F(t) = \frac{1}{2} \frac{\partial c_g}{\partial u} \vg^2(t) =
\frac{\pi\epsilon_0 \vg^2(t)}{\sqrt{\hg^2-r^2}\;\mathrm{arccosh}^2
{\left(\hg/r\right)}}.
\end{equation}
Often, the gate voltage consists of two parts: a static part
$\vgdc$ and a time-dependent part $\vgac \cos(\omega t)$ to
drive the nanotube at frequency $f = \omega/2\pi$. The
experimental condition $\vgac \ll \vgdc$ ensures that terms
proportional to $(\vgac)^2$ are negligible. The force is then
the sum of a static and driving contribution: $F =
F_{\textrm{dc}} + F_{\textrm{ac}} \cos(\omega t)$, with
$F_{\textrm{dc}} = \pi\epsilon_0
/(\hg^2-r^2)^{1/2}\mathrm{arccosh}^2{\left(\hg/r\right)}\cdot
(\vgdc)^2$ and $F_{\textrm{ac}}(t) = \pi\epsilon_0
/(\hg^2-r^2)^{1/2}\mathrm{arccosh}^2{\left(\hg/r\right)}\cdot
2\vgdc \vgac(t)$.

The bending mode vibrations are described by the
Euler-Bernoulli beam equation with tension included, i.e., Eq.
\ref{eq:nems:beam}. When the amplitude of the oscillation
$\uac$ is small compared to the larger of the tube's radius and
the static displacement, terms proportional to $\uac^2$ are
negligible and the tube is in the linear regime\footnote{This
refers to the dynamical behavior. The static displacement is
actually nonlinear when the tube is in the strong bending limit
\cite{sapmaz_PRB_cnt_theory}.}. Similar to the analysis
presented for the buckled beams, the equation of motion (Cf.
Eq. \ref{eq:nems:beam}) is separated:
\begin{eqnarray}
D\frac{\partial^4\udc}{\partial x^4} - \Tdc \frac{\partial^2 \udc
}{\partial x^2}& & = F_{\textrm{dc}},
 \label{eq:nems:static} \\
-\omega^2\rho A \uac+i\omega \gamma \uac + D\frac{\partial^4\uac
}{\partial x^4} - \Tdc \frac{\partial^2 \uac }{\partial x^2} &
-\Tac \frac{\partial^2 \udc }{\partial x^2}  & =
F_{\textrm{ac}}.\label{eq:nems:dynamic}
\end{eqnarray}
The first equation describes the static displacement of the
tube that is induced by the dc gate voltage. This equation is
independent of $\uac$. On the other hand, the ac displacement,
given by Eq. \ref{eq:nems:dynamic}, depends on the static
displacement and also on the static tension $\Tdc$. Similar to
the case of the buckled beam, the static tension has two
contributions, as indicated by Eq. \ref{eq:nems:dcbuckled}: The
first one is the residual tension due to the clamping as the
length of the suspended part is not necessarily equal to the
length when it would not be clamped. For example, a nanotube
resonator could be strained during the growth process or lay
slightly curved on the substrate before suspending it. The
second contribution is the displacement-induced tension: The
gate electrode pulls the resonator towards it, thereby
elongating it. Moreover, the oscillator experiences a
time-dependent variation in its length and $\Tac$ is the part
of the tension that is linear in $\uac$. Both effects are
included in Eq. \ref{eq:nems:dynamic}. As the tension contains
the static displacement, it has to be solved self-consistently
\cite{sapmaz_PRB_cnt_theory, lefevre_PRL_scaling,
poot_PSSB_modelling_CNT} with Eq. \ref{eq:nems:static} to find
the static displacement. The resulting static tension, ac
tension and dc displacement are then inserted into Eq.
\ref{eq:nems:dynamic} to find the eigenfrequencies $\omega_n$
and the response function $\uac(x, \omega)$
\cite{westra_PRL_coupled}.

\begin{figure}[tb]{
\centering
\includegraphics{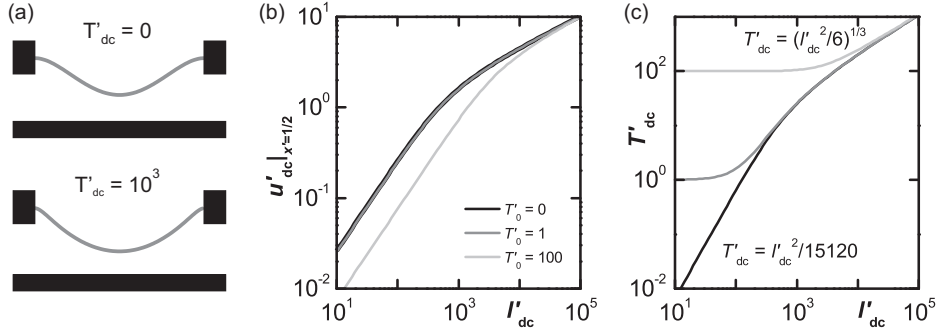}
\caption{(a) static displacement profiles for $\Tdc' = 0$ (top),
and $\Tdc' = 10^3$ (bottom). (b,c) The calculated static
displacement at the center of the nanotube (b) and the corresponding
tension (c) for various value of the residual tension $T_0'$. The
limits for the tension for small and large static forces are
indicated. \label{fig:nems:udc_Tdc}}}
\end{figure}

To analyze the system of Eqs. \ref{eq:nems:tension},
\ref{eq:nems:static} and \ref{eq:nems:dynamic}, it is useful to
take a closer look at their scaling behavior
\cite{lefevre_PRL_scaling}. In this section, we use the
convention that primed variables indicate scaled
(dimensionless) variables. An obvious way to normalize the
coordinate $x$ is to divide it by the tube length: $x' =
x/\ell$, so that the equation for the static displacement Eq.
\ref{eq:nems:static} becomes:
\begin{equation}
\frac{\partial^4 \udc } {\partial x'^4} - \frac{\ell^2\Tdc}{D}
\frac{\partial^2 \udc } {\partial x'^2} =
\frac{\ell^4F_{\textrm{dc}}}{D} \equiv l_{\mathrm{dc}}.
\end{equation}
On the right hand side, a natural length scale for the static
displacement, $l_{\mathrm{dc}}$, appears. However, scaling the
displacement with $l_{\mathrm{dc}}$ is not handy because
$l_{\mathrm{dc}}$ equals zero at zero gate voltage. Therefore
$\udc$ (and $\uac$) are scaled by the radius of the tube: $\udc'
= \udc /r$. Moreover, the tension has become dimensionless,
resulting in an equation for the static displacement where the
number of parameters has been reduced from 5 to 2:
\begin{equation}
\frac{\partial^4 \udc' } {\partial x'^4} - \Tdc' \frac{\partial^2
\udc' } {\partial x'^2} = l_{\mathrm{dc}}',
\label{eq:nems:static_dl}
\end{equation}
where
\begin{equation}
\displaystyle \Tdc' = \frac{\ell^2\Tdc}{D} = T'_0 +
\frac{Ar^2}{2I} \int_0^1 \left(\pder{\udc'}{x'}\right)^2 \intd x'
,~~T'_0 = \frac{\ell^2T_0}{D}, \text{~~and~~}
l_{\text{dc}}' = l_{\text{dc}}/r. \label{eq:nems:dctension_dl}
\end{equation}
The definition of $\Tdc'$ with the $\ell^2$ dependence shows
that tension becomes more and more important when the length of
the device increases. Figure \ref{fig:nems:udc_Tdc}a shows the
dc displacement profiles for different values of the static
tension. In the case where the bending rigidity dominates (top
panel) the profile is rounded at the edge, whereas for high
tension (lower panel) the profile is much sharper. The tension
and center deflection \added{are} calculated by
\added{self-consistently} solving Eq.~\ref{eq:nems:static}
\deleted{self-consistently} with
\added{Eq.~}\ref{eq:nems:dctension_dl}\added{, and} are plotted
in Fig. \ref{fig:nems:udc_Tdc}b and c. Two different slopes can
be distinguished in the double-logarithmic plot of Fig.
\ref{fig:nems:udc_Tdc}b. These correspond to the weak and
strong bending regime \cite{sapmaz_PRB_cnt_theory}\added{,
where $\udc$ is proportional to $F_{\textrm{dc}}$ and
$F_{\textrm{dc}}^{1/3}$ repectively}. The two regimes cross at
${\Tdc'^*} = 6\sqrt{70} \approx 50.2,~l_{\mathrm{dc}}'^{*} = 36
\cdot 70^{3/4} \approx 871$. The gate voltage at which
$l_{\mathrm{dc}}' = l_{\mathrm{dc}}'^{*}$ is called the
cross-over voltage, $\vgstar$. AFM measurements of the
gate-induced displacement of multi-walled carbon nanotubes have
confirmed this scaling behavior experimentally
\cite{lefevre_PRL_scaling}.

\begin{figure}[tb]{
\centering
\includegraphics{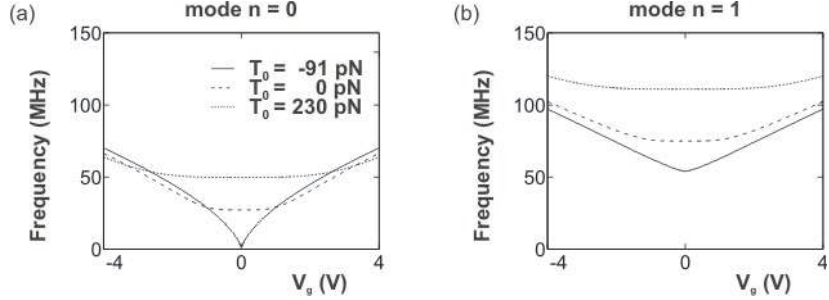}
\caption {Gate-voltage tuning of the eigenfrequencies of the first
(a) and second (b) flexural mode of a $1 \un{\mu m}$-long
suspended carbon nanotube (see Table \ref{tab:nems:nanotubes}).
The residual tensions are $T_0 =-91,~0,~+230 \un{pN}$ for the
solid, dashed and dotted lines respectively, which corresponds to
the dimensionless values $T_0' = -39.4,~0,~+100$.
\label{fig:nems:nanotube_frequency}}}
\end{figure}

To calculate the gate-tuning of the resonance frequency
$f_R(\vgdc)$, the scaling analysis is also applied to the
equation for the ac displacement, Eq. \ref{eq:nems:dynamic}.
One immediately finds the length scale $l_{\mathrm{ac}} =
\ell^4 F_{\textrm{ac}}/D$ for the ac force and $\Tac' =
\Tac\ell^2/EI$. As in Sec. \ref{sssec:nems:beams}, $\omega_{ch}
= (D / \rho A)^{1/2} / \ell^2$ is again the characteristic
frequency scale for the bending mode vibrations.
\replaced{Next}{To find $f_R(\vgdc)$}, one \deleted{first} has
to solve Eqs. \ref{eq:nems:static_dl} and
\ref{eq:nems:dctension_dl} to obtain the static tension and dc
displacement. Then, the boundary conditions are imposed to find
the resonance frequencies. Figure
\ref{fig:nems:nanotube_frequency}a shows the calculated
eigenfrequencies, plotted against the static pulling force for
different residual tensions. The higher the residual tension
is, the higher the resonance frequencies are at \added{a given}
$\vgdc$. The value $T_0' = -39.4 \approx T_c$ indicates that
the thin resonator is close to buckling; this is visible by the
nearly vanishing resonance frequency of the first mode at low
$\vgdc$. When the static force is increased, all resonance
frequencies increase and the differences between the curves due
to the different residual tensions become smaller.

To relate the dimensionless quantities in this Subsection to
physical ones, the system dimensions are needed. Table
\ref{tab:nems:nanotubes} shows the estimated sizes and the
calculated values of several parameters for two typical
nanotube devices, one with a length $\ell = 1 \un{\mu m}$ and
one with a length $\ell = 200 \un{nm}$. Note, that with our
definition of $u_n$, the mass and spring constant appearing in
the zero-point motion $u_0$ (Eq. \ref{eq:basic:zpm}) and the
equipartition theorem (Eq. \ref{eq:basic:equipartition}) are
equal to the total mass $m$ and $k_R = m\omega_n^2$
respectively. There is no need to introduce an effective mass;
see Sec. \ref{ssec:basic:modes_to_HO}. The table indicates that
for a long device tension is important and the difference
between $f_R$ and $f_0$ is large. For a short resonator, the
mechanical properties are mainly determined by the bending
rigidity and $f_R \approx f_0$. The \added{higher} value of
$\vgstar$ indicates that a large\added{r} gate voltage has to
be applied to tune \replaced{this}{the} resonator to the strong
bending regime.
\begin{table}[t]
\centering \caption{Data for two suspended carbon nanotubes
with different lengths \added{$\ell$}. The values of the
parameters are calculated with \replaced{a}{the} nanotube
radius $r = 1.5 \un{nm}$, \deleted{length $\ell$} and distance
to the gate electrode $\hg = 500 \un{nm}$. The resonance
frequency $f_R$, tension $\Tdc$ and static displacement
$u_{\textrm{dc}}(\ell/2)$ are evaluated for a gate voltage
$\vgdc = 4 \un{V}$ and zero residual tension. $m$, $f_0$ and
$D$ are the mass, resonance frequency of the nanotube without
tension and the bending rigidity respectively. Furthermore,
$k_{\mathrm{dc}}$ and $k_R = m\omega_R^2$ are the static and
dynamic spring constants \cite{poot_thesis,
tombler_nature_piezoresistance_SWNT,
minot_PRL_piezoresistance_SWNT}, $\cg$ is the capacitance to
the gate and $u_0$ the zero-point motion (see Sec.
\ref{sec:intro} for its definition and the discussion in Sec.
\ref{sec:basic}). \label{tab:nems:nanotubes}}
\begin{tabular}{crrl}
\addlinespace \toprule
$\ell$                    & 1.00                      & 0.20                      & $\unitt{\mu m}$ \\
\midrule
$m$                       & 6.4                       & 1.3                       & $10^{-21}\unit{kg}$ \\
$D$                       & 4.8                       & 4.8                       & $10^{-24}\unit{Nm^2}$ \\
$f_0$                     & 97.0                      & 2424                      & $\unitt{MHz}$ \\
$f_R$                     & 238.6                     & 2424                      & $\unitt{MHz}$ \\
$k_R$                     & 14.5                      & 299                       & $10^{-3}\unit{N/m}$ \\
$u_0$                     & 2.3                       & 1.6                       & $\unitt{pm}$ \\
$\cg$                     & 8.56                      & 1.71                      & $\unitt{aF}$ \\
$\pderl{\cg}{u}$          & 2.63                      & 0.53                      & $\unitt{zF/nm}$ \\
$F_{\textrm{dc}}\ell$     & 21                        & 4.2                       & $\unitt{pN}$ \\
$l_{\textrm{dc}}$         & 4412                      & 7.1                       & $\unitt{nm}$ \\
$u_{\textrm{dc}}(\ell/2)$ & 4.1                       & 0.018                     & $\unitt{nm}$ \\
$k_{\mathrm{dc}}$         & 11.9                      & 229                       & $10^{-3}\unit{N/m}$ \\
$T_{\mathrm{dc}}$         & 0.35                      & $1.8\cdot 10^{-4}$        & $\unitt{nN}$ \\
$T_{\mathrm{dc}}'$        & 73                        & $1.5\cdot 10^{-3}$        &  \\
$\vgstar$                 & 2.2                       & 54.4                      & $\unitt{V}$ \\
$l_{\textrm{dc}}'$        &$2.9 \cdot 10^3$           & 4.7                       &  \\
\bottomrule
\end{tabular}
\end{table}

As a final note, we consider the difference between a nanobeam
resonator with a high frequency and one with a low frequency
that is tuned using a gate voltage to the same frequency as the
former \cite{poot_PSSB_modelling_CNT}. The current associated
with the motion in mixing experiments \added{(see Sec.
\ref{sssec:detectors:mixing})} \cite{sazonova_nature,
witkamp_NL_bendingmode, lassagne_NL_masssensing,
chiu_NL_masssensing, lassagne_science_coupled_cnt,
wang_science_nanotube_phase_transitions} is proportional to the
\added{length-}averaged displacement amplitude \added{$\mu_n
u_n(\omega)$}; it is thus proportional to the length scale
$l_{\text{ac}}$ that determines the amplitude. As a
consequence, the mechanical signal drops rapidly with
decreasing length, making the measurement of single-walled
carbon nanotubes with $f_0 > 1 \un{GHz}$ (corresponding to a
device with $\ell \lesssim 0.2 \un{\mu m}$ and $r = 1 \un{nm}$)
challenging as the signal is about $100 \times$ smaller
compared to a $f_0 = 100 \un{MHz}$ device with $\ell \approx
0.6 \un{\mu m}$. The latter tube can also operate at $1
\un{GHz}$ frequency by tuning it with a gate-induced tension of
$\Tdc' = 5 \cdot 10^3$. In this case the signal decreases too,
but only by a factor of $10$. A tension of $\Tdc' = 5 \cdot
10^3$ corresponds to a strain of about $0.2\%$, which is larger
than the values in the abovementioned experiments, but is still
smaller than the strain at which single-walled carbon nanotubes
break \cite{walters_APL_suspended, bozovic_PRB_strained_SWNT,
zhao_PRB_nanotube_strength}.

\subsubsection{Nanodrums} \label{ssec:nems:nanodrums}
\begin{figure}[tb]{
\centering
\includegraphics{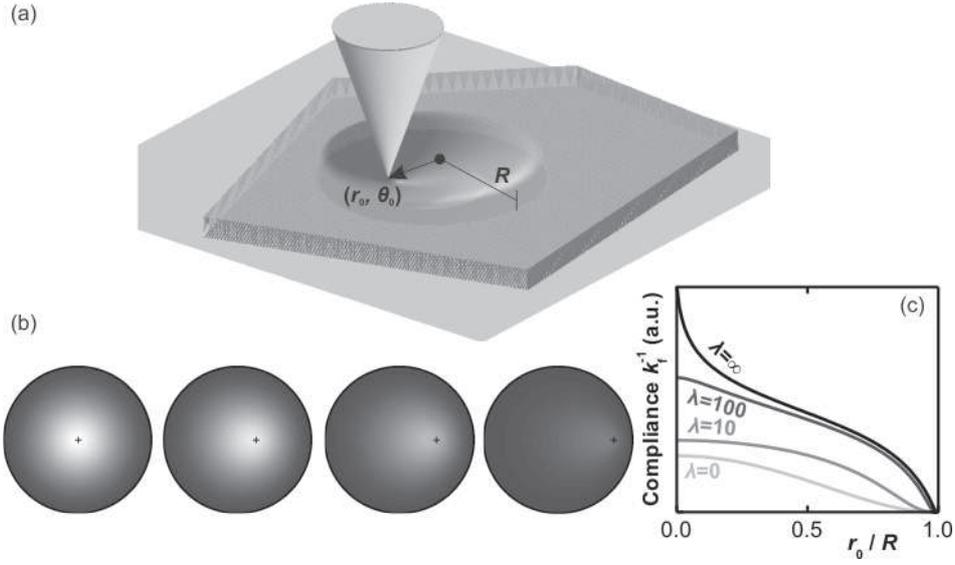}
\caption[Modelled deflections of a nanodrum]{(a) Schematic
overview of the nanodrum. A few-layer graphene flake is suspended
over a circular hole with radius $R$. A force is applied at the
point $(r_0, \theta_0)$ using an AFM tip. This results in a
deflection of the nanodrum (b) Colormaps of the calculated
deflection profile (Eq. \ref{eq:nems:nanodrum_simple}) of a
nanodrum with vanishing tension. The force is applied at the
location of the cross and the color scale is identical in all four
panels: white corresponds to a large deflection and dark gray to
no deflection. (c) The calculated radial compliance profile for
different values of the tension, with $\lambda^2 = TR^2/D$.}
\label{fig:nems:nanodrum}}
\end{figure}

So far, we have only considered one-dimensional resonators such as
beams and strings. An example of a two dimensional resonator is a
graphene nanodrum. This device consists of a hole that is etched
in a substrate and that is covered by a (few-layer) graphene
flake. Surprisingly, an one-atom layer can be suspended and holes
with a diameter of $100 \un{\mu m}$ have been reported
\cite{booth_NL_macroscopic_graphene}. A much smaller version of
these devices (a nanodrum) has been used to study the
(im)permeability of graphene to gases \cite{bunch_NL_impermeable}
and to measure the bending rigidity of and tension in the flake
using an atomic force microscope \cite{poot_APL_nanodrums,
lee_science_AFM_monolayer}. In the latter experiments, an atomic
force microscope tip is used to apply a force $F_{\mathrm{tip}}$
to the flake as illustrated in Fig. \ref{fig:nems:nanodrum}a. This
versatile technique has also been applied to other geometries
\cite{frank_JVSB_graphene_AFM, traversi_NJP_AFM_graphene_PMMA} and
nanomaterials \cite{gomez_navarro_NL_GO_AFM, li_NT_BN_AFM,
song_NL_BN_AFM}. The point $(r_0, \theta_0)$ where the force is
applied can be varied and the resulting deflection of the nanodrum
is measured. The restoring force that opposes the applied force
has several contributions. First of all, there is the bending
rigidity of the flake $D$, and secondly, tension may be present in
the flake.

Since graphite is highly anisotropic, the analysis of the
bending rigidity of an isotropic material in Sec.
\ref{ssec:nems:eulerbernoulli} has to be
generalized.\footnote{The analysis is in principle also valid
for other layers two-dimensional membranes; for a thin
isotropic membrane Eq.~\ref{eq:nems:nanodrum_complete} and the
derivation following it are still valid if the appropriate
bending rigidity $D$ is taken.} Using the compliance tensor in
Eq. \ref{eq:nems:compliance_graphite} the rigidity for bending
along the sheets (see Fig. \ref{fig:nems:graphenesheet}b) is
calculated:
\begin{equation}
D = E_\boxempty h^3/12(1-\nu^2_\boxempty),
\label{eq:nems:rigidity_graphite}
\end{equation}
which only contains the in-plane elastic constants. However, when
the number of graphene layers becomes small, corrections to Eq.
\ref{eq:nems:rigidity_graphite} have to be made. Consider the
situation in Fig. \ref{fig:nems:graphenesheet}b where a few-layer
graphene sheet is bent with a radius of curvature $R_c$. In the
continuum case, the bending energy is given by $U_B/\ell W \equiv
\half D /R_c^2 = \half \int_{-h/2}^{h/2} E_\boxempty (z/R_c)^2
\intd z$, when taking $\nu = 0$ for simplicity. In the case of a
small number of layers $N$, the continuum approximation in the
z-direction is no longer valid and the stress is located only at
the position of the sheets $z_i = c (i - [N+1]/2)$, where $c =
0.335 \un{nm}$ is the inter-layer spacing (see p.
\pageref{p:nems:graphite}). The integral over $z$ is replaced by a
sum and the bending rigidity becomes:
\begin{equation}
D_N = \frac{E_\boxempty h^3}{12(1-\nu^2_\boxempty)}\frac{N-1}{N},
\label{eq:nems:rigidity_fewlayer}
\end{equation}
where the thickness is set by the number of layers $h = Nc$,
\added{which reduced to Eq. \ref{eq:nems:rigidity_graphite} in
the limit $N \rightarrow \infty$.} According to Eq.
\ref{eq:nems:rigidity_fewlayer} the bending rigidity vanishes
for a single layer. However, molecular dynamics simulations
have shown that a single layer of graphene still has a finite
bending rigidity: of the order of one $\un{eV}$
\cite{liu_APL_bending_rigidity, yakobson_PRL_rigidity_graphene,
castro-neto_RMP_graphene_review} (compare this to the value for
a double layer ($N = 2$) calculated with Eq.
\ref{eq:nems:rigidity_fewlayer}: $D_2 = 54 \un{eV}$). The
rigidity of a single layer comes from the fact that electrons
in the delocalized $\pi$-orbitals, located below and above the
sheet, repel each other when the sheet is bent
\cite{castro-neto_RMP_graphene_review}.

The deflection of the nanodrum satisfies the equation for the
deflection of a plate, Eq. \ref{eq:nems:plate}, and when the
force applied by the AFM tip is assumed to be located at a
single point $(x_0, y_0)$, one gets \cite{LL_elasticity,
wan_TSF_nanodrum_theory, norouzi_PRE_bilayerafmtheory,
atalaya_NL_continuum_graphene}:
\begin{equation}
\left(D\nabla^4 - \pder{}{x_\alpha} T_{\alpha\beta}
\pder{}{x_\beta}\right) u(x, y; x_0, y_0) = F_{\mathrm{tip}}
\delta(x-x_0)\delta(y-y_0). \label{eq:nems:nanodrum_complete}
\end{equation}
Here, the $\nabla$-operator and the partial derivatives
$\pderl{}{x_i}$ are working in the xy-plane only, as the
z-dependence is absorbed in the bending rigidity (see Sec.
\ref{ssec:nems:eulerbernoulli}) and where the tension
$T_{\alpha\beta} = \int_0^h \sigma_{\alpha\beta} \intd z$. This
equation is difficult to solve in its most general form, but
fortunately some simplifications can be made: The tension
tensor can have both normal and shear components. It is,
however, always possible to find two orthogonal directions
where the shear components are zero \cite{chung_continuum}.
When we assume that the tension is uniform then these
directions are independent of position, so without loss of
generality the x and y-axis are taken along the principle
directions of the tension. When the difference in tension in
the x and y direction, $\Delta T = (T_{xx} - T_{yy})/2$, is
small, first the solution for isotropic tension
($T_{\alpha\beta} \approx T
\delta_{\replaced{\alpha\beta}{ij}}$) is obtained and then the
correction due to the finite $\Delta T$ can be calculated
\cite{poot_thesis}. Here, we will focus on the situation where
$\Delta T = 0$. For a circular membrane, it is convenient to
use polar coordinates and the equation for the displacement
reads\footnote{$\nabla^2 = \pder{^2}{x^2} + \pder{^2}{y^2} =
\pder{^2}{r^2} +
\frac{1}{r}\pder{}{r} + \frac{1}{r^2}\pder{^2}{\theta^2}$ and \\
$\nabla^4  = \pder {^4}{x^4} + \pder{^4}{y^4} = \pder{^4}{r^4} +
\frac{2}{r}\pder{^3}{r^3} - \frac{1}{r^2}\pder{^2}{r^2} +
\frac{1}{r^3}\pder{}{r} + \frac{4}{r^4}\pder{^2}{\theta^2} +
\frac{1}{r^4}\pder{^4}{\theta^4} + \frac{2}{r^2}\pder{^4}{r^2
\partial \theta^2} - \frac{2}{r^3}\pder{^3}{r
\partial \theta^2}$.}:
\begin{equation}
\big(D\nabla^4 - T \nabla^2\big) u(r, \theta; r_0, \theta_0) =
\frac{F_\mathrm{tip}}{r}\delta(r-r_\mathrm{0})\delta(\theta-\theta_\mathrm{0}).
\label{eq:nems:nanodrum_simple}
\end{equation}
The solution is written as:
\begin{equation}
u(r, \theta; r_\mathrm{0}, \theta_\mathrm{0}) = \sum_{m =
0}^\infty R_m(r; r_0) \cos(m\theta-m\theta_0).
 \label{eq:nems:solution_sum}
\end{equation}
Inserting this into Eq. \ref{eq:nems:nanodrum_simple} yields for
the radial coefficients:
\begin{eqnarray}
R_0(r; r_0) = & A_0 I_0(\lambda r/R) + B_0 K_0(\lambda r/R) + \nonumber \\
& C_0 \ln(r/R) + D_0 + R^{(p)}_0(r; r_0), \\
R_m(r; r_0) = & A_m I_m(\lambda r/R) + B_m K_m(\lambda r/R) + \nonumber \\
& C_m (r/R)^{-m}+ D_m (r/R)^m + R^{(p)}_m(r; r_0) & (m > 0),
\label{eq:nems:radial_sum}
\end{eqnarray}
where $I_m$ and $K_m$ are the Bessel functions of the first and
second kind respectively, and $\lambda = \sqrt{TR^2/D}$ is a
dimensionless parameter that indicates the importance of the
tension in comparison with the bending rigidity of the flake.

The flake with radius $R$ is clamped at the edge of the
circular hole so the boundary conditions are $u(R) = 0$ and
$\tderl{u}{r}|_{r = R} = 0$. Furthermore, the deflection at the
center is finite and smooth (i.e., $\tderl{u}{r}|_{r = 0} =
0$). The set of coefficients $\{A_m, B_m, C_m, D_m\}$ can be
calculated analytically. Figure \ref{fig:nems:nanodrum}b shows
the deflection profiles calculated where the force is applied
at different distances $r_0$ from the center. The deflection of
the flake is clearly reduced when the AFM tip is moved away
from center of the nanodrum. This indicates that its local
compliance $k_f^{-1}(r_0, \theta_0) = \pderl{u(r_0, \theta_0;
r_0, \theta_0)}{F_{\mathrm{tip}}}$ decreases. As the tension is
assumed to be isotropic, $k_f^{-1}$ is independent of
$\theta_0$ and its radial profile, shown in Fig.
\ref{fig:nems:nanodrum}c for different values of the tension,
contains all the information. In analogy with the displacement
profile of a bending and tension-dominated carbon nanotube
(Fig. \ref{fig:nems:udc_Tdc}a), the profile is rounded at the
edge of the hole for vanishing tension ($\lambda = 0$), whereas
for large tension ($\lambda \rightarrow \infty$) the compliance
profile becomes much shaper\footnote{In the limit $\lambda
\rightarrow \infty$, the $\nabla^4$ term in Eq.
\ref{eq:nems:nanodrum_simple} vanishes and only a second order
differential equation remains. Therefore, the boundary
conditions $\tderl{u}{r}|_{r = 0,R} = 0$ are discarded.} at the
edge and diverges at the center for a point force. In practice
the tip has a finite radius of curvature which prevents that
the spring constant of the flake $k_f(r_0 = 0)$ vanishes
\cite{lee_science_AFM_monolayer}. By comparing the
experimentally measured profile with the calculated ones, the
values for $D$ and $T$ can be determined
\cite{poot_APL_nanodrums}. Knowing the tension (and in
principle also the bending rigidity, but this is negligible for
thin flakes) an estimate of the resonance frequency can be
made. The fundamental eigenfrequency of a circular drumhead is
$f_0 = \sqrt{T/\rho_{2d}}/2\pi R \cdot \nu_0^2$, where $\nu_0
\approx 2.4048$ is the first zero of the zero-order Bessel
function of the first kind. Measurements have indicated a
typical value of $T \sim 0.3 \un{N/m}$ for exfoliated single
layer graphene flakes \cite{lee_science_AFM_monolayer}. With
this value and the value for the two-dimensional mass density
of single-layer graphene, $\rho_{2d} = 6.8 \cdot 10^{-7}
\un{kg/m^2}$ (see page \pageref{p:nems:rho2d}) we obtain a
resonance frequency of $100 \un{MHz}$ for a nanodrum with a
diameter of $2R = 5 \un{\mu m}$. By decreasing the size of the
hole, the frequency can be increased. For a hole with a
diameter of $500 \un{nm}$, the frequency already exceeds 1 GHz.
The latter resonator has a mass of only $m = 5\cdot 10^{-19}
\un{kg}$ and a very large zero-point motion of $u_0 = 0.4
\un{pm}$. This makes graphene nanodrums ideal devices for QEMS.

\section{Backaction and cooling} \label{sec:basic}
In this Section we consider the influence of the detector, the
so-called backaction and we discuss ways to cool the resonator
to the ground state. To describe these effects in detail
concepts such as thermal noise, Brownian motion, and the
effective resonator temperature are introduced. Next we will
discuss detector backaction and show that it plays an important
role, especially when continuously probing the resonator
properties. Backaction is therefore often an unwanted, but
unavoidable element in measurements on mechanical systems. It
determines how precise the position can be monitored;
ultimately, quantum mechanics poses a limit on continuous
linear detection, the standard quantum limit. We will derive
this limit in different ways. The standard quantum limit can be
circumvented by employing different detection schemes, such as
square-law detection and backaction evading measurements. These
will however not be discussed in depth in this Review.

Backaction does not just impose limits, it can also work to
one's advantage: It can squeeze the resonator motion
\cite{rugar_PRL_squeezing, almog_PRL_squeezing,
huo_APL_squeeze, zhang_PRA_quantum_feedback,
jaehne_PRA_squeeze}, couple and synchronize multiple resonators
\cite{filatrella_PRE_synchronization,
cross_PRL_synchronization}, and backaction can cool the
resonator as we will show. Besides this self-cooling by
backaction, other active cooling schemes have been developed.
In this Section we will discuss two different cooling schemes,
namely active feedback cooling and sideband cooling.  It is
interesting to note that, at present, cooling has only been
performed on top-down devices because the experiments to
observe thermal motion of bottom-up devices are more
challenging.

\subsection{From modes to harmonic oscillators}
\label{ssec:basic:modes_to_HO} Before starting the discussion
of backaction, first an important point has to be addressed. In
the previous Section we have given the equations of motion for
several nanomechanical systems. Solving these equations gives
the frequency and the displacement profile of a particular
mode. An important conclusion of describing \added{small
displacements around the equilibrium position of} NEMS is that
each mode can be viewed as a harmonic oscillator. To show this,
we start with expanding the displacement in the
basis\footnote{It can be shown that these functions form a
basis for functions that satisfy the homogeneous boundary
conditions \cite{chang_QJMAM_eigenbasis,
poot_PSSB_modelling_CNT}.} formed by the eigenfunctions $\xi_n$
\cite{cleland_JAP_noise, postma_APL_dynamicrange_NW,
poot_PSSB_modelling_CNT, westra_PRL_coupled}:
\begin{equation}
u(x,t) = \sum_n u^{(n)}(t) \xi_n(x).
\end{equation}
Inserting this into the equation of motion and taking the inner
product with $\xi_n$ yields the displacement of mode $n$. For
example, for the nanobeams discussed in Sec.
\ref{sssec:nems:nanotubes}, Eq. \ref{eq:nems:dynamic} yields:
\begin{equation}
\left(\omega_n'^2-\omega'^2 +i\omega' \gamma' \right)\uac^{(n)} =
l_{\mathrm{ac}}\mu_n; \hspace{1cm} \mu_n = \int_0^1 \!\! \xi_n(x')
\intd x'. \label{eq:nems:expand_ho}
\end{equation}
The left hand side shows that the frequency response of each
mode is equal to the response function of a damped driven
harmonic oscillator. The same conclusion is reached for the
other examples: all their modes can be describes as harmonic
oscillators. The mathematical reason behind this is that the
equation of motion for small deformations of a mechanical
system can be written in the form $m_{\text{eff}}\ddot
u(\vc{r},t) = -\gamma \dot u(\vc{r},t) +
\mathcal{L}[u(\vc{r},t)]$ for some Hermitian operator
$\mathcal{L}$. Its eigenfunctions are the mechanical modes and
these form a complete orthogonal basis. After expanding the
displacement in this basis, a set of uncoupled harmonic
oscillators results. An important question that one should ask
after the transformation from the spatial modes to the harmonic
oscillators is: What is the effective mass $m_{\text{eff}}$ of
the oscillator? This question might seem trivial at first, but
the concept of the effective mass has given rise to much
confusion in the past years. From the equation of motion it
follows that the effective mass of mode $n$ equals
$m_{\text{eff},n} = \int_V \rho \xi_n^2 \intd V$. The effective
mass thus depends on the normalization of the basis functions,
or equivalently on the definition of the mode displacement
$u_n$. In this Review we have adopted the convention that the
basis functions $\xi_n$ are orthonormal, i.e. $\ell^{-1}
\int_0^\ell \xi_n^2 \intd x = 1$. In this case\footnote{For
simplicity it is assumed that the mass density is constant.}
the effective mass equals the total mass of the system (i.e.,
$m_{\text{eff}\added{,n}} = m$). On the other hand, one can
also use different normalizations, for example using the
average displacement ($\ell^{-1} \int_0^\ell \xi_n \intd x =
1$) or using the maximum displacement ($\max \xi_n = 1$). In
these cases, the effective mass differs from the total mass and
it depends on the exact mode profile. To illustrate the
confusion this may create, consider the flexural modes of a
beam resonator: With the latter two definitions, every mode has
a different effective mass. Moreover, tuning a flexural
resonator from the bending to stretching-dominated regime (Sec.
\ref{sssec:nems:nanotubes}) changes the effective mass of its
modes. These complications are avoided by using orthonormal
basis functions, where $\ell^{-1} \int_0^\ell \xi_n^2 \intd x =
1$.

Although some groups study the interaction between different
coupled modes in nano and micromechanical systems
\cite{karabalin_PRB_coupled, shim_science_coupled_resonators,
okamoto_APEX_optical_tuning, westra_PRL_coupled,
dunn_APL_modecoupling, lin_natphot_coherent_mixing}, most
experimental and theoretical work focuses on a single mode only
and the harmonic oscillator describes the dynamics of the
entire mechanical system. In this case, we make no distinction
between the resonator (i.e., the entire mechanical system) and
the mode that is studied. \added{However, we note again that
the actual displacement of the resonator at a certain position
$x$ is given by $u^{(n)}\xi_n(x)$ and not by $u^{(n)}$ itself.}

Before reviewing the properties of the classical and quantum
harmonic oscillator, we stress, however, that knowledge about
the displacement profile remains important when analyzing NEMS
experiments. In particular, different detectors or driving
forces may couple differently to the displacement profile and
could detect therefore different modes. In most cases both the
driving and detection mechanisms couple to the average
displacement of the resonator. In this case, anti-symmetric
modes are not visible in nanomechanical experiments as these
have a vanishing value of the length-averaged displacement
$\mu_n$ (see Eq. \ref{eq:nems:expand_ho} and Table
\ref{tab:nems:mode_freqs}\label{pg:basic:average_displacement}).
An example is a nanotube resonator with frequency mixing
readout (Secs. \ref{sssec:nems:nanotubes} and
\ref{sssec:detectors:mixing}) that is \added{either} coupled to
a back gate or to a local gate; the former couples uniformly to
the whole nanotube, so that the detected signal is proportional
to $\mu_n$ and consequently only symmetric modes can be
detected. A local gate may, depending on its position, couple
to all modes.

\subsection{The harmonic oscillator}
The harmonic oscillator is probably the most extensively
studied system in physics. Nearly everything that returns to
its equilibrium position after being displaced can be described
by a harmonic oscillator. Examples range from the suspension of
a car, traffic-induced vibrations of a bridge, and the voltage
in an electrical LC network, to light in an optical cavity.
\added{For large amplitudes, the oscillator can become
nonlinear. We will not study that situation in this Report, but
instead we refer the reader to the large body of literature on
this subject, see e.g. Refs.
\cite{dykman_soviet_nonlinear_medium, strogatz}.} We will now
proceed with describing the classical and quantum harmonic
oscillator in more detail and reviewing their basic properties.

\subsubsection{The classical harmonic oscillator}
In a harmonic oscillator, the potential energy depends
quadratically on the displacement $u$ from the equilibrium
position:
\begin{equation}
V(u) = \half k_R u^2,
\end{equation}
where $k_R$ is the spring constant. The parabolic shape of the
potential results in a force that is proportional to the
displacement. When damping and a driving force $F(t)$ are
included, the equation of motion reads:
\begin{equation}
m\ddot u = - k_R u - m \gamma_R \dot u + F(t),
\label{eq:basic:eqofmotion}
\end{equation}
for a resonator with mass $m$ and damping rate $\gamma_R$. When
the oscillator is displaced and released, it will oscillate at
frequency $\omega_R$ with a slowly decreasing amplitude due to the
damping. The quality factor $Q = \omega_R/\gamma_R$ indicates how
many times the resonator moves back and forth before its energy
has decreased by a factor e.

\begin{figure}[tb]{
\centering
\includegraphics{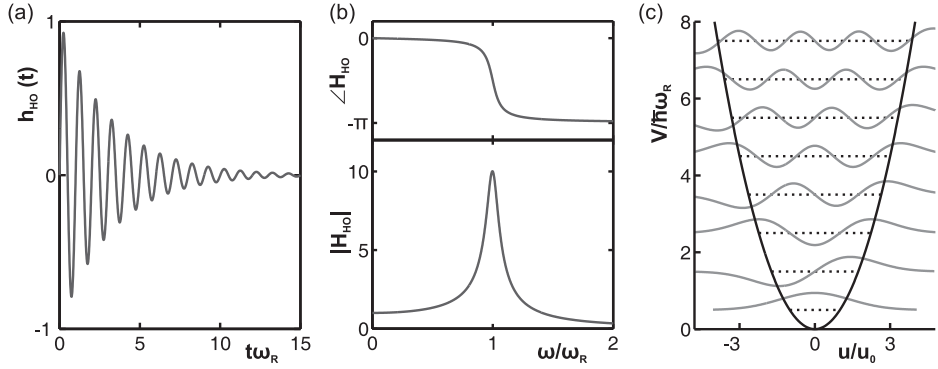}
\caption[Green's function, frequency response and wave functions
of a harmonic oscillator] {(a) The Green's function $h_{HO}(t)$
and (b) frequency response $H_{HO}(\omega)$ of a harmonic
oscillator with $Q = 10$. (c) Eigenenergies $E_n$ (dashed) and the
corresponding wave functions $\psi_n(u)$ (solid) of the harmonic
oscillator for $n = 0 .. 7$.
\label{fig:basic:harmonic_oscillator}}}
\end{figure}

The harmonic oscillator responds linearly to an applied force; in
other words, it is a linear system. Any linear system is
characterized by its impulse response or Green's function
\cite{oppenheim_signals_systems}. For the harmonic oscillator, the
impulse response $h_{HO}(t)$, is the solution to Eq.
\ref{eq:basic:eqofmotion} with $F(t) = k_R\delta(\omega_R t)$:
\begin{equation}
h_{HO}(t) = \sin(\omega_R t) e^{-\frac{\omega_R t}{2Q}}
\Theta(\omega_R t), \label{eq:basic:green}
\end{equation}
where $\Theta(t)$ is the Heaviside stepfunction. The impulse
response function\footnote{This is the Green's function for a
high-Q resonator. For lower Q-values, the resonator oscillates
at a slightly lower frequency $\omega'_R = \omega_R
\sqrt{1-({1}/{2Q})^2}$ and the impulse response of an
underdamped oscillator (i.e., one that has $Q > 1/2$) is:
$h_{HO}(t) = \sin(\omega_R' t) \exp(-\omega_R t/2Q) \cdot
[1-(2Q)^{-2}]^{-1/2} \Theta(t)$. An overdamped resonator ($Q <
1/2$) returns to $u = 0$ without any oscillations and has a
different impulse response. Throughout this Review it is
assumed that $Q \gg 1$ so that $\omega_R' \approx \omega_R$ and
$h_{HO}$ is given by Eq. \ref{eq:basic:green}. Note, that Eq.
\ref{eq:basic:ho_response} is valid for all (positive) values
of $Q$.} describes how the resonator reacts to a kick at time
$t = 0$ and is plotted in Fig.
\ref{fig:basic:harmonic_oscillator}a. The initial displacement
is zero, but the kick gives it the resonator a finite velocity
at $t = 0$. The resonator then oscillates back and forth with a
period $2\pi/\omega_R$ and these oscillations slowly die out
due to the dissipation. With the impulse-response function the
time evolution of the displacement for a force with arbitrary
time-dependence $F(t)$ can be obtained directly:
\begin{equation}
u(t)= h_{HO}(t) \conv F(t)/k_R, \label{eq:basic:convsol}
\end{equation}
where the symbol $\conv$ denotes convolution.

In many experiments the oscillator is driven with a periodic force
$F(t) = F_0 \cos(\omega t)$. After a short ($\sim \gamma_R^{-1}$)
transient, the resonator oscillates with the same frequency as the
driving force. This motion is not necessary in-phase with the
driving signal. Both the amplitude and phase of the motion are
quantified by the transfer function $H_{HO}(\omega)$ that is
obtained by taking the Fourier transformation\footnote{By
convention \cite{oppenheim_signals_systems}, the Fourier
transformation is defined as: $X(\omega) = \mathcal{F}[x(t)] =
\int_{-\infty}^{+\infty} x(t) \exp (-i\omega t) \intd t$ so that
the inverse transformation is given by: $x(t) =
\mathcal{F}^{-1}[X(\omega)] =
\frac{1}{2\pi}\int_{-\infty}^{+\infty} X(\omega) \exp(+i\omega t)
\intd \omega$.} of the equation of motion, Eq.
\ref{eq:basic:eqofmotion}:
\begin{equation}
H_{HO}(\omega) = k_R \frac{u(\omega)}{F(\omega)} = \frac{\omega^2
_R}{\omega_R^2-\omega^2 + i\omega\omega_R/Q}.
\label{eq:basic:ho_response}
\end{equation}
The magnitude $|H_{HO}(\omega)|$ and phase $\angle
H_{HO}(\omega)$ are plotted in Fig.
\ref{fig:basic:harmonic_oscillator}b. When the driving
frequency is far \replaced{below}{from} the resonance
frequency, the oscillator \replaced{adiabatically follows the
applied force: $u(t) = F(t)/k_R \Leftrightarrow H_{HO} = 1$,
and so}{hardly moves and} both $|H_{HO}|$ and $\angle H_{HO}$
are small. The small motion is then almost in phase with the
driving signal. The amplitude grows when sweeping the frequency
towards the natural frequency. Exactly on resonance, the
amplitude has its maximum $|H_{HO}| = Q$. The phase response
shows that at the resonance frequency, the displacement lags
the driving force by $-\pi/2$. When further increasing the
driving frequency, the oscillator can no longer follow the
driving force: the amplitude drops and the lag approaches
$-\pi$. The motion is then $180^\text{o}$ out of phase with the
applied force. The width of the resonance peak is related to
the damping: the full width at half maximum of the resonance
equals $\gamma_R = \omega_R/Q$.

\subsubsection{The harmonic oscillator in quantum mechanics}
\label{ssec:basic:quantum} In quantum mechanics the harmonic
oscillator is described by the Hamiltonian $\hat H = \hat p
^2/2m + \halfl m\omega_R^2 \hat u^2$ \cite{griffiths_qm}, as
the classical displacement coordinate $u$ and momentum $p = m
\dot u$ have to be replaced by the {\it operators} $\hat u$ and
$\hat p = -i\hbar \cdot
\partial/\partial u$. The displacement is described by a
wave function $\psi(u)$ that satisfies the time-independent
Schr\"odinger equation:
\begin{equation}
\hat H \psi = -\frac{\hbar^2}{2m}\pder{^2\psi}{u^2} + \half m
\omega_R^2 \hat u ^2 \psi = E\psi.
\end{equation}
This equation is solved by introducing the creation and
annihilation operators: $\hat a^ \dagger = \left(m\omega_R\hat
u - i\hat p \right)/\sqrt{2m\hbar\omega_R}$ and $\hat a =
\left( m\omega_R\hat u + i\hat p
\right)/\sqrt{2m\hbar\omega_R}$ respectively. The Hamiltonian
then becomes $\hat H = \hbar \omega_R (\hat n + \half)$, where
$\hat n = \hat a^ \dagger \hat a$ is the number operator that
counts the number of phonons in the oscillator. The
eigenenergies are $E_n = \hbar\omega_R(n + \half)$, with
eigenstates $\ket{n}$. The corresponding wave functions
$\psi_n(u)$ are plotted in Fig.
\ref{fig:basic:harmonic_oscillator}c. The lowest ($n = 0$)
eigenstate has a non-zero energy $E_0 = \half\hbar\omega_R$,
the so-called zero-point energy. Even when the oscillator
relaxes completely, it still moves around the potential minimum
at $u = 0$. The probability density of finding the resonator at
position $u$, is given by $|\psi_0(u)|^2$ when the resonator is
in the ground state. The zero-point motion $u_0$ is the
standard deviation of this probability density:
\begin{equation}
u_0 \equiv \left(\int_\infty^\infty u^2 |\psi_0(u)|^2 \intd u
\right)^{1/2} = \bra{0} \hat u^2 \ket{0} ^ {1/2} =
\sqrt{\frac{\hbar}{2m\omega_R}}.\label{eq:basic:zpm}
\end{equation}
The zero-point motion is an important length scale that determines
the quantum limit on continuous linear position measurement and is
also related to the effective resonator temperature as the
following Sections will show.

\subsection{Thermal and quantum noise}
\label{ssec:basic:noise} 
\begin{figure}[tb]{
\centering
\includegraphics{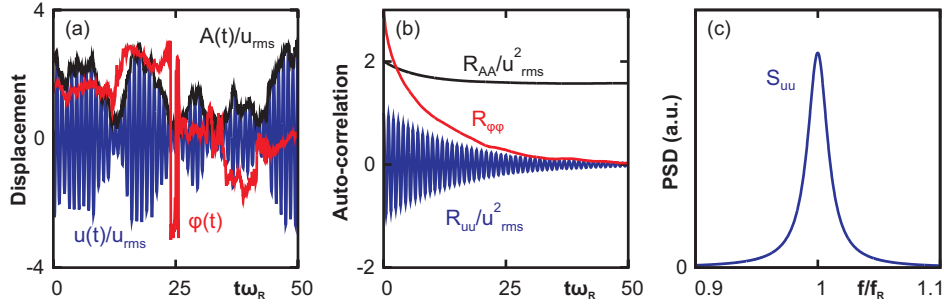}
\caption[Response of a resonator driven by thermal noise] {(a)
Simulated time-trace of the displacement (blue), amplitude (black)
and phase (red) of a resonator that is driven by Gaussian white
noise for $Q = 50$. The phase $\varphi$ is in radians and both the
displacement $u$ and amplitude $A$ are normalized by the
root-mean-square displacement $\langle u^2 \rangle ^{1/2}$. The
corresponding auto-correlation functions and the
displacement-noise power spectral density are shown in (b) and (c)
respectively. The area under the curve in (c) equals the variance
of the displacement. \label{fig:basic:oscillator_noise}}}
\end{figure}

In the previous Section, Sec.
\ref{ssec:basic:quantum} it was shown that a resonator always
moves because it contains at least the zero-point energy. In
practise, except at the lowest temperatures, the zero-point
motion is overwhelmed by thermal noise. Thermal noise is
generated by the environment of the resonator. As an example,
consider a resonator in air. At room temperature, the air
molecules have an average velocity of about $~500 \un{m/s}$.
The molecules randomly hit the resonator and every collision
gives the resonator a kick. These kicks occur independently of
each other, so the resonator experiences a stochastic force
$F_n(t)$. Other thermal noise sources are phonons in the
substrate that couple to the resonator via the clamping points,
fluctuating amounts of charge on nearby impurities and so on.
The environment of the resonator is thus a source of random
fluctuations on the oscillator. The force noise can be
described by an autocorrelation function $R_{F_n F_n}(t) =
\expect[F_n(t')F_n(t'+t)]$ (the symbol $\expect$ denotes the
expectation value) or by its power spectral
density\footnote{The engineering convention for the
single-sided power spectral density, $\overline S_{XX}(\omega)
= S_{XX}(\omega) + S_{XX}(-\omega)$, is used. Here,
$S_{XX}(\omega) = \mathcal{F}[R_{XX}]$ is the double-sided PSD
and $R_{XX}$ is its autocorrelation function. The variance of
$X$ is given by $\langle X^2 \rangle = R_{XX}(0) = (2\pi)^{-1}
\cdot \int_{-\infty}^\infty S_{XX}(\omega) \intd \omega =
(2\pi)^{-1} \cdot \int_0^\infty \overline S_{XX}(\omega) \intd
\omega$.} (PSD) $\overline S_{F_n F_n}(\omega)$
\cite{shanmugan_random}. For white noise, the latter is
independent of frequency and $F_n(t)$ has an infinite variance.
For a given $F_n(t)$ the realized displacement is easily found
using the Green's function, i.e., with Eq.
\ref{eq:basic:convsol}. Figure
\ref{fig:basic:oscillator_noise}a shows a simulated time-trace
of this so-called Brownian motion. The resonator oscillates
back and forth with frequency $\omega_R$ while its phase and
amplitude vary on a much longer timescale. The displacement can
be written as $u(t) = A(t) \cos[\omega_R t + \varphi(t)]$ (see
\appref{app:complex}) and the time-traces of the amplitude $A$
and phase $\varphi$ are plotted in Fig.
\ref{fig:basic:oscillator_noise}a as well. Figure
\ref{fig:basic:oscillator_noise}b shows the calculated
autocorrelation functions of the displacement, amplitude and
phase. The displacement autocorrelation $R_{uu}(t)$ displays
oscillations with period $2\pi/\omega_R$, whereas $R_{AA}$ and
$R_{\varphi\varphi}$ do not contain these rapid oscillations.
All three functions fall off at timescales $\sim Q/\omega_R$.
Note, that $R_{AA}(t)$ does not go to zero for long times,
because $\expect[A] > 0$ as $A(t)$ is always positive.

The displacement PSD is proportional to the force noise PSD and is
given by \cite{shanmugan_random}:
\begin{equation}
\overline S_{uu}(\omega) = k_R^{-2}|H_{HO}(\omega)|^2 \overline
S_{F_n F_n}.
\end{equation}
When $\overline S_{F_n F_n}$ is white in the bandwidth of the
resonator, which is typically assumed, $S_{uu}$ has the
characteristic bell shape shown\footnote{This \added{peak
shape} is often said to be \deleted{a} Lorentzian, although
formally that is not correct. The peak is proportional to
$[(\omega_R^2 - \omega^2)^2 + \omega_R^4/Q^2]^{-1}$, which can
be approximated by $\omega_R^{-2}/[(\omega_R - \omega)^2 +
\omega_R^2/Q^2]$ for $\omega \approx \omega_R$. The latter is
indeed a Lorentzian, but the approximation is only valid for
frequencies close to the resonance frequency of a high-Q
resonator.} in Fig. \ref{fig:basic:oscillator_noise}c.
Moreover, the PSD can be used to obtain the variance of the
displacement:
\begin{equation}
\langle u^2 \rangle = \frac{1}{2\pi} \int_0^\infty \overline
S_{uu}(\omega) \intd \omega = \frac{1}{4}  \frac{Q\omega_R}
{k_R^2} \overline S_{F_n F_n} = \frac{\pi}{2}  \frac{Q f_R}
{k_R^2} \overline S_{F_n F_n}. \label{eq:basic:variance}
\end{equation}
The force noise PSD is related to the temperature, and in
equilibrium the resonator temperature equals the environmental
temperature. The equipartition theorem
\cite{kittel_thermal_physics} relates the variance of the
displacement to the equilibrium temperature:
\begin{equation}
\half k_R \langle u^2 \rangle = \half m \langle \dot u^2 \rangle =
\half \kb T. \label{eq:basic:equipartition}
\end{equation}
The thermal energy $\kb T$ is distributed equally between the
potential energy and the kinetic energy. By combining Eqs.
\ref{eq:basic:variance} and \ref{eq:basic:equipartition}, a
relation between the force noise PSD and the properties of the
resonator is found:
\begin{equation}
\overline S_{F_n F_n}(\omega) = 4 \kb T m \omega_R/Q.
\label{eq:basic:fluctuation_dissipation}
\end{equation}
This so-called fluctuation-dissipation theorem
\cite{kittel_thermal_physics, nyquist_PR_thermalnoise,
weber_PR_fluctdiss} shows that on one hand the force noise PSD can
directly be obtained from the resonator properties and
temperature, without knowing its microscopic origin. On the other
hand, the force noise determines the dissipation (i.e., quality
factor) of the resonator.

\replaced{At}{In} equilibrium, the temperature of the resonator
is proportional to the variance of its Brownian motion as
indicated by Eq. \ref{eq:basic:equipartition}. However, out of
equilibrium the force noise is no longer given by Eq.
\ref{eq:basic:fluctuation_dissipation}, and the resonator
temperature can be different from $T$. The effective resonator
temperature $T_R$ is defined as:
\begin{equation}
T_R \equiv \frac{k_R \langle u^2 \rangle}{\kb}  = \frac{k_R}{\kb}
\int_0^\infty \overline S_{uu}(\omega) \frac{\intd \omega}{2\pi} =
\frac{k_R}{ \kb} \int_0^\infty |H_R|^2 \overline S_{FF}(\omega)
\frac{\intd \omega}{2\pi}, \label{eq:basic:TR}
\end{equation}
which yields $T_R = T$ in equilibrium. When the force noise is
larger than that of Eq. \ref{eq:basic:fluctuation_dissipation},
the effective resonator temperature is higher than the
environmental temperature. When $\langle u^2 \rangle$ is smaller
than its equilibrium value, $T_R < T$. Equation \ref{eq:basic:TR}
shows that the resonator temperature can be obtained from the
experimental displacement noise PSD: The resonator temperature is
proportional to the area under the curve (Fig.
\ref{fig:basic:oscillator_noise}c). Note that the suggestive
notation $\intd \omega/2\pi = \intd f$ is used in Eq.
\ref{eq:basic:TR} as in an measurement typically the real
frequency $f$ is on the horizontal axis and not the angular
frequency $\omega$.

\begin{figure}[tb]{
\centering
\includegraphics{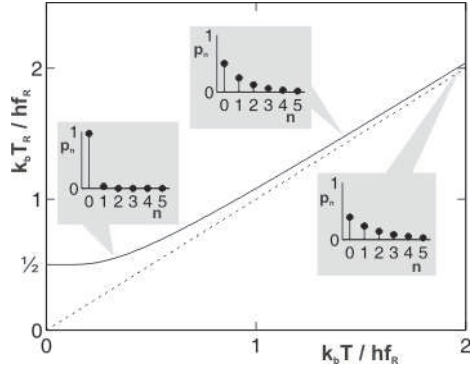}
\caption[Resonator temperature and occupation] {The resonator
temperature $T_R$ plotted against the environmental temperature
$T$. At low temperatures $T$ the resonator temperature saturates
at the zero-point energy: $T_R = \halfl \hbar \omega_R /\kb$ and
at high temperatures $T_R = T$ (dashed line). The insets show the
occupation probability $P_n$ of the energy levels at $\kb T /
\hbar \omega_R = 0.3,~1/\ln 2 \approx 1.44$ and 2.0.}
\label{fig:basic:TR_occupation}}
\end{figure}

When the resonator is cooled to very low temperatures where $\kb T
\sim \hbar \omega_R$, the classical description breaks down as the
quantized energy-level structure (Fig.
\ref{fig:basic:harmonic_oscillator}c) becomes important.
Semi-classically, the thermal and quantum effects are combined by
replacing the force noise of Eq.
\ref{eq:basic:fluctuation_dissipation} with the Callen and Welton
equation \cite{callen_physrev_noise}:
\begin{equation}
\overline S_{F_n F_n}(\omega) = \frac{4 m \omega}{Q} \cdot \half
\hbar \omega \coth \left(\frac{\hbar \omega}{2 \kb T} \right),
\end{equation}
so that for $Q \gg 1$:
\begin{equation}
\langle u^2 \rangle = u_0^2 \cdot \coth \left(\frac{\hbar
\omega_R}{2 \kb T} \right) \hspace{3mm} \Leftrightarrow
\hspace{3mm} T = \frac{\hbar \omega_R}{\kb} \ln^{-1}
\left( \frac{\langle u^2 \rangle + u_0^2 } {\langle u^2 \rangle -
u_0^2}\right).\label{eq:basic:coth}
\end{equation}
Then by inserting the first part of Eq. \ref{eq:basic:coth} in
Eq. \ref{eq:basic:TR} the resonator temperature is
obtained\footnote{Note, that some authors use Eq.
\ref{eq:basic:coth} instead of Eq. \ref{eq:basic:TR} as the
definition of $T_R$, which implies that when $\langle u^2
\rangle = u_0^2$, $T_R =0$. This in contrast to the definition
used here where $T_R = \added{\half} \hbar \omega_R /\kb$ for
$\langle u^2 \rangle = u_0^2$. In the latter case $T_R$ is not
the actual temperature, but it is a measure for the (quantum or
thermal) fluctuations.}. The dependence of $T_R$ on the
environmental temperature is shown in Fig.
\ref{fig:basic:TR_occupation}. At high temperatures ($\kb T \gg
\hbar \omega_R$) the resonator temperature is the temperature
of the environment: $T_R = T$. At zero temperature the
resonator temperature is determined by the quantum
fluctuations: $T_R = \halfl \hbar \omega_R /\kb$.

In thermal equilibrium, the energy levels of a harmonic oscillator
have occupation probabilities that are given by
\cite{kittel_thermal_physics}:
\begin{equation}
P_n = \left(e^{-\frac{\hbar \omega_R}{\kb T}} \right)^n \left(1-
e^{-\frac{\hbar \omega_R}{\kb T}}\right).
\end{equation}
The average thermal occupation is $\overline n = \sum_{n =
0}^\infty n P_n = [\exp (\hbar \omega_R/\kb T) -1]^{-1}$
\cite{kittel_thermal_physics}, which equals $\kb T_R/ \hbar
\omega_R - \half$. The insets in Fig.
\ref{fig:basic:TR_occupation} show the occupation probabilities at
three different temperatures $T$. At low temperature the resonator
is in the ground state most of the time. At $\kb T = \ln 2 ~ \hbar
\omega_R$ the probability finding the resonator in the ground
state is exactly 50\% and the average occupation is $\overline n =
1$. At any non-zero temperature, there is always a finite
probability to find the resonator in an excited state. With the
statement that ``the resonator is cooled to its ground state'' one
actually means $\overline n \lesssim 1$
\label{pg:basic:ground_state}. In Sec. \ref{ssec:basic:cooling} we
will explain in detail how this goal can be achieved using
different cooling techniques, but now we will focus on the
detection of the resonator position, in particular on the role of
the detector, backaction and the standard quantum limit.

\subsection{Backaction and quantum limits on position detection}
\label{ssec:basic:sensitivity} Since the discovery of the
Heisenberg uncertainty principle in 1927 it is known that
quantum mechanics imposes limitations on the uncertainty with
which quantities can be measured. This was first discovered for
single measurements of conjugate variables, such as the
position $u$ and momentum $p$ of a particle, or the components
$\sigma_x, \sigma_y$ and $\sigma_z$ \added{of the spin} of a
spin-1/2 particle.

When \replaced{a}{the} system is initially in a superposition
of the eigenstates of the operator corresponding to the
quantity that is measured, a strong measurement gives one of
the eigenvalues as the outcome \cite{griffiths_qm}. The
probability for measuring a value $\mu_i$ is $|c_i|^2$ when
$\ket{\psi} = \sum_i c_i \ket{\mu_i}$ was the expansion of the
original state in the basis of eigenstates of the operator
$\hat \mu$, with $\hat \mu \ket{\mu_i} = \mu_i \ket{\mu_i}$. At
the same time, the wave function collapses into the state
corresponding to the measured value $\mu_i$: $\ket{\psi}
\rightarrow \ket{\mu_i}$. To obtain the probability of a
certain outcome of a measurement of a different quantity $\nu$,
the state $\ket{\mu_i}$ has to be expanded in the basis
$\ket{\nu_i}$. The uncertainties in $\mu$ and $\nu$ satisfy:
$\Delta \mu \cdot \Delta \nu \ge \big| [\hat \mu, \hat \nu]/2i
\big|$, where $[\hat \mu, \hat \nu] = \hat \mu \hat \nu - \hat
\nu \hat \mu$ is the commutator of $\hat \mu$ and $\hat \nu$.
For $\hat \mu = \hat u$ and $\hat \nu = \hat p$, this yield the
Heisenberg uncertainty principle for position and momentum:
$\Delta u \cdot \Delta p \ge \hbar/2$. Note that quantum
mechanics does not forbid to determine the position with
arbitrary accuracy in a single measurement.

Most measurements are, however, not single, strong measurements,
but weak continuous measurements instead
\cite{jacobs_CP_continuous_measurements}. As a measurement of the
position disturbs the momentum of the resonator, a subsequent
measurement of the position after a time $\Delta t$ inevitably is
influenced by the previous measurement. This backaction is
therefore important in continuous linear displacement detectors.
In an experiment, backaction results in three different effects on
the resonator: a frequency shift, a change in damping, and a
change in the resonator temperature. If the resonator temperature
is lower than the bath temperature, backaction has led to
self-cooling, whereas a higher temperature indicates that there is
a net energy flow from the detector to the resonator.

A way to circumvent backaction is to perform measurements on
the position {\it squared}. As we show in \appref{app:squared},
such a square-law detector probes the energy states of the
resonator and these are not disturbed by the measurement
\added{itself} as the energy \added{operator equals, and hence}
commutes with, the Hamiltonian of the system. Therefore, this
measurement scheme is called a quantum non-demolition (QND) or
a backaction evading (BE or BAE) measurement
\cite{braginsky_science_QND}.

In the following section, continuous (linear) detectors and
their backaction are discussed in detail and the quantum limits
on position detection are explored. The analysis presented is
largely based on the work by Clerk and co-workers
\cite{clerk_PRB_ql_linear, clerk_overview}. We start with an
analysis of generic linear detectors and what the effects of
backaction are. Then we discuss three different routes to
arrive at the quantum limit: the Haus-Caves derivation, the
power-spectral density method, and the optimal estimator
approach. These all have different ranges of applicability and
rigor but lead to the same conclusion: the quantum limited
resolution for continuously monitoring the resonator position
is approximately the zero-point motion $u_0$.

The sensitivity and resolution of a detector will be important
concepts in the following discussion. These notions, however,
sometimes lead to confusion and, before defining them
mathematically, we will first discuss the similarities
differences between them. Both are a measure of the imprecision
with which the position is measured. A noisy detector has a bad
sensitivity and a bad position resolution, whereas a good
detector has a good sensitivity and resolution.
Counterintuitively, this means that the latter detector has
{\it lower sensitivity values}, although one often says that it
has a {\it higher sensitivity}. This is the first point that
leads to confusion. The second is the difference between
sensitivity and resolution. This can be understood as follows:
Consider a resonator that is standing still and that is coupled
to a detector that inevitably introduces noise in its output
signal. When measuring the position for a short period, the
inferred position has a large uncertainty due to the detector
noise. By measuring longer, the noise averages out and the
uncertainty decreases. This uncertainty is the resolution of
the detector and is measured in units of length. It thus
depends on the duration of the measurement $\Delta t$. For
white detector noise, the resolution improves (i.e. its value
decreases) as $1/\sqrt{\Delta t}$. The proportionality constant
between the resolution and the measuring time is the
sensitivity, and it has the units of m/$\sqrt{\text{Hz}}$. In
the discussion of Fig. \ref{fig:basic:timetrace_spectrum} it
will be shown that the sensitivity is easily extracted from the
noise spectrum of the detector output.

\subsubsection{Continuous linear detectors}
\label{sssec:basic:lineardetectors} 
\begin{figure}[tb]
\centering
\includegraphics{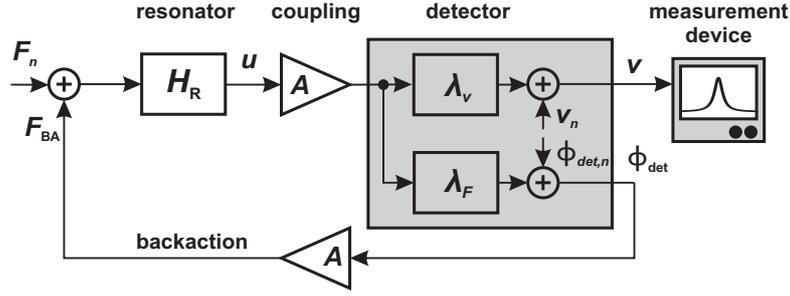}
\caption[Action and backaction in a linear detector] {The detector
is coupled (via $A$) to the resonator. Its output $v(t)$ depends
linearly on $u(t)$ with a response function $\lambda_v$ but also
contains detector noise $v_n$. The detector exerts a backaction
force $F_{BA} = A \Phi_{det}$ on the resonator that contains a
stochastic part $F_{BA,n}$ and a linear response to $u(t)$. Both
contributions add up with the thermal or quantum force noise
$F_n$. The resonator displacement $u$ is obtained via the transfer
function $H_R = H_{HO}/k_0$. \label{fig:basic:linear_detector}}
\end{figure}

\begin{figure}[tb]{
\centering
\includegraphics{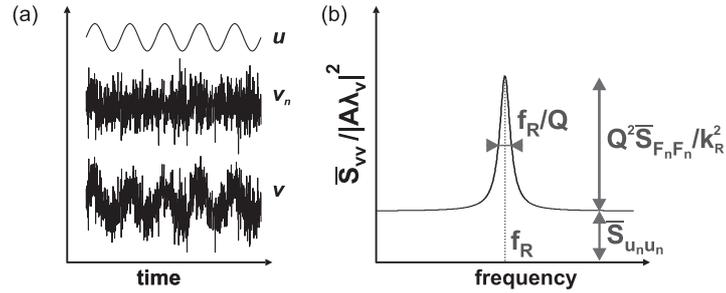}
\caption{The resonator displacement $u$ is measured by the
detector. The detector also adds imprecision noise $v_n$. The sum
of this noise and the physical displacement $v$ is recorded. (a)
shows schematically time traces of the resonator displacement $u$,
imprecision noise $v_n$ and the apparent displacement $v$. Often,
one is interested in the noise spectrum of the detector output,
especially when the resonator displacement is its Brownian motion.
This can be done using a spectrum analyzer and (b) shows that from
a resulting spectrum the resonance frequency $f_R$, the quality
factor $Q$, the detector noise floor $\overline S_{u_n u_n}$ and
the force noise $\overline S_{F_n F_n}$ are readily extracted. The
signal-to-noise ratio is the height of the resonance peak divided
by the height of the noise floor. It is therefore advantageous to
have a large quality factor as this leads to a larger signal-to
noise ratio.} \label{fig:basic:timetrace_spectrum}}
\end{figure}

A continuous linear position
detector gives an output that depends linearly on the current and
past position of the resonator. Figure
\ref{fig:basic:linear_detector} shows the scheme of a generic
linear detector. Note again, that the analysis is for a generic
continuous linear detector and quantum effects only come into play
in Sec. \ref{sssec:basic:haus-caves}. We will discuss all its
elements step by step. The output signal of the
detector\footnote{Continuous linear detectors are usually
(implicitly) assumed to be time-invariant
\cite{caves_PRD_QL_amplifiers}. Unless stated otherwise, this is
also assumed in this Report. Examples where the linear detector is
not time-invariant are frequency-converting, stroboscopic and
quadrature measurements \cite{caves_PRD_QL_amplifiers,
braginsky_science_QND, caves_RMP, braginsky_quantum_measurement}.}
is related to the displacement by:
\begin{equation}
v(t) = A \lambda_v(t) \conv u(t) + v_n(t).\label{eq:basic:convout}
\end{equation}
Here, $\lambda_v$ is the responsivity of the detector, with
$\lambda_v(t) = 0$ for $t < 0$ \added{due to causality}. $v_n$
is the detector noise and $A$ is a dimensionless coupling
strength. The coupling between the resonator and the detector
is an important element when discussing different detectors in
Sec. \ref{sec:detectors}. The choice of the coupling $A$ is
slightly arbitrary as it could also be incorporated in
$\lambda_v$. However, in most detectors it is possible to make
a distinction between the coupling to the resonator and the
output. Different types of detector will be discussed
extensively in Sec. \ref{sec:detectors}, but in the case of an
optical interferometer, $v$ represents the number of photons
arriving at the photon counter, in a single-electron transistor
it is the current through the island, and in a dc
superconducting-interference-device detector it represents the
output voltage. The difference between $u$, $v$ and $v_n$ is
illustrated in Fig. \ref{fig:basic:timetrace_spectrum}a. Note,
that often the detector responds instantaneously to the
displacement. In that case $A\lambda_v(t) =
\pderl{v}{u}~\delta(t)$ and $A\lambda(\omega) = \pderl{v}{u}$.
The frequency response of the resonator is then
flat\footnote{When the output of the detector has a delay time
$\tau$, $A\lambda_v(t) = \pderl{v}{u}~\delta(t - \tau)$ and
$A\lambda_v(\omega) = \pderl{v}{u}~\exp(-i \omega \tau)$.}.

\added{To calculate the sensitivity of the detector, the noise
at the output of the detector, $v_n$, is referred back to the
input using the known response function $\lambda_v$ and gain
$A$. This yields the displacement noise $u_n$ at the detector
input. The equivalent input noise PSD is $\overline S_{u_n u_n}
= \overline S_{v_n v_n}/|A\lambda_v|^2$. This power spectral
density is an important parameter that characterizes the
detector. In nanoelectromechanical experiments, this noise
floor is usually determined by the classical noise in the
electronics of the measurements setup. In optical experiments,
however, the noise floor can be shot-noise limited; quantum
mechanics now sets the imprecision of the experiment. We will
discuss the quantum limit in the next subsection. Note that for
a flat frequency response $u_n$ is simply given by
$v_n/A\lambda_v$.}

The detector does not only add noise to the measured signal, it
also exerts a force $F_{BA}(t)$ on the resonator. This is the
so-called backaction force. Backaction, in its most general
definition, is the influence of a measurement or detector on an
object. The detector backaction is a force on the resonator.
This can be seen from the following argument: when there is no
coupling between the resonator and the detector, i.e., $A = 0$,
the Hamiltonian describing the total system is the sum of that
of the Hamiltonian of the oscillator and of the detector
Hamiltonian. When these are coupled ($A \neq 0$), there is an
interaction Hamiltonian of the form $H_{int} = -A \Phi_{det}
u$. The backaction force is then $F_{BA} = - \pderl{H_{int}}{u}
= A \Phi_{det}$ \cite{clerk_PRB_ql_linear}. The backaction
force is thus also proportional to the detector-resonator
coupling $A$. The quantity $\Phi_{det}$ is related to one of
the internal variables of the detector; for example in an
optical cavity it is proportional to the number of photons, in
a SET to the electron occupation, and in a SQUID to the
circulating current.

The backaction force has three different contributions:
\begin{itemize}
\item A deterministic force that is independent of the
    displacement. This changes the equilibrium position of
    the resonator. Without loss of generality it can thus
    be set to zero and it is not considered here.
\item A force that responds linearly to the displacement:
    $F_{BA,u} = A \lambda_F(t) \conv Au(t)$. This changes
    the effective resonator response from $H_R$ to $H_R'$,
    where $H_R'^{-1} = H_R^{-1}+A^2 \lambda_F(\omega)$. An
    example of this force is the optical spring that we
    will encounter in Sec. \ref{ssec:basic:sideband}. This
    part of the backaction can lead to cooling, see Sec.
    \ref{sssec:basic:backactioncooling} \cite{naik_nature}.
\item A stochastic force $F_{BA, n} \equiv A \Phi_{det,n}$
    that is caused by the fluctuations in the detector,
    $\Phi_{det,n}$. Note, that this force noise and the
    imprecision noise $v_n$ can be correlated, i.e.,
    $\overline S_{\Phi_{det,n} v_n} \neq 0$. In any case,
    this contribution tries to heat the resonator
    \added{since it adds up to the original thermal force
    noise.}.

\end{itemize}
In summary, this whole process of action and backaction is the one
shown schematically in Fig. \ref{fig:basic:linear_detector}: The
resonator position is coupled to the input of the detector, which
adds imprecision noise $v_n$ and exerts force noise on the
resonator. For small coupling $A$, the statistical properties of
$\Phi_{det, n}$ and $v_n$ are independent of the resonator
displacement \cite{clerk_PRB_ql_linear, clerk_overview}.

\deleted{To calculate the sensitivity of the detector, the
noise at the output of the detector, $v_n$, is referred back to
the input using the known response function $\lambda_v$ and
gain $A$. This yields the displacement noise $u_n$ at the
detector input. The equivalent input noise PSD is $\overline
S_{u_n u_n} = \overline S_{v_n v_n}/|A\lambda_v|^2$. This power
spectral density is an important parameter that characterizes
the detector. In nanoelectromechanical experiments, this noise
floor is usually determined by the classical noise in the
electronics of the measurements setup. In optical experiments,
however, the noise floor can be shot-noise limited; quantum
mechanics now sets the imprecision of the experiment. We will
discuss the quantum limit in the next subsection.}

One way to analyze the data is to measure the detector output
using a spectrum analyzer. This way, information about the
resonance frequency, quality factor, and imprecision and force
noise PSD can easily be obtained, as illustrated in Fig.
\ref{fig:basic:timetrace_spectrum}b. However, in a linear
detection scheme one is usually interested in measuring the
resonator position as accurate as possible, without perturbing
the resonator considerably. Using optimal-control and
estimation theory, the best estimate $\hat u$ for the resonator
displacement in the absence of the detector, $u_i$, is found.
The resolution of the detector is $\Delta u =
\left(\expect{[u_i^2 -\hat u^2]}\right)^{1/2}$, as explained in
detail in \appref{app:optimal}. It quantifies the difference
between the displacement that the resonator would have had when
it was not measured and the one reconstructed from the detector
output. By rewriting the resolution as $\Delta u =
\left(\expect{[(u^2 - \hat u^2) + (u_i^2 - u^2)
]}\right)^{1/2}$ it becomes clear that there are two
contributions \cite{tittonen_PRA_torsional_resonator}: the
first one\deleted{ quantifies}, which we name $\Delta u_n$,
indicates how well the realized displacement is reconstructed
from the detector output, whereas the second term, $\Delta
u_{BA}$, quantifies the difference between $u_i$ and the
realized displacement, i.e., how much the motion is perturbed.
Heuristically, we can understand that the first term is due to
the imprecision of the detector, whereas the second is due to
the backaction.

\begin{figure}[tb]
\centering
\includegraphics{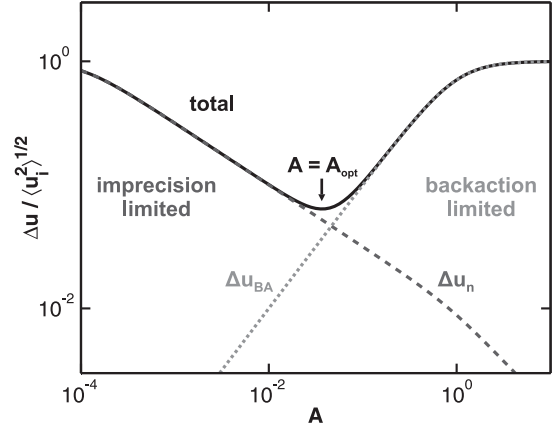}
\caption[Resolution vs coupling] {(a) The position resolution of
the detector in units of the amplitude of the Brownian motion \added{$\langle u_i^2 \rangle^{1/2}$}
(classical limit) for different coupling strengths $A$. The solid
line is the total resolution; the dotted line is the contribution
of the backaction force noise (Eq. \ref{eq:basic:resolution_ba})
and the dashed line the contribution of the imprecision noise (Eq.
\ref{eq:basic:resolution_imp}). The total resolution is optimized
at $A = A_{\text{opt}} \approx 0.0355$. The resolution has been
calculated for $Q_0 = 10^4$, $\lambda_v = 1$, $\lambda_F = 0$,
$\overline S_{v_n v_n} \cdot k_0^2 = \overline S_{\Phi_{det,n}
\Phi_{det,n}} = \overline S_{F_n F_n}$ and $\overline
S_{\Phi_{det,n} v_n} = 0$. \label{fig:basic:resolution_coupling}}
\end{figure}

The resolution is plotted in Fig.
\ref{fig:basic:resolution_coupling}a as a function of the
coupling strength $A$. In experiments this coupling strength is
an important parameter and Sec. \ref{sec:detectors} we will
give typical numbers for the different detection schemes. We
now discuss the general features of Fig.
\ref{fig:basic:resolution_coupling} in more detail. With a low
coupling $\Delta u$ is large (i.e., the detector has a low
resolution) because of the large imprecision noise contribution
$\Delta u_n$ (dashed line) of the detector. The backaction
contribution is very small. An increase of the coupling reduces
$\Delta u$ because by increasing the coupling, the mechanical
signal becomes larger whereas $S_{v_n v_n}$ remains the same,
so the relative contribution of the imprecision noise
decreases. The increase of the coupling also raises the
backaction contribution, which still remains small. The
resolution improves with increasing $A$ up to the point where
the optimal value $A = A_{\text{opt}}$ is reached. A further
increase of $A$ makes the backaction force noise dominant,
driving the resonator significantly, thus yielding a higher
$\Delta u$.

\subsubsection{The Haus-Caves derivation of the quantum
limit}\label{sssec:basic:haus-caves} The system analysis of the
linear detector discussed above is valid for any -- quantum
limited or not -- linear detector. An elegant way of deriving
the quantum limit of a continuous linear position detector was
given by Haus and Mullen \cite{haus_PR_quantum_limit} and was
extended by Caves \cite{caves_PRD_QL_amplifiers}. They consider
the situation where the input and output signal of the detector
are carried by single bosonic modes, $\hat a_u$ and $\hat a_v$
respectively. When the ``photon number gain'' of the detector
is $G = \langle \hat a_v^\dagger \hat a_v \rangle / \langle
\hat a_u^\dagger \hat a_u \rangle$, one might think that the
modes are related to each other by $\hat a_v = \sqrt G \hat
a_u$. This is, however, not valid as this gives $[\hat a_v,
\hat a_v^\dagger] = G$ instead of the correct value, 1
\cite{caves_PRD_QL_amplifiers, clerk_overview}. The actual
relation is $\hat a_v = \sqrt G \hat a_u + \hat v_n$. Here,
$\hat v_n$ represents the noise added by the amplifier. This
operator has a vanishing expectation value ($\langle \hat v_n
\rangle = 0$) and is uncorrelated with the input signal ($[\hat
a_u, \hat v_n] = [\hat a_u^\dagger, \hat v_n] = 0$). Requiring
$[\hat a_v, \hat a_v^\dagger] = 1$ yields $[\hat v_n, \hat
v_n^\dagger] = 1-G$ for the commutator and, more importantly,
$\Delta a_v ^2 \geq G \Delta a_u^2 + \halfl |G-1|$ for the
noise in the number of quanta of the output mode
\cite{clerk_overview}. The first term is the amplified
\replaced{input signal}{zero-point motion of the input mode}
(i.e., the resonator motion) and the second one is the noise
added by the amplifier. In the limit of large gain\footnote{In
the opposite limit where the detector does not have any
\added{net} gain, i.e., $G = 1$, no additional noise is
required by quantum mechanics.} ($G \gg 1$), the equivalent
input noise of the detector is $\Delta (a_u^{eqv})^2 \equiv
\Delta a_v ^2 / G - \Delta a_u^2 \geq \halfl$. This means that
a quantum-limited detector adds at least half a vibrational
quantum of noise to the signal.

As pointed out in Ref. \cite{clerk_overview} most practical
detectors cannot easily be coupled to a single bosonic mode
that carries the information of the resonator to the detector,
because there is also a mode that travels from the detector
towards the resonator. Therefore, the linear-system analysis at
the beginning of this Section is used to further explore the
quantum limits on continuous linear position detection.

\subsubsection{A quantum-limited detector}
In the previous discussion, no constraints were enforced on the
detector noises $\Phi_{det,n}$ and $v_n$. If both noise
contributions could be made small enough, the resolution would be
arbitrarily good. This is unfortunately not possible. It can be
shown that the power spectral densities must
satisfy\footnote{Here, it is assumed that the measurement of
$v(t)$ does not result in an additional force noise on the
resonator and that the detector has a large power gain. For more
details, see Ref. \cite{clerk_overview}.}:
\begin{equation}
\overline S_{v_n v_n}(\omega) \cdot \overline S_{\Phi_{det,n}
F_{det,n}}(\omega) - \left| \overline S_{\Phi_{det,n} v_n}(\omega)
\right|^2 \ge \left| \hbar \lambda_v(\omega) \right|^2,
\end{equation}
or, equivalently, when this is referred to the input:
\begin{equation}
\overline S_{u_n u_n}(\omega) \cdot \overline S_{F_{BA,n}
F_{BA,n}}(\omega) - \left| \overline S_{F_{BA,n} u_n}(\omega)
\right|^2 \ge \hbar ^2.
\end{equation}
These constraints enforce the quantum limit of the linear detector
and should be considered as the continuous-detector equivalent of
the Heisenberg uncertainty principle: Accurately measuring the
position results severe force noise and vice versa.

\begin{figure}[tb]
\centering
\includegraphics{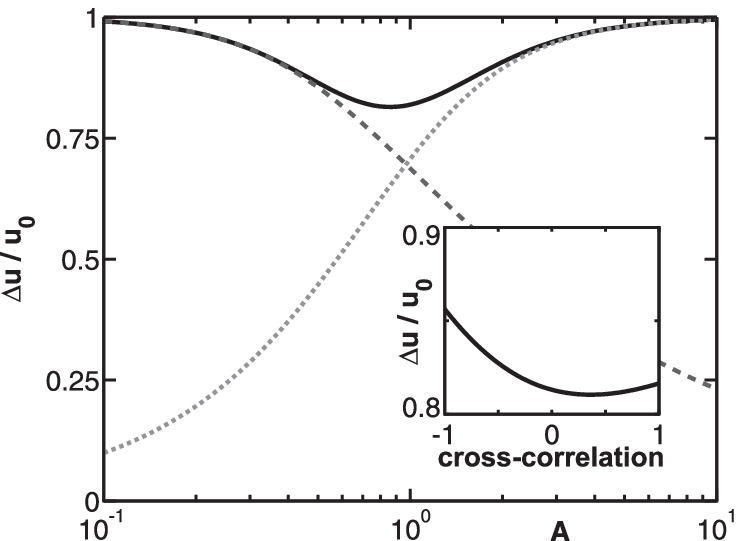}
\caption[Quantum limit of the resolution] {The quantum limit for a
continuous linear position detector using the optimal estimation
method. The resolution of the detector in units of the zero-point
motion is plotted for different coupling strengths $A$. The solid
line is the total resolution, $\Delta u$; the dotted and dashed
line are the contribution of backaction, $\Delta u_{BA}$, and
imprecision noise, $\Delta u_n$, respectively.
The inset shows that the total resolution depends on the
cross-correlation coefficient $S_{F_{BA,n} u_{BA,n}}/(S_{F_{BA,n}
F_{BA,n}} \cdot S_{u_n u_n})^{1/2}$. The main panel is calculated
with an optimal cross-correlation of 0.36. For all values of $A$
the total added noise is slightly less than $u_0$.
\label{fig:basic:resolution_QL}}
\end{figure}

Clerk {\it et al.} \cite{clerk_overview} continue now by
finding the gain where the total added noise at the input,
i.e., $\overline S_{u_n u_n}(\omega) + |H_R'|^2\overline
S_{F_{BA,n} F_{BA,n}}(\omega)$, is minimized. As they already
point out, this is not entirely correct because at every
frequency a different optimal gain is required. Usually, only
the optimal gain at $\omega = \omega_0$ is used and then the
magnitude of the signal and detector noise are equal at that
frequency. In that case, the imprecision noise and the
backaction-induced displacement provide exactly half of the
total added noise \cite{clerk_overview, lahaye_science,
teufel_natnano_beyond_SQL}. However, by optimizing the total
{\it resolution}, the true optimal gain is found, see Fig.
\ref{fig:basic:resolution_QL}. The resolution is optimized at
$A = 0.87$ and reaches a value of $0.81$ times the zero-point
motion.

All three methods (the Haus-Caves derivation, the total added
noise at the input, and the optimal estimator) indicate that the
detector adds about the same amount of noise as the zero-point
fluctuations of the resonator itself.

\subsection{Cooling}
\label{ssec:basic:cooling} 
\begin{figure}[tb]{
\centering
\includegraphics{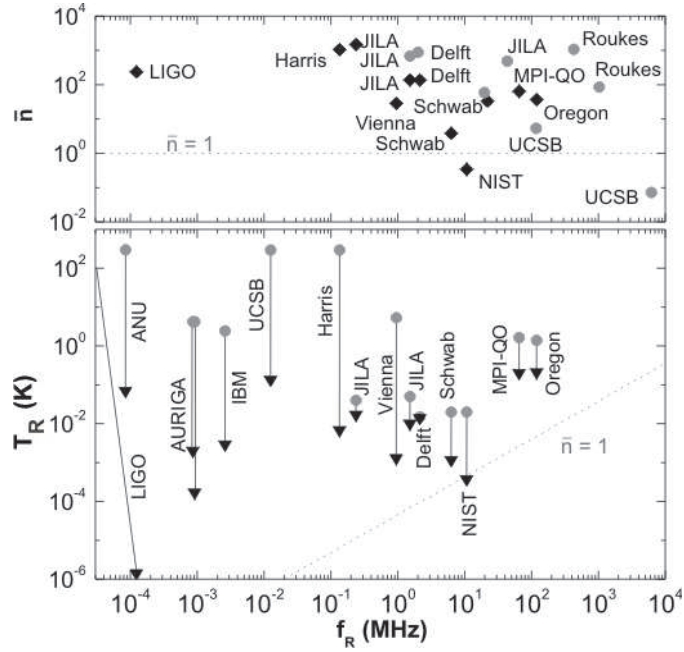}
\caption[Cooling of resonators] {Overview of mechanical resonators
with low temperature or occupation, compiled from Table
\ref{tab:basic:cooling}. The top panel shows the experiments with
the lowest occupation numbers. These are reached using
conventional cooling (gray) and active feedback and sideband
cooling (black). The dotted line is located at $\overline n = 1$. As
discussed on page \pageref{pg:basic:ground_state}, once the
thermal occupation is below this value, the resonator is cooled to
the ground state.
The bottom panel shows the starting temperature $T$ (gray) and the
final temperature $T_R^{\min}$ (black) achieved by groups that
have actively cooled their resonator below $100 \un{mK}$. Note
the diagonal line in the LIGO experiment which is
due to a strong optical spring effect (see Sec.
\ref{ssec:basic:sideband})\replaced{, so b}{. B}oth the
resonator temperature and frequency change when the laser power is
increased \cite{corbitt_PRL_cooling_mK}.
\label{fig:basic:basic_cooling}}}
\end{figure}

To prepare a nanomechanical system
in the ground state, the thermal occupation of its normal modes
should be minimal. The most direct approach is to mount an
ultra-high frequency ($f_R > 1 \un{GHz}$) resonator in a
dilution refrigerator ($T < 50 \un{mK}$) so that  $\overline n
\leq 1$. Such a resonator will, however, have a very small
zero-point motion and the readout of tiny high-frequency
signals at millikelvin temperatures is difficult. An
alternative approach is to perform the experiments with
lower-frequency resonators and/or at higher temperatures. The
thermal occupation is then higher than one and cooling
techniques have to be used to reduce the temperature of the
resonator $T_R$ well below the environmental temperature $T$.
Figure \ref{fig:basic:basic_cooling} and Table
\ref{tab:basic:cooling} show how recent experiments are
approaching the limit $n \leq 1$ over a range of frequencies
that spans seven orders of magnitude.
\rowcolors{1}{white}{bkgnd}

\begin{table}[htbp] \centering
\caption{Overview of resonator temperature and cooling of
recent experiments with micro- and nanomechanical resonators.
The table shows the resonance frequency $f_R$, the temperature
of the environment $T$, the minimum resonator temperature
$T^{\min}_R$ and the corresponding number of quanta. The
numbers of the first column correspond to the experiments
listed in Table \ref{tab:intro:short}.}
\begin{tabular}{rccccccr}
\addlinespace \toprule
                          & $\mathbf{f_R (\unitt{MHz})}$ & $\mathbf{T \unitt{(K)}}$  & $\mathbf{T^{\min}_R \unitt{(K)}}$ & $\mathbf{\bar n}$         & {\bf Cooling method}      & {\bf Factor}              & {\bf Ref.} \\
\midrule
1                         & $1.0 \cdot 10^{3}$        & 4.2                       & 4.2                       & $85$                      &                           &                           & \cite{huang_nature_GHz} \\
2                         & $117$                     & 0.030                     & 0.030                     & $5.3$                     &                           &                           & \cite{knobel_nature_set} \\
3                         & $20$                      & 0.035                     & 0.056                     & $59$                      &                           &                           & \cite{lahaye_science} \\
4                         & $7.3 \cdot 10^{-3}$       & 295                       & 18                        & $5.1 \cdot 10^{7}$        & Photothermal              & 16.3                      & \cite{metzger_nature_cooling} \\
5                         & $22$                      & 0.030                     & 0.035                     & $33$                      & Backaction                &                           & \cite{naik_nature} \\
6                         & $0.81$                    & 295                       & 10                        & $2.6 \cdot 10^{5}$        & Sideband                  & 29.3                      & \cite{arcizet_nature_cavity} \\
7                         & $0.013$                   & 295                       & 0.14                      & $2.2 \cdot 10^{5}$        & Feedback                  & $2.2 \cdot 10^{3} $       & \cite{kleckner_nature_feedback} \\
8                         & $0.28$                    & 295                       & 10                        & $7.5 \cdot 10^{5}$        & Sideband                  & 29.3                      & \cite{gigan_nature_cavity} \\
9                         & $0.81$                    & 295                       & 5.0                       & $1.3 \cdot 10^{5}$        & Feedback                  & 58.6                      & \cite{arcizet_PRL_sensitive_monitoring} \\
10                        & $58$                      & 300                       & 11                        & $4.0 \cdot 10^{3}$        & Sideband                  & 27.1                      & \cite{schliesser_PRL_cavity_cooling} \\
11                        & $127$                     & 295                       & 295                       & $4.8 \cdot 10^{4}$        &                           &                           & \cite{li_natnano_piezo} \\
12                        & $1.7 \cdot 10^{-4}$       & 295                       & 0.80                      & $9.7 \cdot 10^{7}$        & Sideband                  & 366.5                     & \cite{corbitt_PRL_opticalspring} \\
13                        & $1.3 \cdot 10^{-5}$       & 295                       & $6.9 \cdot 10^{-3} $      & $1.1 \cdot 10^{7}$        & Sideband                  & $4.2 \cdot 10^{4} $       & \cite{corbitt_PRL_cooling_mK} \\
14                        & $0.55$                    & 300                       & 175                       & $6.7 \cdot 10^{6}$        & Photothermal              & 1.7                       & \cite{favero_APL_cooling_micromirror} \\
15                        & $43$                      & 0.25                      & 1.0                       & $483$                     &                           &                           & \cite{flowers_PRL_APC} \\
16                        & $2.6 \cdot 10^{-3}$       & 2.2                       & $2.9 \cdot 10^{-3} $      & $2.3 \cdot 10^{4}$        & Feedback                  & 85.7                      & \cite{poggio_PRL_feedback} \\
17                        & $7.0 \cdot 10^{-3}$       & 295                       & 45                        & $1.3 \cdot 10^{8}$        & Sideband                  & 6.5                       & \cite{brown_PRL_circuit_cooling} \\
18                        & $0.71$                    & 295                       & 295                       & $8.6 \cdot 10^{6}$        &                           &                           & \cite{caniard_PRL_backaction_cancelation} \\
19                        & $0.13$                    & 294                       & $6.8 \cdot 10^{-3} $      & $1.1 \cdot 10^{3}$        & Sideband                  & $4.3 \cdot 10^{4} $       & \cite{thompson_nature_cavity_membrane} \\
20                        & $0.56$                    & 35                        & 0.29                      & $1.1 \cdot 10^{4}$        & Sideband                  & 120.0                     & \cite{groeblacher_EPL_sideband} \\
21                        & $8.5 \cdot 10^{-5}$       & 300                       & 0.070                     & $1.7 \cdot 10^{7}$        & Feedback                  & $4.3 \cdot 10^{3} $       & \cite{mow_PRL_feedback_cooling} \\
22                        & $74$                      & 295                       & 21                        & $5.9 \cdot 10^{3}$        & Sideband                  & 14.0                      & \cite{schliesser_natphys_sideband} \\
23                        & $0.24$                    & 0.040                     & 0.017                     & $1.5 \cdot 10^{3}$        & Sideband                  & 2.3                       & \cite{regal_natphys_cavity} \\
24                        & $428$                     & 22                        & 22                        & $1.1 \cdot 10^{3}$        &                           &                           & \cite{feng_natnano_selfsustaining} \\
25                        & $5.0 \cdot 10^{-3}$       & 4.2                       & 4.2                       & $1.8 \cdot 10^{7}$        &                           &                           & \cite{poggio_natphys_QPC} \\
26                        & $2.0$                     & 0.020                     & 0.084                     & $874$                     &                           &                           & \cite{etaki_natphys_squid} \\
27                        & $8.7 \cdot 10^{-4}$       & 4.2                       & $2.0 \cdot 10^{-3} $      & $4.8 \cdot 10^{4}$        & Feedback                  & $2.1 \cdot 10^{3} $       & \cite{vinante_PRL_feedback} \\
28                        & $9.1 \cdot 10^{-4}$       & 4.2                       & $1.7 \cdot 10^{-4} $      & $3.9 \cdot 10^{3}$        & Feedback                  & $2.5 \cdot 10^{4} $       & \cite{vinante_PRL_feedback} \\
29                        & $1.0 \cdot 10^{3}$        & 295                       & 295                       & $5.9 \cdot 10^{3}$        &                           &                           & \cite{liu_natnano_timedomain} \\
30                        & $8.9$                     & 295                       & 295                       & $6.9 \cdot 10^{5}$        &                           &                           & \cite{li_nature_harnessing_forces} \\
31                        & $1.5$                     & 0.050                     & 0.050                     & $682$                     &                           &                           & \cite{teufel_PRL_stripline_cooling} \\
32                        & $1.5$                     & 0.050                     & 0.010                     & $136$                     & Sideband                  & 5.0                       & \cite{teufel_PRL_stripline_cooling} \\
33                        & $8.9$                     & 295                       & 295                       & $6.9 \cdot 10^{5}$        &                           &                           & \cite{unterreithmeier_nature_dielectric} \\
34                        & $6.3$                     & 300                       & 58                        & $1.9 \cdot 10^{5}$        & Feedback                  & 5.1                       & \cite{lee_PRL_feedback_toroid} \\
35                        & $13.9$                    &         295               & 295                       & $4.4 \cdot 10^{5}$        &                           &                           & \cite{li_natnano_broadband_transmission} \\
36                        & $0.95$                    & 5.3                       & $1.3 \cdot 10^{-3} $      & $29$                      & Sideband                  & $4.1 \cdot 10^{3} $       & \cite{groeblacher_natphys_cooling} \\
37                        & $65$                      & 1.65                      & 0.20                      & $64$                      & Sideband                  & 8.2                       & \cite{schliesser_natphys_lowoccupation} \\
38                        & $8.2$                     & 360                       & 7.35                      & $1.9 \cdot 10^{4}$        & Sideband                  & 49.0                      & \cite{eichenfield_nature_photonic_crystal} \\
39                        & $119$                     & 1.4                       & 0.21                      & $37$                      & Sideband                  & 6.6                       & \cite{park_natphys_sideband} \\
40                        & $1.2 \cdot 10^{-4}$       & 300                       & $1.4 \cdot 10^{-6} $      & $237$                     & Feedback                  & $2.1 \cdot 10^{8} $       & \cite{abbott_NJP_kg_groundstate} \\
41                        & $8.5$                     & 300                       & 12.5                      & $3.1 \cdot 10^{4}$        & Sideband                  & 23.9                      & \cite{lin_PRL_disks} \\
42                        & $1.04$                    & 0.015                     & 0.130                     & $2.6 \cdot 10^{3}$        & Sideband                  &                           & \cite{teufel_natnano_beyond_SQL} \\
43                        & $8.1$                     & 300                       & 300                       & $7.7 \cdot 10^{5}$        &                           &                           & \cite{anetsberger_natphys_nearfield} \\
44                        & $6.3$                     & 0.020                     & $1.2 \cdot 10^{-3} $      & $3.8$                     & Sideband                  & 17.3                      & \cite{rocheleau_nature_lown} \\
45                        & $8.3$                     & 300                       & 300                       & $7.5 \cdot 10^{5}$        &                           &                           & \cite{anetsberger_PRA_far_below_SQL} \\
46                        & $6.2 \cdot 10^{3}$        & 0.025                     & 0.025                     & $0.07$                    &                           &                           & \cite{oconnell_nature_quantum_piezo_resonator} \\
47                        & $2.1$                     & 0.015                     & 0.014                     & $136$                     & Feedback                  & 1.1                       & \cite{poot_APL_feedback} \\
48                        & $10.7$                    & 0.020                     & $3.7 \cdot 10^{-4} $      & $0.34$                    & Sideband                  & 113.8                     & \cite{teufel_arXiv_groundstate} \\
\bottomrule
\end{tabular}
\label{tab:basic:cooling}
\end{table}

\rowcolors{1}{white}{white}

In most experiments either active feedback or sideband cooling is
employed. In the former case, which is discussed in detail in Sec.
\ref{ssec:basic:activefeedback}, the position of the resonator is
measured and the detector signal is fed back to the resonator to
damp its motion. With sideband cooling (Sec.
\ref{ssec:basic:sideband}) the resonator is embedded in an optical
or microwave cavity. Phonons can be removed from the resonator by
up-converting a red-detuned photon to the resonance of the cavity,
thus cooling the resonator. Finally, also other cooling mechanisms
like bolometric (photothermal) \cite{metzger_nature_cooling,
favero_APL_cooling_micromirror, jourdan_PRL_cooling} and
backaction cooling \cite{naik_nature} (see Sec
\ref{ssec:basic:sensitivity}) are used. We will now discuss these
cooling mechanisms in more detail.

\subsubsection{Backaction cooling}
\label{sssec:basic:backactioncooling} To measure the
displacement, the resonator is coupled to a detector. As shown
in Sec. \ref{ssec:basic:sensitivity}, this influences the
resonator, and, in particular, this backaction adds force noise
and it can damp the motion, which can lead to cooling. The
damping rate of the resonator is then increased from its
intrinsic value $\gamma_0 = \omega_0/Q_0$ to $\gamma_R =
\gamma_0 + \gamma_{BA}$, where $\gamma_{BA} = -\omega_0 ^2 A^2
\imag[\lambda_F]/\omega$ is the damping induced by the detector
\cite{clerk_PRB_ql_linear}. The resonator temperature
is\footnote{This assumes that the effective spring constant
$k_0 - k_0  A^2 \real[\lambda_F]$ does not change considerably.
As shown in Sec. \ref{ssec:basic:activefeedback}, it is in
general much more difficult to alter the spring constant than
to alter the damping.} \cite{naik_nature,
teufel_PRL_stripline_cooling, clerk_NJP_qemsbath}:
\begin{equation}
T_R = \frac{\gamma_0 T + \gamma_{BA} T_{BA}}{\gamma_0 +
\gamma_{BA}}. \label{eq:basic:Tresonator}
\end{equation}
The resonator temperature is thus the weighted average of the
environmental temperature $T$ and the so-called backaction
temperature of the detector $T_{BA}$. For strong
resonator-detector coupling ($\gamma_{BA} \gg \gamma_0$) the
effective temperature is $T_R = T_{BA}$. When this is below the
environmental temperature, the resonator is cooled by the
backaction. Eq. \ref{eq:basic:TR} shows that the backaction
temperature is determined by the force noise exerted on the
resonator: $T_{BA} = S_{F_{BA,n} F_{BA,n}}/4\kb m \gamma_{BA}$.
\added{Both $S_{F_{BA,n} F_{BA,n}}$ and $\gamma_{BA}$ are
proportional to $A^2$ so that $T_{BA}$ is independent of the
resonator-detector coupling $A$. In other words, the backaction
temperature is an intrinsic property of the detector.}

\replaced{Although it might not be immediately clear,}{Note,
that} this cooling mechanism corresponds to the usual notion of
cooling: Cooling is done by coupling something to something
else that is colder. In the case of backaction cooling the cold
object is the detector. Because the resonator is not {\it
actively} cooled, but only brought into contact with the
detector, backaction cooling is therefore also called ``self
cooling'' or ``passive feedback cooling''.

\subsubsection{Active feedback cooling}
\label{ssec:basic:activefeedback} It was shown in Sec.
\ref{ssec:basic:noise} that the resonator temperature is
proportional to its random motion. By reducing that motion the
resonator gets cooled. One way to do this is using feedback.
When the position of the resonator is measured and fed back to
it via an external feedback loop, the motion can be amplified
or suppressed. Feedback control was already used to regulate
the motion of soft cantilevers
\cite{garbini_JAP_optimal_control, bruland_JAP_optimal_control,
mertz_APL_feedback} for magnetic resonance force microscopy
\cite{rugar_nature_MRFM}, when it was realized that this
technique can also be used to cool a mechanical resonator
towards its ground state \cite{hopkins_PRB_feedback}. Actually,
the lowest resonator temperature to date ($T_R = 1.4 \un{\mu
K}$, see Table \ref{tab:basic:cooling}) has been reached using
this cooling method \cite{abbott_NJP_kg_groundstate}.

\begin{figure}[tb]{
\centering
\includegraphics{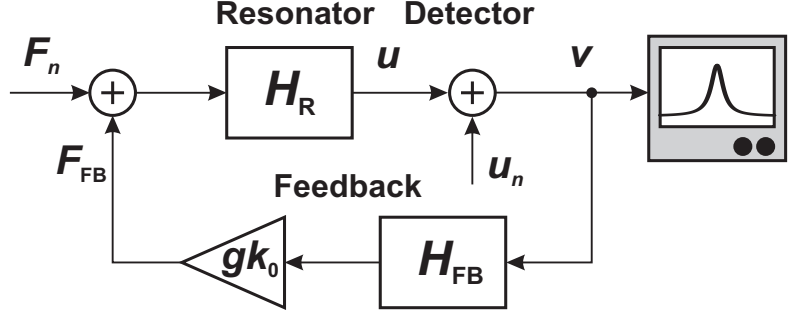}
\caption{Linear system representation of \deleted{active} the active
feedback cooling scheme. The resonator displacement $u$ is
converted by detector to its output signal $v$. This adds
imprecision noise $u_n$ and the sum of this noise and the physical
displacement is the signal that is measured, for example using a
spectrum analyzer. This signal $v$ is also fed back to the
resonator to attenuate the Brownian motion. In the feedback loop a
filter with response $H_{FB}$ and a variable gain $g \cdot k_0$
are included. The resulting feedback force $F_{FB}$ adds up with
the (thermal) force noise $F_n$. The resonator's response to the
applied forces is determined by its transfer function $H_R =
H_{HO}/k_0$.}\label{fig:basic:feedback_cooling}}
\end{figure}

Feedback systems are usually analyzed within the linear system
representation. Figure \ref{fig:basic:feedback_cooling} shows a
schematic of the process. The resonator, with frequency
$\omega_0/2\pi$ and Q-factor $Q_0$, is driven by the thermal
force noise $F_n(t)$ and its displacement $u(t)$ is detected.
The detector output contains not only the displacement but also
imprecision noise $u_n(t)$.\footnote{In this section we assume
that the detector has unit gain, i.e., $A\lambda_v(\omega) =
1$. Then the imprecision noise at the output, $v_n$, and the
noise referred to the input, $u_n$, are the same.} The apparent
position $v$ is the signal at the output of the detector and
this is thus the sum of the physical displacement and the
detector noise: $v = u + u_n$. The information contained in $v$
is used to apply a force $F_{FB}$ to the resonator that damps
its thermal motion\footnote{In principle, also the backaction
force noise of the detector has to be included
\cite{garbini_JAP_optimal_control, vinante_PRL_feedback,
mancini_PRL_feedback}. Although this term was never important
in active feedback cooling experiments so far, the effect can
be included by using the damping and resonator temperature of
the coupled resonator and detector instead of the intrinsic
ones of the resonator alone.}. The relation between the
feedback force and the apparent position is described by the
linear system or filter with transfer function
$H_{FB}(\omega)$. The output of the filter is multiplied by a
selectable gain\footnote{The gain $g$ and filter $H_{FB}$ are
defined such that $|H_{FB}(\omega_0)| = 1$. Both $g$ and
$H_{FB}$ are dimensionless.} $g$. One can think of it as a knob
to crank up the gain of an amplifier. This forms a closed-loop
system \cite{oppenheim_signals_systems,
garbini_JAP_optimal_control} with the following equations of
motion:
\begin{eqnarray}
m\ddot u(t) + m\omega_0 \dot u(t)  /Q_0 + m\omega_0^2 u(t) =
F_n(t) + F_{FB}(t), \\
F_{FB}(t) = m\omega_0^2 g \cdot h_{FB}(t) \conv [u(t) + u_n(t)].
\end{eqnarray}
The presence of the feedback results in a different
displacement for a given thermal noise realization $F_n(t)$.
The feedback thus changes the resonator response from $H_R$ to
the closed-loop transfer function $H_R'$, given by:
\begin{equation}
H_R'^{-1} = H_R^{-1} - g k_0 H_{FB}, \text{~~~or~~~} H_R' =
\frac{k_0^{-1}}{1 - \left(\frac{\omega}{\omega_0}\right)^2 +
\frac{i}{Q_0} \frac{\omega}{\omega_0} - g
H_{FB}}.\label{eq:basic:HR_acc}
\end{equation}
Comparing this with the response of the resonator itself, cf.
Eq. \ref{eq:basic:ho_response}, shows that the real part of
$H_{FB}$ modifies the resonance frequency from $\omega_0$ to
$\omega_R = \omega_0 \sqrt{1 - g \real[H_{FB}(\omega)]}$,
whereas the imaginary part alters the damping rate from
$\gamma_0$ to $\gamma_R =  \gamma_0 - g \omega_0^2
\imag[H_{FB}/\omega]$. Using this closed-loop transfer
function, the PSDs of the physical (i.e., the real resonator
displacement) and observed displacement (i.e., the detector
output) are obtained:
\begin{eqnarray}
\overline S_{uu}(\omega) & = &  \frac{\overline S_{F_n
F_n}/(m\omega_0^2)^2 + g^2 |H_{FB}|^2 \overline S_{u_n u_n}}
{\left| 1 - \left(\frac{\omega}{\omega_0}\right)^2 + \frac{i}{Q_0}
\frac{\omega}{\omega_0} - gH_{FB}(\omega)
\right|^2},\label{eq:basic:Suu_active}
\\
\overline S_{vv}(\omega) & = &  \frac{\overline S_{F_n
F_n}/(m\omega_0^2)^2 + \left|1-
\left(\frac{\omega}{\omega_0}\right)^2 + \frac{i}{Q_0}
\frac{\omega}{\omega_0} \right| ^2 \overline S_{u_n u_n}} {\left|
1 - \left(\frac{\omega}{\omega_0}\right)^2 + \frac{i}{Q_0}
\frac{\omega}{\omega_0} - gH_{FB}(\omega) \right|^2}.
\end{eqnarray}
The resonator displacement PSD $\overline S_{uu}$ shows that
the resonator indeed responds to the force noise with the
modified transfer function $H_R'$ instead of $H_R$. The force
noise that drives the resonator (i.e., the numerator of the
right-hand side of Eq. \ref{eq:basic:Suu_active}) still
contains the original contribution $S_{F_n F_n}$, but now it
also has a contribution due to the imprecision noise that is
fed back to the resonator. Since the latter is always
\replaced{positive}{greater than zero}, the feedback loop adds
additional force noise to the resonator. The apparent position
PSD is also modified: It is not simply the sum of $\overline
S_{uu}$ and $\overline S_{u_n u_n}$ because the feedback
creates correlations between the imprecision noise $u_n(t)$ and
the actual resonator position $u(t)$.

\begin{figure}[tb]
\centering
\includegraphics{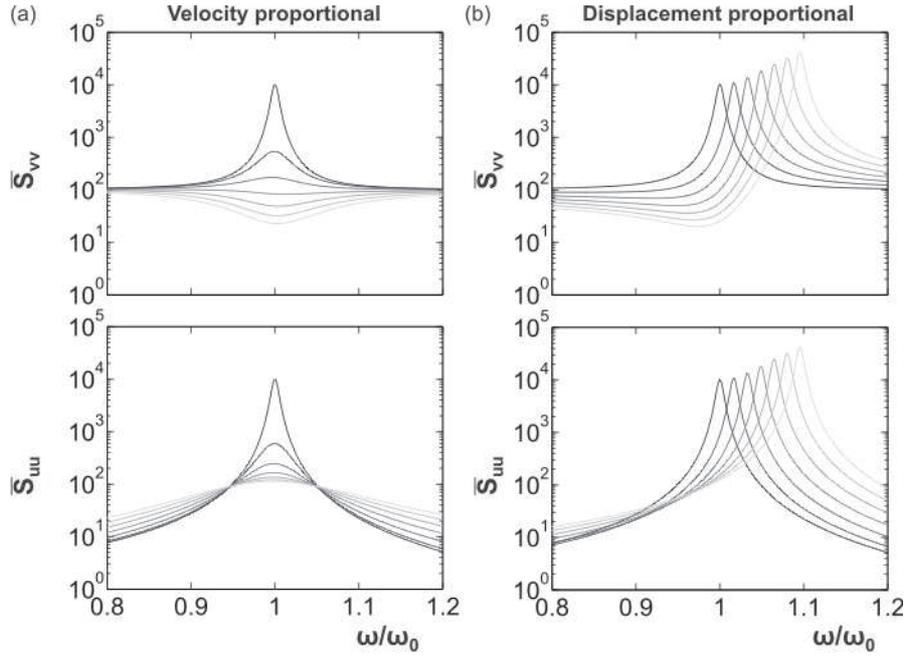}
\caption[Cooling with displacement- and velocity-proportional
feedback] {Feedback cooling of a resonator with $Q_0 = 100$ using
velocity-proportional (a) and displacement-proportional (b)
feedback. The top panels show the PSD of the observed displacement
(i.e., that of the detector output) $\overline S_{vv}$ and the
bottom panels show the real (physical) displacement PSD,
$\overline S_{uu}$. The gain is stepped from $g = 0$ (black) to $g
= 0.2$ (light gray). These plots are calculated with $\overline
S_{u_n u_n} = 10^2$ and all PSDs are scaled by $\overline S_{F_n
F_n}/(m\omega_0^2)^2$.} \label{fig:basic:prop_velocity}
\end{figure}
\begin{figure}[tb]
\centering
\includegraphics{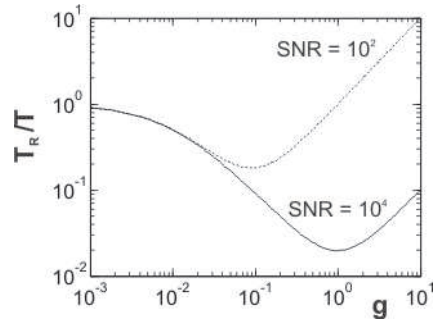}
\caption{Resonator temperature for velocity-proportional feedback
vs feedback gain $g$ for a resonator with $Q_0 = 100$. The dotted
line is for $k_0^2 \overline S_{u_n u_n} / \overline S_{F_n F_n} =
10^2$ which gives a signal-to-noise ratio of $10^2$ and the solid
line is for $k_0^2 \overline S_{u_n u_n} / \overline S_{F_nF_n} =
1$, with  $\textnormal{SNR} = 10^4$. For low gains the two curves
overlap. When $g \gtrsim 0.1$, the detector noise already starts
to heat the resonator for the low-SNR curve, whereas the high-SNR
curve still \replaced{decreases}{goes to lower resonator temperatures}. As predicted by
Eq. \ref{eq:basic:Tmin} the minimum temperature is lower in the
latter case, and this occurs at a higher gain.
\label{fig:basic:TR_gain}}
\end{figure}

So far, the analysis was general for any linear feedback system
and different implementations of the feedback filter $H_{FB}$
are possible, each with their advantages and drawbacks. Using
optimal control theory, the best feedback filter can in
principle be found \cite{garbini_JAP_optimal_control}. In
practise, often simpler, but therefore suboptimal, filters are
used. The PSDs of the true and apparent resonator displacement
are plotted in Fig. \ref{fig:basic:prop_velocity} for the two
simplest feedback schemes:
\begin{itemize}
\item{Velocity-proportional feedback where the measured
    displacement is used to apply a velocity-dependent
    force on the resonator with $h_{FB}(t) = -
    \omega_0^{-1} \cdot \pderl{}{t},~ H_{FB} = -i \omega /
    \omega_0$. In this case the damping rate increases from
    $\gamma_0$ to $\gamma_0\cdot(1+gQ_0)$ as indicated by
    Eq. \ref{eq:basic:HR_acc}. Figure
    \ref{fig:basic:prop_velocity}a shows that at low gains,
    $\overline S_{uu}$ is lowered and thus that the
    resonator is cooled. However, when the gain is
    increased further, the tails of $\overline S_{uu}$
    start to rise as the detector noise (the second term in
    the numerator of Eq. \ref{eq:basic:Suu_active}) is fed
    back into the resonator. Above a certain value $g =
    g_{\min}$, too much detector noise is fed back to the
    resonator and the resonator temperature increases
    again. Figure \ref{fig:basic:TR_gain} shows the
    resonator temperature as a function of the feedback
    gain. The minimum resonator temperature in the limit $g
    \gg Q_0^{-1}$ is \cite{poggio_PRL_feedback}:
    \begin{equation}
    T_{R, \min} = \sqrt{\frac{m\omega_0^3 T}{\kb Q_0} \overline S_{u_n
    u_n}} = \frac{2T}{\sqrt{\textnormal{SNR}}}, \textnormal{~for~} g =
    g_{\min} = \sqrt{\textnormal{SNR}}/Q_0.\label{eq:basic:Tmin}
    \end{equation}
    The minimum resonator temperature is thus set by the
    signal-to-noise ratio ($\textnormal{SNR} \equiv
    \overline S_{uu}(\omega_0)_{g = 0}/\overline S_{u_n
    u_n}$) \replaced{of}{between} the original thermal
    noise peak and the detector noise floor, as was
    illustrated in
    Fig.~\ref{fig:basic:timetrace_spectrum}b. Finally, note
    that, unlike for backaction cooling, the resonator
    temperature does not saturate at a certain value when
    $g \rightarrow \infty$. Eventually more and more noise
    is added and the resonator temperature keeps on
    increasing with increasing gain.}
\item{Displacement-proportional feedback where the
    displacement is directly fed back to the resonator,
    which is characterized by $h_{FB}(t) = -
    \delta(t),~H_{FB} = -1$. This changes the spring
    constant from $m\omega_0^2$ to $m\omega_0^2(1+g)$ and
    the resonance frequency to $\omega_0 \cdot
    (1+g)^{1/2}$. \added{The stiffening of the resonator
    reduces its thermal motion and hence its temperature,
    but} to achieve the same cooling factor as with the
    velocity-proportional feedback the gain should be $Q_0$
    times larger. This, however, also feeds back much more
    detector noise to the resonator in the usual situation
    where $Q_0 \gg 1$. Cooling can therefore only be
    achieved when the SNR is large. Figure
    \ref{fig:basic:prop_velocity}b shows that only heating
    instead of cooling is achieved for the choice of the
    parameters used to perform the calculation. For high-Q
    resonators velocity-proportional feedback is superior
    to displacement-proportional feedback.}
\end{itemize}

It is important to note that the feedback creates correlations
between the resonator displacement and the detector imprecision
noise, which lead to a substantial change in the shape of the
noise spectra. Figure \ref{fig:basic:prop_velocity} shows
calculated noise power spectral densities for different gains.
Without feedback (i.e., $g = 0$) the spectrum of the apparent
motion is simply the sum of that of the harmonic oscillator,
$\overline S_{u u}$, and a constant background level due to the
noise, $\overline S_{u_n u_n}$. The Figure shows that, even
though the peaks in $\overline S_{uu}$ only become broader or
shift in frequency, the peaks in $\overline S_{vv}$ becomes
distorted when the feedback is applied. The resonator
temperature is no longer simply given by the area under the
peak in $S_{vv}$ due to the abovementioned correlations between
$u_n$ and $u$. For large gains it is even possible that the
peak \added{in the spectrum $\overline S_{vv}$} changes into a
dip \deleted{in the spectrum $\overline S_{vv}$} (Fig.
\ref{fig:basic:prop_velocity}a). A way to circumvent this
problem of determining the resonator temperature, is to use a
second detector to measure the resonator motion
\cite{lee_PRL_feedback_toroid}. The noise of the second
detector is not correlated with that of the resonator and the
measured PSD is again the sum of a constant background and the
resonator motion.

Active feedback cooling experiments have mainly been done on
resonators in the $\un{kHz}$ range, where one can simply
measure the position, differentiate and feed the resulting
signal back to the resonator. Cohadon {\it et al.} demonstrated
the first active cooling of a mirror using feedback control
\cite{cohadon_PRL_feedback}. They use a high-finesse cavity
with a coated plano-convex resonator as the end mirror. A
feedback force is applied using a 500 mW laser beam. An
acousto-optical modulator is used to control the exerted
radiation pressure on the resonator. Next, Kleckner and
Bouwmeester actively cooled a 12.5 kHz AFM cantilever from room
temperature to $0.135 \un{K}$ \cite{kleckner_nature_feedback}.
A tiny plane mirror attached to the cantilever served as the
movable end mirror of the cavity and again a second high-power
laser was used to apply the velocity proportional force.
Arcizet {\it et al.} cooled a millimetre-scale resonator in an
optical cavity to 5 K by applying an electrostatic feedback
force on the resonator \cite{arcizet_PRL_sensitive_monitoring}.
By using a piezo element to apply feedback to an ultrasoft
silicon cantilever cantilever cooling from 2.2 K to 5 mK was
demonstrated by Poggio and coworkers
\cite{poggio_PRL_feedback}. Feedback cooling has also been
demonstrated with the gravitational-wave detectors AURIGA
\cite{vinante_PRL_feedback} and LIGO
\cite{abbott_NJP_kg_groundstate}. In the former experiment the
2 ton heavy detector was cooled from 4 K to 0.17 mK. The motion
of the bar is measured capacitively using a resonant electric
circuit coupled to a SQUID amplifier. The output of the
amplifier is put through a low-pass filter to create the
$\pi/2$ phase shift required for velocity proportional feedback
and then injected into the electronic circuit. In the latter
experiment the center-of-mass motion of the four mirrors of a
Fabry-Perot cavity with 4 km long arms is reduced from room
temperature to only $1.4 \un{\mu k}$. This is done by adjusting
the servo control of the mirrors, which creates a
velocity-proportional feedback force in the right frequency
range without affecting the very high signal-to-noise ratio.

When the resonator frequencies are high, say above 1 MHz,
delays in the feedback circuit start to play a role. The force
is then applied when the resonator has already advanced and a
purely velocity-proportional feedback will have a
displacement-proportional component, degrading the cooling
performance. The effect of a delay is even more dramatic when
it equals half the resonator period, so that the Brownian
motion is actually amplified instead of attenuated.
Furthermore, the bandwidth (or sampling speed in the case of a
digital filter) of $H_{FB}$ should be at least a few times
$\omega_R$, which is often not an issue for $\omega_R \lesssim
1 \un{MHz}$. However, active feedback cooling has been reported
for MHz resonators \replaced{by}{using} either adjusting the
delay \cite{lee_PRL_feedback_toroid} of the signal, or using a
mixer circuit to down-convert the mechanical signal to a lower
frequency \cite{poot_APL_feedback}. Finally, note that feedback
can also have other purposes than cooling. Examples are the
regulation of the motion of soft cantilevers as mentioned at
the beginning of this section, and feedback can be used to
modify or null the nonlinearity of a resonator
\cite{nichol_APL_NL_feedback_NW}.

\subsubsection{Sideband cooling} \label{ssec:basic:sideband}
\begin{figure}[tb]
\centering
\includegraphics{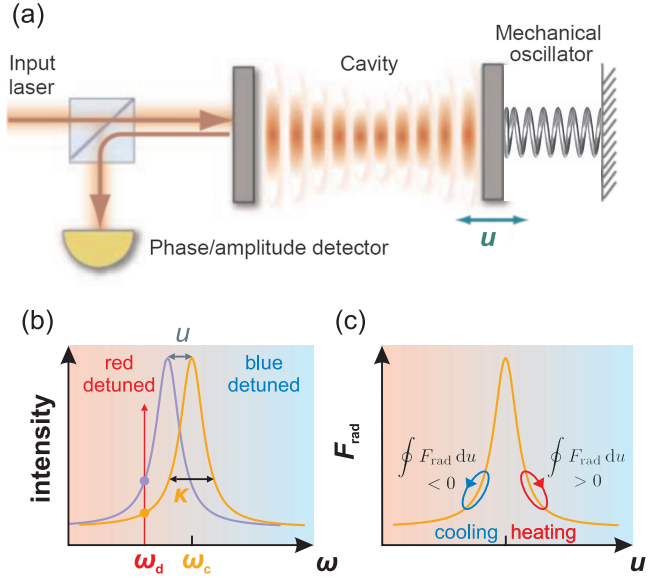}
\caption[Schematic overview of an optical cavity and sideband
cooling] {(a) schematic overview of an optical cavity. The cavity
is driven via an input laser with frequency $\omega_d$. The left
mirror is fixed, but the right mirror is a mechanical resonator
that can move. The resonator displacement $u$ determines the
length of the cavity and thus the cavity resonance frequency
$\omega_c$. A part of the circulating power is transmitted by the
left mirror to a detector. The linewidth of the cavity is
$\kappa$. From T.~J. Kippenberg, K.~J. Vahala, Science 321 (2008)
1172--1176. Reprinted with permission from AAAS. (b) When the
cavity is driven on its resonance, the intensity inside the cavity
is largest, a detuning reduces the intensity. A displacement of
the resonator shifts the cavity resonance (\replaced{purple}{light orange}) and
changes the intensity of the light inside the cavity (orange and
\replaced{purple}{light orange} dots). (c) The displacement dependence of the
radiation pressure $F_{\text{rad}}$. When the resonator
oscillates, the force reacts with a delay due to a finite value of
$\kappa$ as indicated by the ellipsoidal trajectories.
\label{fig:basic:cavity}}
\end{figure}

Another commonly used cooling technique is sideband cooling
\cite{kippenberg_science_optomechanics_overview,
marquardt_physics_optomechanics_overview,
teufel_NJP_cooling_prospects}. In this technique the resonator
is embedded in an optical \cite{arcizet_nature_cavity,
schliesser_natphys_lowoccupation,
schliesser_PRL_cavity_cooling, schliesser_natphys_sideband,
gigan_nature_cavity, corbitt_PRL_opticalspring,
thompson_nature_cavity_membrane, groeblacher_EPL_sideband,
groeblacher_natphys_cooling, park_natphys_sideband,
eichenfield_nature_photonic_crystal} or microwave cavity
\cite{brown_PRL_circuit_cooling, regal_natphys_cavity,
teufel_PRL_stripline_cooling, rocheleau_nature_lown,
teufel_arXiv_groundstate}. These detection schemes will be
discussed in detail in Sec. \ref{sec:detectors}, but here a
brief introduction is given. Figure \ref{fig:basic:cavity}a
shows a schematic drawing of an optical cavity where the right
mirror is the mechanical resonator. Both mirrors have a low
transmission so that a photon is reflected many times before it
can go out through the left mirror, and then toward the
detector. Such a cavity has many different optical eigenmodes,
but here we focus on a single one and denote its resonance
frequency by $\omega_c$. In analogy with the Q-factor and
linewidth $\gamma_R$ of a mechanical resonator, the cavity has
an optical Q-factor $Q_{\text{opt}}$ and linewidth $\kappa$. A
laser sends light with frequency $\omega_d$ into the cavity.
This frequency can be different from the cavity resonance
frequency $\omega_c$; the light is then detuned. Because the
resonance frequency of the cavity is determined by the cavity
length, a displacement of the resonator changes $\omega_c$. As
illustrated in Fig. \ref{fig:basic:cavity}b this leads to a
change in the intensity (and phase) of the light in the cavity,
which results in a change in the detector output. Optical
cavities used this way are very sensitive position detectors
for two reasons: first, it enhances the intensity of the light
by a factor $Q_{\text{opt}}$, and secondly it makes the
intensity depend strongly on the displacement
\cite{marquardt_physics_optomechanics_overview}.

Each photon in the cavity carries a momentum $p_{ph} =
\hbar\omega_d/c$ whose direction is reversed when it reflects
off the mirror. Here $c$ is the speed of light. The resonator
thus experiences a kick of $2p_{ph}$ every time a photon
reflects, the so-called radiation pressure. A single round trip
of a photon in a cavity of length $L$ takes a time
$2L/c$\replaced{, whereas t}{. T}he average time that a photon
spends in the cavity is $\kappa^{-1}$\replaced{. T}{ and
therefore t}he total transfer of momentum per photon is
\added{thus} $p_{tot} = \hbar\omega_c/\kappa L$. The total
force exerted on the resonator is also proportional to the
number of photons present in the cavity, $n_c$, and equals
$F_{\text{rad}} = \hbar\omega_c n_c/ L$. Note, that $n_c$ is
proportional to $\kappa^{-1}$ and to the input power.

The radiation pressure depends on the displacement of the
resonator. This is illustrated in Fig. \ref{fig:basic:cavity}c:
A small change in $u$ changes the cavity frequency $\omega_c$,
which in turn leads to a proportional change in radiation
pressure. This thus changes the effective spring constant of
the resonator to $k_R = k_0 - \pderl{F_{\text{rad}}}{u}$; the
so-called optical spring \cite{braginski_JETP_ponderomotive,
dorsel_PRL_bistability, buonanno_PRD_opticalspring,
sheard_PRA_opticalspring, corbitt_PRA_opticalspring,
corbitt_PRL_opticalspring, hossein-zadeh_OE_optical_spring,
wiederhecker_OE_broadband_tuning}. Similar to
displacement-proportional feedback this can cool the resonator
\cite{corbitt_PRL_opticalspring, abbott_NJP_kg_groundstate,
rosenberg_natphot_routing} and can even lead to bistability of
the resonator position~\cite{dorsel_PRL_bistability}. A much
stronger cooling effect is, however, the fact that the number
of photons does not respond immediately to a change in
displacement, but that they can only slowly leak out of the
cavity at a rate $\sim \kappa$. \added{This was first realized
\cite{braginski_JETP_ponderomotive} and demonstrated
\cite{braginski_JETP_microwave_dissipation} by Braginski{\v \i}
and coworkers.} When $\kappa \gg \omega_0$, a change in the
displacement changes the number of photons instantaneously and
$F_{\text{rad}}$ follows the orange curve in Fig.
\ref{fig:basic:cavity}c. However, a finite value of $\kappa$
causes a delay in the response of the radiation pressure as
indicated by the ellipsoids in Fig. \ref{fig:basic:cavity}c. In
the case of red-detuned driving ($\omega_d < \omega_c$; the
ellipse is traversed counterclockwise) work is done by the
resonator so that it loses energy, whereas for blue-detuned
driving ($\omega_d > \omega_c$; clockwise trajectory) the
resonator gains energy. The increased damping for red detuning
cools the resonator, as the backaction temperature associated
with the detector is very low
\cite{wilson-rae_PRL_ground_state_cooling,
marquardt_PRL_sideband_cooling, teufel_NJP_cooling_prospects}.
This process is called ``dynamical backaction'' and the cooling
mechanism is called ``sideband cooling'' because the cavity is
driven off-resonance, i.e., on a sideband\footnote{A recent
proposal uses a displacement-dependent cavity damping instead
of the usual displacement-dependent cavity frequency
\cite{elste_PRL_cavity_cooling_damping}.}
\cite{dykman_soviet_heating_cooling}. Note, that sideband
cooling can also be described in the language of backaction
cooling (Sec. \ref{sssec:basic:backactioncooling}) as the
optical force responds with a delay to the displacement. This
is thus a detector response $\lambda_v$ (Fig.
\ref{fig:basic:linear_detector}) with a real (the optical
spring) and an imaginary part (the delay).

The ultimate limit on the resonator temperature that can be
reached with sideband cooling has been studied using the
radiation-pressure Hamiltonian in the rotating-wave approximation
\cite{wilson-rae_PRL_ground_state_cooling,
marquardt_PRL_sideband_cooling}:
\begin{equation}
\hat H = \hbar \Delta \hat c^ \dagger \hat c + \hbar \omega_0 \hat
a^ \dagger \hat a + \hbar G_{OM} \hat c^
\dagger \hat c (\hat a^ \dagger +  \hat
a).\label{eq:basic:hamiltonian}
\end{equation}
Here, $G_{OM} = u_0 \pderl{\omega_c}{u}$ is the optomechanical
coupling rate and $\hat c^\dagger$ ($\hat c$) is the creation
(annihilation) operator for a cavity photon (from (to) a photon
with frequency $\omega_d$). $\Delta = \omega_d - \omega_c$ is
the detuning of the laser light with respect to the cavity
resonance frequency. The quantum mechanical picture of sideband
cooling is that a phonon together with a red-detuned driving
photon can excite a photon in the cavity. This is a likely
process because it up-converts the red-detuned driving photon
to a frequency closer to the cavity resonance. This removal of
phonons cools the resonator. The opposite process is also
possible: a cavity photon can emit a phonon and a
lower-frequency photon, thereby heating the resonator. The rate
of these two processes depends on the density-of-states of the
cavity at $\omega_d + \omega_0$ and $\omega_d - \omega_0$
respectively. If the detuning is at exactly at the mechanical
frequency $\Delta = - \omega_0$ and the cavity linewidth is
small, the lowest temperatures are obtained. The process is
analogues to the doppler cooling of cold atoms. In the
good-cavity limit ($\omega_m \gtrsim \kappa$) the lowest
resonator occupation is $\overline n_{\min} = (\kappa /
4\omega_0)^2 \ll 1$ \cite{wilson-rae_PRL_ground_state_cooling,
marquardt_PRL_sideband_cooling,
marquardt_physics_optomechanics_overview}. Although the
good-cavity limit (also called the resolved-sideband regime)
\replaced{was reached a few years ago}{has been reached in
recent experiments} \cite{schliesser_natphys_sideband,
teufel_PRL_stripline_cooling, schliesser_natphys_lowoccupation,
groeblacher_natphys_cooling, park_natphys_sideband,
groeblacher_nature_strong_coupling, rocheleau_nature_lown},
ground-state cooling \replaced{was not immediately}{has not yet
been} demonstrated, due to the fact that the driving
\replaced{could not}{cannot} be increased to sufficiently high
powers. The cooling power should be large enough to remove the
heat coming from the environment to reach the ground state and
the cooling power is proportional to the input power
\cite{teufel_NJP_cooling_prospects}. Table
\ref{tab:basic:cooling} and Fig. \ref{fig:basic:basic_cooling}
show the final thermal occupation numbers that have been
reached up to now.

The effects of sideband cooling (i.e., frequency shift, change
in damping, and cooling) were first considered in the context
of gravitational wave detectors, where the effect was
more-or-less viewed as a technical point with limited
applications, see Refs. \cite{braginski_JETP_ponderomotive,
braginski_JETP_microwave_dissipation, linthorne_JPD_sideband}
and references therein. The first experiments with the aim of
cooling mechanical resonators towards the ground state were
reported in 2006 by Arcizet \cite{arcizet_nature_cavity} and
Gigan \cite{gigan_nature_cavity} cooled resonators with
frequencies of a few hundred kHz by modest factors of 30 and 10
respectively. These measurements were done using free-space
optical cavities which were in the unresolved sideband regime
(i.e., where $\omega_R \ll \kappa$). Schliesser and co-workers
used a mircotoroidal resonator vibrating at 58 MHz as cavity
(see Sec. \ref{sssec:detectors:micro_cavities}) and
demonstrated cooling to 11 K
\cite{schliesser_PRL_cavity_cooling}, which corresponds to an
occupation number of $\overline n = 4300$. A much larger,
gram-scale mirror was cooled to a much lower temperature $T_R =
6.3 \un{mK}$ by the LIGO team \cite{corbitt_PRL_opticalspring,
corbitt_PRL_cooling_mK}. The resonator frequency is 1 kHz in
this case (this is with the optical spring included, the bare
resonance frequency is only 12.7 Hz). The thermal occupation
$\overline n \sim 10^5$ is therefore higher than the
abovementioned experiments by Schliesser. A major step forward
was made in 2008 by the Kippenberg group, who were the first to
reach the resolved sideband regime
\cite{schliesser_natphys_sideband}. Their 73.5 MHz
microtoroidal resonator has a cavity line width of only 3.2
MHz, placing this device deep in the good-cavity limit. The
resonator was cooled to $\sim 19 \un{K}$.

It was realized that a way to further cool resonators is to
start at a lower temperature, so less phonons have to be
refrigerated away to reach the groundstate. This approach was
pursued by the group of Lehnert using superconducting stripline
resonators (Sec. \ref{sssec:detectors:stripline}). Their first
experiment was in the unresolved-sideband regime
\cite{regal_natphys_cavity}, but this was soon superseded by
measurements in the good-cavity limit
\cite{teufel_PRL_stripline_cooling}, where a 0.237 MHz
resonator was cooled from a 50 mK base temperature to 10 mK.
Superconduting striplines are of course always operated at low
temperature, but recently a lot of effort has been done to
place optical cavities at cryogenic temperatures. Gr\"oblacher
{\it et al.} precooled their microresonator to 5 K and
subsequently cooled the 1 MHz resonator to 1.3 mK, where
$\overline n = 32$ \cite{groeblacher_natphys_cooling}. Around
the same time, Schliesser {\it et al.} cooled a resonator
placed in a cryostat from 1.7 K to 0.2 K where $\overline n =
63$. \replaced{Very recently, the first actively cooled device
with an occupation number smaller than one was demonstrated by
Teufel {\it et al.} \cite{teufel_arXiv_groundstate}. They
employed sideband cooling in a resonant superconducting
circuit.}{So far, the lowest occupation factor $\overline n =
3.8$ of an actively cooled resonator has been achieved by
sideband cooling using a superconducting stripline
\cite{rocheleau_nature_lown}.}

\subsubsection{Concluding remarks}
In the previous Subsections backaction cooling, active feedback
cooling, and sideband cooling have been discussed. Although it
might appear that these methods are unrelated, the converse is
true: From the linear system representation of backaction (Fig.
\ref{fig:basic:linear_detector}) and that of active feedback
cooling (Fig. \ref{fig:basic:feedback_cooling}), it is clear
that they are closely related. In the former case the delayed
force, needed to achieve the largest cooling factors, is caused
directly by the detector, whereas in the latter it is actively
exerted by the experimenter. The coupling $A$ and the gain $g$
play the same role in the respective pictures, and so do the
detector force response $\lambda_F$ and the feedback filter
$H_{FB}$. Also, the lowest temperature that can be reached is
determined by the noise, $\Phi_{det,n}$ and $u_n$ respectively.
Note, however, that for backaction cooling the resonator
approaches $T_{BA}$ when the coupling $A \rightarrow \infty$,
whereas in the case of active feedback cooling the resonator
temperature diverges for $g \rightarrow \infty$, and the
minimum temperature occurs at a finite value of the feedback
gain. It was also shown that one can describe sideband cooling
in a similar fashion: namely as the (delayed) response of the
optical force to the displacement, and the same is true for
bolometric forces where the delayed force is due to the finite
heat capacity \cite{metzger_nature_cooling,
favero_APL_cooling_micromirror, jourdan_PRL_cooling,
metzger_PRL_instability}. Finally, we note that there are
cooling mechanisms in non-cavity systems, such as in
superconducting single-electron transistors \cite{naik_nature,
blencowe_NJP_dynamics_SET} and double dots
\cite{zippilli_PRB_cooling_doubledot} that are formally
identical to sideband cooling.

Another important point is illustrated in the experiments done
by Teufel and coworkers (lines 31, 32 in Tables
\ref{tab:intro:short} and \ref{tab:basic:cooling}). When
cooling the resonator, in this case using sideband cooling
\cite{teufel_PRL_stripline_cooling}, the quality factor
decreases. The resolution of the detector then degrades when
the sensitivity stays the same because the resonator bandwidth
$\gamma_R$ increases. In other words, one has less time to
average-out the imprecision noise. The opposite occurs when the
damping of the resonator is reduced: the resolution is
improved, but the resonator temperature increases significantly
\cite{teufel_natnano_beyond_SQL}. In particular, this happens
in a cavity optomechanical system for a blue-detuned laser
drive, but this effect has also been observed in
single-electron transistors \cite{naik_nature}, superconducting
interference devices \cite{poot_PRL_backaction} and in many
other detectors. In this regime energy is transferred from the
cavity to the mechanical system. The oscillation amplitude
grows and the resonance becomes sharper as the resonator energy
increases. When the coupling is strong enough, the damping rate
can vanish ($\gamma_0 + \gamma_{BA} = 0$) or even become
negative. Beyond this instability the resonator exhibits self
oscillations that are only bound by nonlinearities in either
the resonator or the detector
\cite{braginski_JETP_ponderomotive}, and this can lead to
complex nonlinear dynamics. Theoretical work on the classical
effects in this regime has been documented in Refs.
\cite{braginsky_PLA_instability, meystre_JOSAB_interferometers,
marquardt_PRL_multistability,
ludwig_NJP_optomechanical_instability}. Experimentally this
instability has been observed in a variety of micro and
nanomechanical systems, including microtoroidal resonators
\cite{kippenberg_PRL_toroid, carmon_PRL_instability,
schliesser_PRL_cavity_cooling}, Fabry-Perot cavities
\cite{arcizet_nature_cavity,
corbitt_PRA_opticalspring,corbitt_PRL_cooling_mK,
metzger_PRL_instability}, superconducting stripline resonators
\cite{teufel_PRL_stripline_cooling} and microspheres
\cite{ma_OL_microsphere}. Devices that are self oscillating can
be useful from an application point of view as ultra-high
frequency, low phase-noise oscillators
\cite{feng_natnano_selfsustaining}\added{, or as memory
elements \cite{bagheri_natnano_memory}}.

\section{Detection methods} \label{sec:detectors} 
\begin{figure}[tb]
\centering
\includegraphics{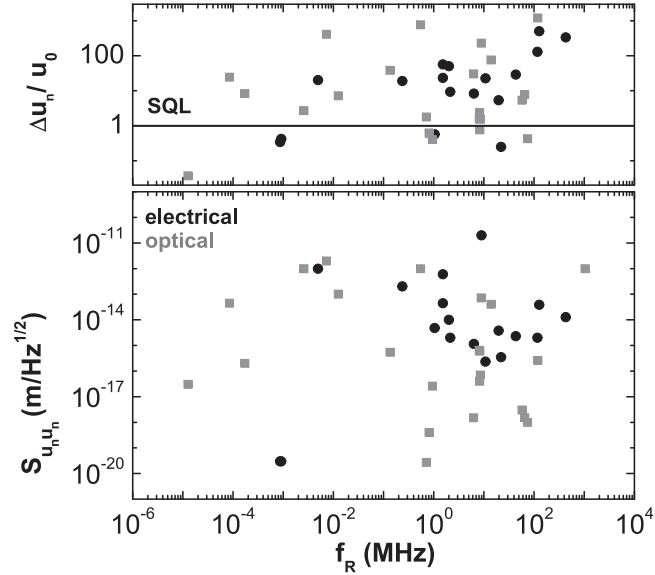}
\caption{Resolution (top panel) and sensitivity (bottom panel) of
the experiments that are listed in Table
\ref{tab:detectors:sensitivity}. Experiments with optical
(electrical) detection are shown in gray (black) respectively. The
solid line indicates the standard quantum limit. As discussed in
Sec. \ref{sec:basic}, detectors can have an imprecision resolution
below the quantum limit at the price of a large backaction force
noise. There is a trend that the resolution and sensitivity degrades
(their value increases) with increasing resonator frequency. \label{fig:detectors:sensitivity_u0}}
\end{figure}

In this section
we discuss the main detection methods for mechanical motion in
nanoscale \replaced{systems}{devices} or in mechanical devices
which aim at reaching the quantum regime. Most methods involve
a linear displacement detector, i.e., the output of the
detector depends linearly on the displacement of the resonator.
As shown in the previous sections, such schemes inevitably
introduce backaction on the resonator position, which
consequently leads to the fundamental limit on the sensitivity
set by quantum mechanics. We will only briefly comment on
quadratic (square-law) detectors, as they have not been studied
to the same extent as linear detectors. In recent years,
tremendous improvement in the sensitivity has been obtained
using a variety of different detection methods. Using optical
cavities, displacements sensitivities as small as $10^{-20}
\un{m/\rHz}$ (see Table \ref{tab:detectors:sensitivity}) have
been reached; using mesoscopic electromechanical devices the
sensitivity can be as good as $10^{-16} \un{m/\rHz}$. Figure
\ref{fig:detectors:sensitivity_u0}a shows the resolution
$\Delta u_n$ due to the imprecision noise $\overline S_{u_n
u_n}$ of the experiments listed in the Table. (For the exact
definition of $\Delta u_n$ in this context we refer to Eq.
\ref{eq:basic:resolution_equiv} in \appref{app:optimal}). Note,
that this quantity does not include the effects of backaction
force noise and can therefore be substantially smaller than the
standard quantum limit as the Figure shows. This is most
clearly demonstrated by the relatively low-frequency resonators
that are read-out optically
\cite{arcizet_PRL_sensitive_monitoring, arcizet_nature_cavity,
poggio_PRL_feedback, vinante_PRL_feedback,
groeblacher_natphys_cooling}. Although the backaction force
noise is present to ensure the SQL and thereby heating the
resonator; at room temperature, this contribution is masked by
the much larger thermal motion. Also in solid-state devices at
low temperature, $\Delta u_n < u_0$ has been achieved
\cite{naik_nature, teufel_natnano_beyond_SQL,
rocheleau_nature_lown} and there an increase in resonator
temperature due to backaction is seen \cite{naik_nature}.
\rowcolors{1}{white}{bkgnd}

\begin{table}[htbp]
\centering \caption{Overview of the sensitivity and resolution of
recent experiments. The table shows the type of detector that was
used, and the displacement-imprecision noise $S_{u_n u_n}$ from
which the resolution $\Delta u_n$ is calculated. The ratio of the
resolution to the zero-point motion $u_0$ is also stated. The
numbers of the first column correspond to the experiments listed
in Table \ref{tab:intro:short} and Table \ref{tab:basic:cooling}.}
\begin{tabular}{rccccr}
\addlinespace \toprule
                          & {\bf Detector}            & {\bf $\mathbf{S_{u_n u_n} \un{fm / \rHz}}$} & $\mathbf{\Delta u_n \unitt{(fm)}}$ & $\mathbf{\Delta u_n / u_0}$ & {\bf Ref.} \\
\midrule
1                    & Magn. mot.           &                      &                      &                      & \cite{huang_nature_GHz} \\
2                    & SET                  & $2.0$                & $658$                & $130$                & \cite{knobel_nature_set} \\
3                    & SET                  & $3.8$                & $113$                & $5.4$                & \cite{lahaye_science} \\
4                    & Opt. cav.            & $2.0 \cdot 10^{3}$   & $4.8 \cdot 10^{3}$   & $415$                & \cite{metzger_nature_cooling} \\
5                    & SET                  & $0.35$               & $5.9$                & $0.25$               & \cite{naik_nature} \\
6                    & Opt. cav.            & $4.0 \cdot 10^{-4}$  & $4.5 \cdot 10^{-3}$  & $0.62$               & \cite{arcizet_nature_cavity} \\
7                    & Opt. cav.            & $100$                & $38$                 & $7.2$                & \cite{kleckner_nature_feedback} \\
8                    & Opt. cav.            &                      &                      &                      & \cite{gigan_nature_cavity} \\
9                    & Opt. cav.            & $4.0 \cdot 10^{-4}$  & $4.5 \cdot 10^{-3}$  & $0.62$               & \cite{arcizet_PRL_sensitive_monitoring} \\
10                   & Opt. cav.            & $3.0 \cdot 10^{-3}$  & $0.53$               & $5.4$                & \cite{schliesser_PRL_cavity_cooling} \\
11                   & Piezoresist.         & $39$                 & $1.8 \cdot 10^{4}$   & $506$                & \cite{li_natnano_piezo} \\
12                   & Opt. cav.            & $0.2$                & $0.06$               & $8.3$                & \cite{corbitt_PRL_opticalspring} \\
13                   & Opt. cav.            & $0.03$               & $9.5 \cdot 10^{-4}$  & $0.04$               & \cite{corbitt_PRL_cooling_mK} \\
14                   & Opt. cav.            & $1.0 \cdot 10^{3}$   & $2.8 \cdot 10^{4}$   & $778$                & \cite{favero_APL_cooling_micromirror} \\
15                   & APC                  & $2.3$                & $268$                & $29$                 & \cite{flowers_PRL_APC} \\
16                   & Opt. cav.            & $1.0 \cdot 10^{3}$   & $271$                & $2.7$                & \cite{poggio_PRL_feedback} \\
17                   & MW res.              &                      &                      &                      & \cite{brown_PRL_circuit_cooling} \\
18                   & Opt. cav.            & $2.7 \cdot 10^{-5}$  & $2.3 \cdot 10^{-4}$  & $1.8$                & \cite{caniard_PRL_backaction_cancelation} \\
19                   & Opt. cav.            & $0.54$               & $49$                 & $39$                 & \cite{thompson_nature_cavity_membrane} \\
20                   & Opt. cav.            &                      &                      &                      & \cite{groeblacher_EPL_sideband} \\
21                   & Opt. cav.            & $45$                 & $2.5$                & $25$                 & \cite{mow_PRL_feedback_cooling} \\
22                   & Opt. cav.            & $1.0 \cdot 10^{-3}$  & $0.05$               & $0.42$               & \cite{schliesser_natphys_sideband} \\
23                   & Stripline            & $200$                & $2.5 \cdot 10^{3}$   & $19$                 & \cite{regal_natphys_cavity} \\
24                   & Magn. mot.           & $12.8$               & $6.6 \cdot 10^{3}$   & $339$                & \cite{feng_natnano_selfsustaining} \\
25                   & QPC                  & $1.0 \cdot 10^{3}$   & $588$                & $20$                 & \cite{poggio_natphys_QPC} \\
26                   & SQUID                & $10$                 & $132$                & $51$                 & \cite{etaki_natphys_squid} \\
27                   & Capacitive           & $3.0 \cdot 10^{-5}$  & $1.0 \cdot 10^{-6}$  & $0.34$               & \cite{vinante_PRL_feedback} \\
28                   & Capacitive           & $3.0 \cdot 10^{-5}$  & $1.2 \cdot 10^{-6}$  & $0.42$               & \cite{vinante_PRL_feedback} \\
29                   & Opt. cav.            & $1.0 \cdot 10^{3}$   & $9.5 \cdot 10^{6}$   & $4.8 \cdot 10^{5}$   & \cite{liu_natnano_timedomain} \\
30                   & Opt. trans.          & $72$                 & $6.2 \cdot 10^{3}$   & $232$                & \cite{li_nature_harnessing_forces} \\
31                   & Stripline            & $600$                & $1.7 \cdot 10^{3}$   & $57$                 & \cite{teufel_PRL_stripline_cooling} \\
32                   & Stripline            & $45$                 & $696$                & $23$                 & \cite{teufel_PRL_stripline_cooling} \\
33                   & Capacitive           & $2.0 \cdot 10^{4}$   & $1.9 \cdot 10^{5}$   & $8.5 \cdot 10^{3}$   & \cite{unterreithmeier_nature_dielectric} \\
34                   & Opt. cav.            & $1.5 \cdot 10^{-3}$  & $0.2$                & $30$                 & \cite{lee_PRL_feedback_toroid} \\
35                   & Opt. trans.          & $40$                 & $2.8 \cdot 10^{3}$   & $76$                 & \cite{li_natnano_broadband_transmission} \\
36                   & Opt. cav.            & $0.03$               & $0.18$               & $0.4$                & \cite{groeblacher_natphys_cooling} \\
37                   & Opt. cav.            & $1.5 \cdot 10^{-3}$  & $0.34$               & $7.9$                & \cite{schliesser_natphys_lowoccupation} \\
38                   & Opt. trans.          & $0.04$               & $11.7$               & $2.4$                & \cite{eichenfield_nature_photonic_crystal} \\
39                   & Opt. cav.            & $0.26$               & $61$                 & $1.2 \cdot 10^{3}$   & \cite{park_natphys_sideband} \\
40                   & Opt. cav.            & $1.0 \cdot 10^{-3}$  &                      &                      & \cite{abbott_NJP_kg_groundstate} \\
41                   & Opt. cav.            & $0.07$               & $4.0$                & $1.5$                & \cite{lin_PRL_disks} \\
42                   & Stripline            & $4.8$                & $15$                 & $0.57$               & \cite{teufel_natnano_beyond_SQL} \\
43                   & Opt. cav.            & $0.64$               & $23$                 & $1.6$                & \cite{anetsberger_natphys_nearfield} \\
44                   & Stripline            & $1.2$                & $209$                & $8.3$                & \cite{rocheleau_nature_lown} \\
45                   & Opt. cav.            & $0.6$                & $12.5$               & $0.76$               & \cite{anetsberger_PRA_far_below_SQL} \\
46                   & Qubit                &                      &                      &                      & \cite{oconnell_nature_quantum_piezo_resonator} \\
47                   & SQUID                & $2.0$                & $24$                 & $9.3$                & \cite{poot_APL_feedback} \\
48                   & MW res.              & $0.23$               & $92$                 & $23$                 & \cite{teufel_arXiv_groundstate} \\
\bottomrule
\end{tabular}
\label{tab:detectors:sensitivity}
\end{table}

\rowcolors{1}{white}{white}

The most straightforward method to analyze the data measured in an
experiment is to record the output of the linear detector as a
function of time; for a photodetector in a cavity experiment (Sec.
\ref{ssec:detectors:cavity}) this would be an output current and
for a dc SQUID position detector (Sec. \ref{ssec:detectors:flux})
this is the voltage over the SQUID. This signal can be fed to a
spectrum analyzer and from the measured spectrum, the resonance
frequency and Q-factor can directly be obtained in the linear
response. From the thermal noise spectra (cf. Fig.
\ref{fig:basic:timetrace_spectrum}b) other parameters like $S_{u_n
u_n}$, $S_{F_{tot} F_{tot}}$ can be obtained as illustrated in the
Figure.

An important issue in measuring the resonator dynamics is the
available bandwidth of the setup. In particular this holds for
the solid-state devices that often have a high impedance ($ \gg
50 \un{\Omega}$). The combination of high impedances and
unavoidable stray capacitances can lead to RC times that are
smaller then the resonator period $1/f_R$. There are several
ways to circumvent this problem: frequency mixing, impedance
matching using tank circuits, or using low-impedance devices
such as superconducting quantum interference devices (SQUIDs)
and microwave striplines. Yet another way relies on
self-detection which yield dc information about the vibrational
motion, e.g. rectification and spectroscopy measurements. These
(non-linear) methods are mainly used in bottom-up devices.

In the following sections, we will discuss different detection
schemes focusing on the resonator-detector coupling and on the
backaction mechanism. As we concentrate on detection schemes
aimed at the quantum regime, we will, for example, not consider
piezo-resistive detection schemes, which are very common in
MEMS and for applications, but rarely used for nanoscale
experiments \cite{li_natnano_piezo, mahboob_APL_UCF}. The main
disadvantage of this method is that it requires a large current
that is dissipated in the resonator.

\subsection{Cavities}
\rowcolors{1}{white}{bkgnd}
\begin{table}[htbp]
\centering \caption{Overview of the coupling factors reached in
optical and microwave cavity experiments. As explained in the
text $g_{OM} = \pderl{\omega_c}{u}$ quantifies how much the
cavity frequency changes with the displacement and $G_{OM} =
g_{OM} \times u_0$ is the vacuum coupling rate. The cavity
frequency $\omega_c/2\pi$ lies in the GHz range for the
microwave cavities and is $\sim 10^2 \un{THz}$ for the optical
cavities\replaced{.}{ and} $\kappa$ is the cavity linewidth. In
the experiment on the last line, a superconducting qubit is
used instead of a cavity. In this case, the value of vacuum
Rabi rate is listed as an estimate of $G_{OM}$
\cite{oconnell_nature_quantum_piezo_resonator}.}
\begin{tabular}{lccccccc}
\addlinespace \toprule
{\bf Group}           & {$\displaystyle \frac{g_{OM}}{2\pi}~\left(\unitt{\displaystyle \frac{ MHz}{ nm}}\right)$}
                          & {$\displaystyle \frac{G_{OM}}{2\pi}~(\unitt{Hz})$}
                          & {$\displaystyle \frac{\omega_c}{2\pi}~(\unitt{GHz})$}
                          & {$\displaystyle \frac{\kappa}{2\pi}~(\unitt{MHz})$}
                          & {$\mathbf L_{OM} ~(\unitt{m})$} & {\bf Ref.} \\
\midrule
JILA                      & $1.16 \cdot 10^{-3} $     & 0.15                      & 4.91                      & 0.49                      & $4.23 \cdot 10^{-3} $     & \cite{regal_natphys_cavity} \\
JILA                      & $4.60 \cdot 10^{-3} $     & 0.14                      & 5.22                      & 0.23                      & $1.13 \cdot 10^{-3} $     & \cite{teufel_PRL_stripline_cooling} \\
JILA                      & $3.20 \cdot 10^{-2} $     & 0.86                      & 7.49                      & 2.88                      & $234 \cdot 10^{-6} $   & \cite{teufel_natnano_beyond_SQL} \\
Schwab                    & $7.50 \cdot 10^{-3} $     & 0.18                      & 5.01                      & 0.49                      & $668 \cdot 10^{-6} $   & \cite{hertzberg_natphys_BAE} \\
Schwab                    & $8.40 \cdot 10^{-2} $     & 2.12                      & 7.48                      & 0.60                      & $89.0 \cdot 10^{-6} $    & \cite{rocheleau_nature_lown} \\
Aalto                     & 1.0                       & 16.5                      & 7.64                      &                           & $7.64 \cdot 10^{-6} $     & \cite{sulkko_NL_FIB_resonator} \\
NIST                      & 56                        & 226                       & 7.47                      & 0.17                      & $133 \cdot 10^{-9} $   & \cite{teufel_nature_strong_coupling} \\
NIST                      & 49                        & 198                       & 7.54                      & 0.20                      & $154 \cdot 10^{-9} $   & \cite{teufel_arXiv_groundstate} \\
\midrule
Vienna                    & $1.13 \cdot 10^{2} $      & 206                       & $2.82 \cdot 10^{5} $      & 120                       & $2.50 \cdot 10^{-3} $     & \cite{gigan_nature_cavity} \\
LKB Paris                 & $1.17 \cdot 10^{2} $      & 75                        & $2.82 \cdot 10^{5} $      & 2.08                      & $2.40 \cdot 10^{-3} $     & \cite{arcizet_PRL_sensitive_monitoring} \\
LKB Paris                 & $1.48 \cdot 10^{2} $      & 0.020                     & $3.70 \cdot 10^{5} $      & 0.26                      & $2.50 \cdot 10^{-3} $     & \cite{caniard_PRL_backaction_cancelation} \\
Harris                    & 2.1                       & 2.68                      & $2.82 \cdot 10^{5} $      & 0.32                      & 0.13                      & \cite{thompson_nature_cavity_membrane} \\
Vienna                    & 15                        & 6.98                      & $3.85 \cdot 10^{5} $      & 0.20                      & $25.0 \cdot 10^{-3} $    & \cite{groeblacher_natphys_cooling} \\
MPI-QO                    & $1.40 \cdot 10^{4} $      & 599                       & $3.85 \cdot 10^{5} $      & 19                        & $27.5 \cdot 10^{-6} $    & \cite{schliesser_natphys_lowoccupation} \\
Painter                   & $1.23 \cdot 10^{5} $      & $5.99 \cdot 10^{5} $      & $1.94 \cdot 10^{5} $      & 646                       & $1.58 \cdot 10^{-6} $     & \cite{eichenfield_nature_photonic_crystal} \\
Painter                   & $3.38 \cdot 10^{4} $      & $8.79 \cdot 10^{4} $      & $1.97 \cdot 10^{5} $      & 113                       & $5.84 \cdot 10^{-6} $     & \cite{lin_PRL_disks} \\
Vienna                    & 11                        & 2.70                      & $2.82 \cdot 10^{5} $      & 0.22                      & $25.0 \cdot 10^{-3} $    & \cite{groeblacher_nature_strong_coupling} \\
MPI-QO / LMU              & 10                        & 145                       & $1.94 \cdot 10^{5} $      & 4.90                      & $19.4 \cdot 10^{-3} $    & \cite{anetsberger_natphys_nearfield} \\
Cornell                   & $9.35 \cdot 10^{3} $      &                           & $2.00 \cdot 10^{5} $      & 2943                      &                           & \cite{wiederhecker_nature_spoke_disks} \\
MPI-QO / LMU              & 40                        & 660                       & $3.53 \cdot 10^{5} $      &                           & $8.82 \cdot 10^{-3} $     & \cite{anetsberger_PRA_far_below_SQL} \\
MPI-QO                    & $1.20 \cdot 10^{4} $      & $1.08 \cdot 10^{3} $      & $3.87 \cdot 10^{5} $      & 15                        & $32.3 \cdot 10^{-6} $    & \cite{weis_science_transparency} \\
MPI-QO                    &                           & $2.68 \cdot 10^{5} $      & $1.93 \cdot 10^{5} $      & $2.0 \cdot 10^{4}$        &                           & \cite{gavartin_PRL_L3_cavity} \\
Cornell                   & $6.00 \cdot 10^{4} $      & $7.32 \cdot 10^{5} $      & $1.89 \cdot 10^{5} $      & $1.0 \cdot 10^{4}$        & $3.14 \cdot 10^{-6} $     & \cite{wiederhecker_OE_broadband_tuning} \\
\midrule
UCSB                      &                           & $6.20 \cdot 10^{7} $      & 6.17                      & 9.36                      &                           & \cite{oconnell_nature_quantum_piezo_resonator} \\
\bottomrule
\end{tabular}
\label{tab:detectors:coupling}
\end{table}
\rowcolors{1}{white}{white}

\label{ssec:detectors:cavity} The most popular and sensitive
technique in optomechanics involve optical cavities. There are
several reviews about this topic, see e.g. Refs.
\cite{kippenberg_OE_overview,
kippenberg_science_optomechanics_overview,
favero_natphot_overview,
marquardt_physics_optomechanics_overview} and we will only
discuss the basic features of these devices. In a cavity,
photons bounce back and forth many times before they are
absorbed or escape the cavity. During this time, the light
interacts with the resonator and the total interaction strength
depends on the number of times that the photon reflects inside
the cavity\added{; the so-called finesse $\mathcal{F}$}. The
light that eventually comes out of the cavity contains the
displacement signal, enabling position detection. Also, when
the light interacts with the resonator, the photons exert a
backaction force on the resonator; the radiation pressure that
was introduced in Sec. \ref{ssec:basic:sideband}. Different
types of optical cavities exist that we will treat separately
in the next Subsections. Cavities cannot only be realized in
the optical domain where the frequencies are \deleted{in the
order of} hundreds of THz, but they can also be made with
superconducting resonant circuits that operate in the microwave
frequency range (GHz). At the end of this section we will
introduce these microwave cavities and compare them to their
optical equivalents.

An important figure of merit for cavity-optomechanical systems
is the coupling strength, which is usually defined as $g_{OM} =
\pderl{\omega_c}{u}$. It indicates how much the cavity
frequency shifts per unit motion. Table
\ref{tab:detectors:coupling} shows the coupling strength, the
cavity frequency, its damping rate, and some derived quantities
for recent experiments with optical and microwave cavities: The
optomechanical coupling length is defined as $L_{OM} = \omega_c
/ g_{OM}$, which equals the physical cavity length in the case
of a Fabry-P\'erot cavity. $L_{OM}$ is a convenient quantity as
the displacement normalized by it, directly gives the relative
change in cavity frequency: $u/L_{OM} = (\omega_c(u) -
\omega_c(0))/\omega_c(0)$. To obtain a high optomechanical
coupling it is thus advantageous to have a small cavity. In the
experiments reported in Refs. \cite{gigan_nature_cavity,
arcizet_PRL_sensitive_monitoring,
caniard_PRL_backaction_cancelation} the Fabry-P\'erot cavity
length is only a few millimeters. The coupling rate that
appears in the cavity-resonator Hamiltonian (Eq.
\ref{eq:basic:hamiltonian}) has been introduced as $G_{OM} =
g_{OM} u_0$, which indicates how much the cavity moves due to
the zero-point motion of the resonator and this coupling rate
is sometimes called the vacuum coupling rate
\cite{gorodetsky_OE_vacuum_coupling}. The role of this quantity
becomes clear when one considers a quantum state in which both
the mechanical resonator and the cavity are in an eigenstate of
the individual subsystems. In the uncoupled system (i.e., for
$G_{OM} = 0$) the system remains in this state since it is also
an eigenstate of the total Hamiltonian. However, this is no
longer the case in the {\it coupled} system. The state will
then oscillate between states with more energy in the
mechanical resonator and states with more photons in the
cavity. The rate at which this happens is proportional to
$G_{OM}$. Note that this is very similar to Rabi oscillations
in a two-level system coupled to a cavity
\cite{walther_rpp_cQED}. Furthermore, the Hamiltonian in Eq.
\ref{eq:basic:hamiltonian} shows that the optomechanical
coupling is related to the backaction: the force exerted by a
single photon in the cavity is $\hbar g_{OM}$, which can be as
large as $13 \un{fN}$ for the on-chip cavities that we will
encounter in Sec. \ref{sssec:detectors:onchip}. Finally, we
note that the coupling strengths $g_{OM}$ and $G_{OM}$ are
solely determined by the device geometry and are independent of
the optical and mechanical Q-factors.
\begin{figure}[tb]
\centering
\includegraphics{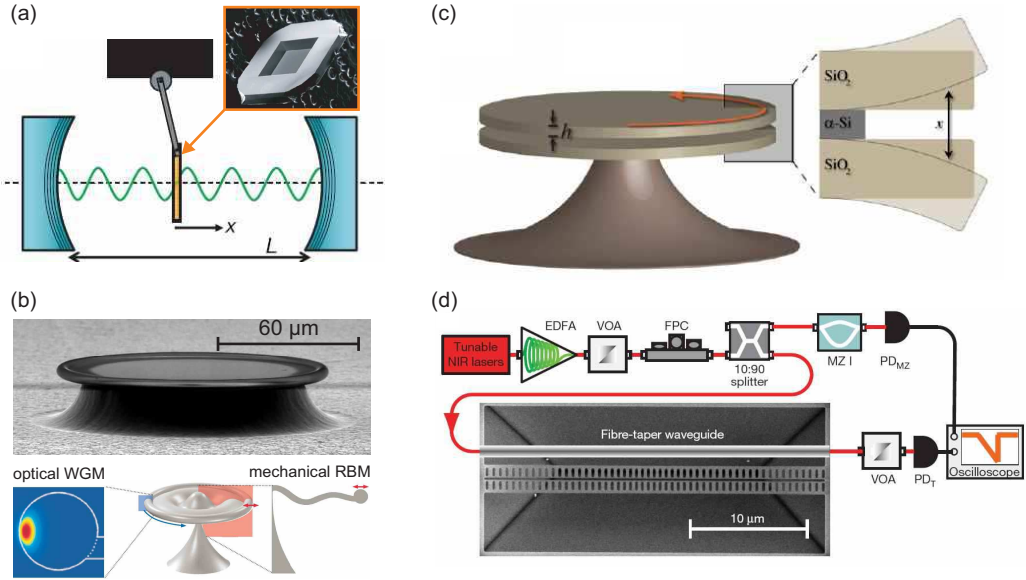}
\caption{Different implementations of optical cavities for
position detection. (a) Schematic overview of an optical cavity
with a movable membrane in it. The cavity consists of two fixed
mirrors at a distance $L$. The membrane at position $x$ can be
located at different intensities of the standing wave in the
cavity (green). The membrane (inset) is made of $50 \un{nm}$ thick
silicon nitride and is held by a silicon substrate. The membrane
reflectivity is $r_c = 0.42$. Reprinted by permission from
Macmillan Publishers Ltd: Nature 452, 72--75, copyright 2008. (b)
Top: scanning electron micrograph of a microtoroid. Bottom:
schematic illustration of the optical whispering gallery mode
(WGM) and the mechanical radial breathing mode (RBM). The WGM
encircles the toroid and is mainly located at the rim. In RBM
vibrations the rim of the toroid moves outward, thereby elongating
the WGM cavity length. Adapted from Refs.
\cite{armani_nature_microtoroid} and
\cite{schliesser_AP_toroid_overview}. Reprinted by permission from
Macmillan Publishers Ltd: Nature 421, 925--928, copyright 2003.
(c) Artist impression of a double-disk cavity. The cavity consists
of two silica disks separated by a layer of amorphous silicon.
Similar to the microtoroid in panel (c), WMGs exist in \added{each of} the two
disks. The\replaced{y}{se} from bonded and antibonded modes (modes with even and
odd parity respectively). The mechanical mode of interest is that
where the gap between the two disks changes. Reprinted figure with
permission from Q. Lin {\it et al}, Phys. Rev. Lett. 103 (2009)
103601. Copyright 2009 by the American Physical Society. (d) The
setup used to study the mechanics and the optical properties of
the zipper cavity. Near-infrared (NIR) laser light goes through an
erbium-doped fiber amplifier (EDFA) and a variable optical
attenuator (VOA). The polarization is adjusted in the
fiber-polarization controller (FPC) and the light is split in two
paths: one through a fiber Mach-–Zehnder interferometer (MZI) and
the other through a tapered fiber that couples the light to the
zipper cavity. Finally, the photons in both paths are detected
with photodetectors (PD). Reprinted by permission from Macmillan
Publishers Ltd: Nature 459, 550--555, copyright 2009.
\label{fig:detectors:cavities}}
\end{figure}

\subsubsection{Fabry-P\'erot cavities}
A schematic picture of a typical \added{Fabry-P\'erot} setup
has already been shown in Fig. \ref{fig:basic:cavity}a and was
discussed in Sec. \ref{ssec:basic:sideband}. Briefly, the
cavity consists of two mirrors, one of which can move, whereas
the other one is fixed. When the moveable mirror is displaced,
the cavity length changes and therefore also the cavity
resonance wavelength. An important property of the cavity is
the reflectivity of the mirrors. A large reflectivity means
that the optical quality factor\footnote{\added{The quality
factor indicate the number of oscillations of the optical field
before it leaves the cavity or before it is dissipated.} In
optics, one usually speaks about the finesse $\mathcal{F}$ of
the cavity\replaced{, which is the number of \emph{reflections}
of the field before it is lost. The finesse}{, which} is
related to the optical quality factor: $\mathcal{F} \equiv
\Delta \omega_c / \kappa = Q \times \Delta \omega_c/ \omega_c$.
Here, $\Delta \omega_c$ is the free spectral range of the
cavity, i.e. the distance between subsequent cavity resonances
($\Delta \omega_c = \pi c /L$). Typical numbers for
$\mathcal{F}$ range from $10^3$ to $10^5$ for high-quality
cavities \cite{marquardt_physics_optomechanics_overview}.
\label{fn:detectors:finesse_Q}} is high, i.e., the light
bounces back and forth many times so that the interaction
between mechanical motion and the photon is strongly enhanced.
A small fraction of the light is allowed to come out of the
cavity and is guided to a photo-detector. These can detect
individual photons and therefore the position detection
sensitivity \replaced{can be}{is} shot noise limited.

There are two main implementations to measure the light that
comes out of the cavity. The first method, shown in Fig.
\ref{fig:basic:cavity}a, uses a polarized beam splitter in
combination with a $\lambda/4$-plate to separate the incoming
laser light from the light that comes out of the cavity. The
advantage of this setup is its compactness. The second setup
consists of two arms with cavities that are illuminated by a
single laser via a beam splitter. Light that comes out of the
two arms interferes destructively at the beam splitter when the
length of the two arms is identical. However when one of the
mirrors displaces, this is no longer the case and photons
arrive at the photo-detector. This implementation is a
so-called Michelson interferometer and is used to search for
gravitational waves \cite{abramovici_science_LIGO}. With two $4
\un{km}$ arms, displacements of kg-scale resonators (i.e.,
mirrors) are detectable to within $10^{-19} \un{m/\rHz}$
\cite{abbott_NJP_kg_groundstate}.

For a Fabry-P\'erot cavity the coupling constant is $g_{OM} =
\omega_c / L$, where $L$ is the cavity length
\cite{schliesser_AP_toroid_overview} as explained in
Sec.~\ref{ssec:detectors:cavity}. For a cm-long cavity and
light with a visible wavelength this gives $g_{OM} \sim 2\pi
\cdot 10 \un{MHz/ nm}$ (see Table
\ref{tab:detectors:coupling}). In the presence of a strong
coherent pump laser, the {\it effective} coupling rate is
enhanced by the square root of the number photons in the cavity
$\sqrt{n_c}$. The optomechanical strong coupling regime where
$G_{OM}\cdot n_c^{1/2} \gg \omega_R, \kappa_c$ was first
reached in a 25 mm long Fabry-P\'erot cavity
\cite{groeblacher_nature_strong_coupling}, as evidenced by the
\deleted{expected} mode splitting \cite{dobrindt_PRL_splitting}
of the mechanical resonator and the detuning of the cavity.
Strong coupling has now also been demonstrated in other
optomechanical systems \cite{teufel_nature_strong_coupling,
weis_science_transparency, safavi_nature_EIT}.

Fabry-P\'erot cavities are very sensitive position detectors
and have already been used for a while to approach the quantum
limit on position detection, see for example Ref.
\cite{tittonen_PRA_torsional_resonator}. Their properties
(shot-noise limited sensitivity, backaction etc.) are therefore
well known. The most sensitive implementation of the
Fabry-P\'erot interferometer reaches the impressive sensitivity
of $10^{-20} \un{m/\rHz}$ for the motion of the differential
motion of the two mirrors forming a 0.25 mm long cavity with a
finesse of 230 000 \cite{caniard_PRL_backaction_cancelation}.
Currently, a lot of effort is put in reaching true strong
coupling where $ G_{OM} > \kappa, \omega_R$ and in the
observation of quantum backaction
\cite{verlot_PRL_correlations, bjorkje_PRA_shot_noise}.

\subsubsection{Movable membrane inside the cavity}
\label{sssec:detectors:membrane} Another use of cavities in
optomechanics, is the setup pioneered by the group of Harris,
where a flexible membrane is positioned inside a rigid cavity
\cite{thompson_nature_cavity_membrane}. A schematic drawing of
this setup is shown in Fig. \ref{fig:detectors:cavities}a. One
of the advantages of this implementation is that the mechanical
resonator and the mirrors of the cavity are separated, so that
a high-quality cavity can be made without degrading the
mechanical properties of the resonator. The membrane, with a
low (field) reflectivity $r_c$, can be positioned at different
locations $x$ inside that cavity, i.e., at nodes or at
anti-nodes of the standing light-field waves. Depending on
where the membrane is placed, the coupling is different: At an
anti-node, the membrane strongly interacts with the cavity,
whereas at a node, it does not. In a way this system can be
viewed as two coupled cavities. For a particular optical mode,
the light is predominantly on one side of the membrane exerting
a radiation pressure on it from that side. From the above it is
clear that the cavity frequency $\omega_c$ should be a periodic
function of the membrane position $x_m$ and so is the
optomechanical coupling. Using a one-dimensional model
\cite{thompson_nature_cavity_membrane} the coupling strength is
obtained: $g_{OM} \approx 4\pi |r_c| c/(L\lambda) \sin(4\pi x_m
/ \lambda)$ assuming $|r_c| \ll 1$. This optomechanical system
thus exhibits a tuneable coupling.

At a node of the optical field ($x_m=n\lambda/2$), the linear
coupling term between the cavity and the resonator vanishes and
a small quadratic term remains. Interestingly, this quadratic
term can be greatly enhanced by slightly tilting ($ \sim 1
\un{mrad}$) the membrane \cite{sankey_natphys_quadratic}. In
this case, avoided crossings of different optical cavity modes
occur and near these avoided crossings the cavity frequency
depends strongly on the position of the membrane squared (i.e.,
square-law position detection). In the realization of Ref.
\cite{sankey_natphys_quadratic} the tilting results in an
increase in the coupling from $30 \un{kHz/nm^2}$ to $ \gtrsim
30 \un{MHz/nm^2}$, which might enable direct measurements of
the quantization of the membrane's energy (see the discussion
in Sec. \ref{sec:basic} and \appref{app:squared})
\cite{jacobs_PRL_continuous_energy,
jacobs_EPL_energy_measurement}.

\subsubsection{Optical cavities on the micro scale}
\label{sssec:detectors:micro_cavities} As we have discussed
before, it is advantageous to have a cavity with a high finesse
(or the related optical quality factor, see footnote
\ref{fn:detectors:finesse_Q}) in combination with a high
mechanical Q-factor. In the experiments described in Sec.
\ref{sssec:detectors:membrane} this was done by physically
separating the mirrors from the mechanical resonator. A
different approach is to use microtoroids (see Fig.
\ref{fig:detectors:cavities}b), which can have optical
Q-factors in excess of $10^8$ or equivalently $\mathcal{F} >
10^6$ \cite{armani_nature_microtoroid}. Light can be coupled
into these devices via free-space evanescent coupling by
positioning a tapered fiber close ($\sim 1 \un{\mu m}$) to it
\cite{cai_PRL_fiber_coupling}. The light travels around the
outer edge of the toroid in, what is called, a
whispering-gallery mode (WGM) (see Fig.~
\ref{fig:detectors:cavities}b, lower left panel). The light in
this mode is strongly coupled to the mechanical vibrations of
the toroid, in particular to its radial breathing mode (RBM).
In this mode, the toroid expands and retracts in the radial
direction, thereby changing its diameter slightly in time. The
mechanical RBM frequencies are of the order of 10 to $100
\un{MHz}$; the mechanical quality factor can be as high as
32000 and depends on the exact device geometry
\cite{anetsberger_natphot_dissipation_toroid}. The coupling
length is in this case the toroid radius, $L_{OM} = R$
\cite{schliesser_AP_toroid_overview}, and the coupling constant
$g_{OM} = \omega_c/R$ can reach $2\pi\cdot 10 \un{GHz/nm}$ (see
Table \ref{tab:detectors:coupling}). This high value can be
understood from the much smaller dimensions of the toroid
cavity: The cm-long Fabry-P\'erot cavities are now replaced a
by toroid with a circumference of a few hundred $\un{\mu m}$.

A very similar system are silica
microspheres~\cite{cai_PRL_fiber_coupling, ma_OL_microsphere,
park_natphys_sideband}, where again the whispering-gallery mode
couples strongly to the mechanical breathing modes with a
frequency of the order of $100 \un{MHz}$ for a typical diameter
of $\sim 30 \un{\mu m}$. The optical quality factor of these
devices exceeds $10^7$ and typical values for the mechanical
Q-factor are $10^4$. Light is coupled into the WGM by focussing
a laser beam close to the sphere. Similar to the fiber taper,
the light enters the sphere via the evanescent field and this
causes a detectable phase shift in the transmitted
light~\cite{park_natphys_sideband}.

Cooling has been extensively studied in the microtoroid systems
\cite{schliesser_PRL_cavity_cooling,
schliesser_natphys_sideband, lee_PRL_feedback_toroid} and an
advantage is that they can be integrated and precooled to
helium temperatures in cryostats
\cite{schliesser_natphys_lowoccupation, park_natphys_sideband}.
In the latter experiments, thermal occupation numbers as low as
$n = 70$ and $37$ have been achieved respectively. The small
scale of these devices also bears another advantage in that it
can be coupled to other mechanical resonator by placing them in
the near vicinity of the toroid. An example is described in
Refs. \cite{anetsberger_natphys_nearfield,
anetsberger_PRA_far_below_SQL}, in which the flexural modes of
a nanomechanical SiN string are probed via the toroid with a
position sensitivity that is two times below the standard
quantum limit. The coupling $g_{OM}$ decreases exponentially
with the distance between the string and the toroid and a
maximum value of $g_{OM} = 2\pi \times 10 \un{MHz /nm}$ has
been reported. Finally, it was demonstrated that microtoroid
can be actuated electrostatically using gradient forces, making
it an optoelectromechanical system \cite{mcrea_PRA_toroid}. A
recent review summarizes the achievements in cavity
optomechanics with whispering-gallery modes
\cite{schliesser_AP_toroid_overview}.

\subsubsection{On-chip optical cavities}
\label{sssec:detectors:onchip} A clever way to further increase
the optomechanical coupling has been realized by Painter and
co-workers \cite{eichenfield_nature_photonic_crystal,
eichenfield_nature_optomechanical_crystals, lin_PRL_disks}. The
basic idea is as follows: When placing two optical waveguides
in close vicinity (submicron scale), symmetric and
anti-symmetric optical modes form with a mode volume of the
order of $\lambda^3$. The optical coupling length, $L_{OM}$,
can be viewed as the length scale over which a photon's
momentum is transferred, which, in this case, is reduced to a
length scale of the order of the optical wavelength, $\lambda$.
Thus, the coupling can be estimated to be $g_{OM} \sim \omega_c
/ \lambda$. Since the wavelength is about $1.5 \un{\mu m}$, the
coupling can be at least an order of magnitude larger than in
the micro-cavities discussed above (see Table
\ref{tab:detectors:coupling}). Two implementations have been
built. In one version~\cite{wiederhecker_nature_spoke_disks,
lin_PRL_disks}, a pair of silica (or SiN) disks separated by
nanometre-scale gaps was used as shown in Fig.
\ref{fig:detectors:cavities}c. The coupling depends on the
separation between the disks and for an air-gap of $138
\un{nm}$, the coupling was found to be $g_{OM} = 2\pi \cdot 33
\un{GHz/nm}$. Efficient cooling of the mechanical mode was
achieved~\cite{lin_PRL_disks}, and static and dynamic
mechanical wavelength routing was
demonstrated~\cite{rosenberg_natphot_routing}. Furthermore,
Wiederhecker {\it et al.} demonstrated attractive and repulsive
forces between the disks~\cite{wiederhecker_nature_spoke_disks}
and optomechanical tuning of the cavity modes over more than 30
nm~\cite{wiederhecker_OE_broadband_tuning} in a similar device.

An even larger coupling is obtained with two stoichiometric
silicon nitride ladder structures with a photonic crystal
structure (``zipper cavities'') as illustrated in Fig.
\ref{fig:detectors:cavities}d. With a separation of $120
\un{nm}$ between the two waveguides, a coupling of $g_{OM} =
2\pi \cdot 123 \un{GHz}$ has been achieved
\cite{eichenfield_nature_photonic_crystal}. The strong coupling
yields a large optical spring effect (see Sec.
\ref{ssec:basic:sideband}), where the resonance frequency is
mainly determined by the laser field instead of by the
structural properties of the resonator. This effect shows up as
a change in resonance frequency if the input power and detuning
are changed. Interestingly, the optical spring only acts on the
differential motion of the beams. The common-mode vibrations
are not affected by the light field. By varying the detuning of
the driving light, Lin {\it et al.} could shift the
differential mode through the resonance frequency of the common
mode. The observed Fano-like lineshape indicates coherent
mixing of the mechanical excitations
\cite{lin_natphot_coherent_mixing}.

Another advantage of optomechanical crystals is the control
over the location of the optical and mechanical modes
\cite{eichenfield_nature_optomechanical_crystals}. Since the
spatial extent of modes differs, the coupling strength can be
engineered. Thus, the simultaneous confinement of optical
(photonic crystals) and mechanical modes (phononic crystal)
leads to strong, controllable light-matter interactions
\cite{alegre_OE_phononic_bandgap}. Finally, note that in
on-chip optomechanical devices the optical gradient force is
typically much larger than the photon pressure
\cite{li_nature_harnessing_forces, li_natphot_bipolar_MZ,
lin_PRL_disks, wiederhecker_nature_spoke_disks,
roels_natnano_gradient_force}.

\subsubsection{Superconducting microwave cavities}
\label{sssec:detectors:stripline}
\begin{figure}[tb]
\centering
\includegraphics{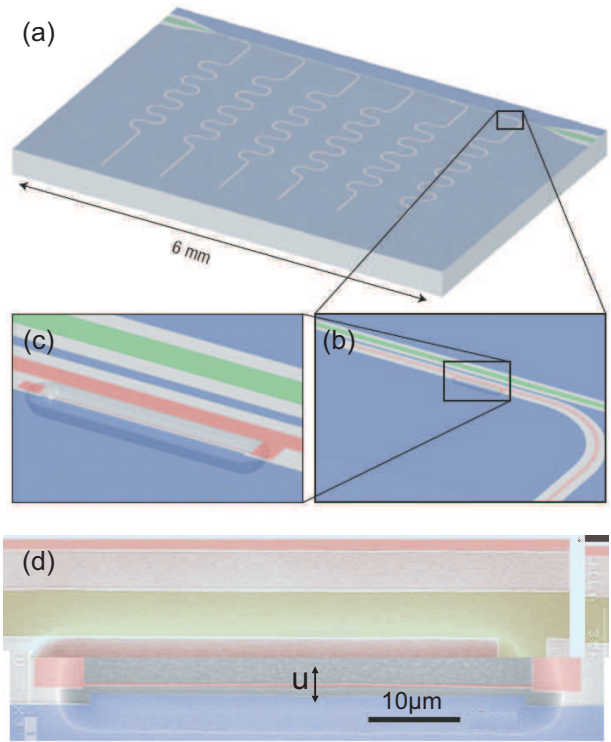}
\caption{Illustration of the stripline resonators used by Regal
{\it et al.} (a) Chip with 6 meandering stripline resonators
(pink) and the straight feedline (green). The $\lambda/4$
resonators have a slightly different length, which allows for
frequency multiplexing. (b) The resonators are coupled to the
feedline by a capacitive elbow coupler. (c) and (d) zooms of the
resonator. The $50 \un{\mu m}$ long aluminum beam resonator is
under tensile stress which is induced by annealing at an elevated
temperature. A movement of the beam changes the capacitance of the
stripline to the ground plane (blue) and thereby the cavity
resonance frequency. Reprinted by permission from Macmillan Publishers Ltd: Nature Physics 4, 555--560, copyright 2008.
\label{fig:detectors:stripline}}
\end{figure}

Instead of using an optical cavity, electromagnetic waves can also
be confined in a superconducting microwave cavity
\cite{day_nature_stripline}. The electromagnetic cavity mode is
sensitive to capacitive changes and this principle has been used
to measure nanomechanical motion~\cite{regal_natphys_cavity,
teufel_natnano_beyond_SQL, rocheleau_nature_lown}. The device used
in the first realization of such a transmission-line position
detector \cite{regal_natphys_cavity} is shown in Fig.
\ref{fig:detectors:stripline}.

The working principle is as follows: In the superconducting
transmission line with an inductance and capacitance per unit
length, radiation propagates. On one side microwaves are
injected via a capacitively coupled feedline; the other end can
be open or shorted to ground (so-called $\lambda/2$ and
$\lambda/4$ microwave cavities respectively). In the first case
standing waves with a voltage anti-node at the end
form\replaced{, whereas}{. On the other hand,} in the latter
case waves with a node at the end form. The mechanical
resonator is connected to the stripline and positioned close to
the ground plane. Its displacement couples to the cavity
capacitance and this in turn changes the cavity resonance
frequency which can accurately be probed via transmission or
reflectivity measurements. In the realization of the group of
Lehnert (i.e., the one shown in Fig.
\ref{fig:detectors:stripline}) this is done via the
transmission of the feedline~\cite{regal_natphys_cavity,
teufel_NJP_cooling_prospects, teufel_PRL_stripline_cooling,
teufel_natnano_beyond_SQL}; in the Schwab-group implementation
the transmission through the cavity is
measured~\cite{rocheleau_nature_lown, hertzberg_natphys_BAE}.

Typical lengths of these on-chip resonators are in the
millimetre range so that the microwave resonance frequency is a
few GHz. The electric quality factor is typically $10^4$; in
the experiments, the dissipation is determined by the coupling
to the feedline and not by internal losses (i.e., they are
overcoupled) \cite{regal_natphys_cavity}. An overview of the
key parameters of these devices is given in Table
\ref{tab:detectors:coupling}. Large optomechanical coupling is
obtained for thin, long doubly clamped beams that are
positioned as close as possible near the ground plane. A
coupling of $g_{OM} = 2\pi \cdot 84 \un{kHz/nm}$ is
reported~\cite{rocheleau_nature_lown} for a 170 nm wide and 140
nm thick beam, that is formed from $60 \un{nm}$ of
stoichiometric, high-stress silicon nitride and 80 nm of
aluminum. The resonator is located 75 nm from the gate
electrode. Further substantial improvements in the coupling
strength are difficult with this particular geometry and
fabrication technology. Sulkko {\it et al.}
\cite{sulkko_NL_FIB_resonator} used a focussed ion beam to
create a very small gap of about $10 \un{nm}$ between a
mechanical resonator and a resonant microwave circuit and
obtained $g_{OM} = 2\pi \cdot 1 \un{MHz/nm}$, thus increasing
the coupling considerably (see Table
\ref{tab:detectors:coupling}).

An important difference with optical cavities is that there are
no single photon detectors available for microwave frequencies.
In optical systems these do exist, and the detection of the
light in itself is quantum-limited. Present commercial
microwave amplifiers, however, always add substantially more
noise than required by quantum mechanics since they are not
shot-noise limited. This problem can largely be overcome by
using a Josephson parametric amplifier. Teufel {\it et al.}
\cite{teufel_natnano_beyond_SQL} have made a nearly
shot-noise-limited microwave interferometer and demonstrated
nanomechanical motion detection with an imprecision below the
standard quantum limit. On the other hand, similar to their
optical equivalents, the photon-pressure backaction of the
stripline position-detectors on the mechanical resonator has
been observed \cite{teufel_PRL_stripline_cooling} and even
backaction-evading measurements have been reported
\cite{hertzberg_natphys_BAE}. In the latter experiment, a
single-quadrature measurement of motion with a sensitivity of
four times the zero-point motion has been demonstrated.
Sideband cooling has been performed in a series of experiments
\cite{teufel_PRL_stripline_cooling, rocheleau_nature_lown,
hertzberg_natphys_BAE}. Very recently Teufel {\it et al.} were
the first ones to demonstrate ground state cooling using a
superconducting LC resonator \cite{teufel_arXiv_groundstate},
in which a movable membrane both acts as the capacitor of the
LC circuit and as the mechanical oscillator (``drum resonator''
geometry). They obtained a thermal occupation of $\overline n =
0.34$, where the drum resonator is in the ground state for 57\%
of the time. In the same type of device they also observed
strong coupling between the mechanical resonator and the LC
resonator \cite{teufel_nature_strong_coupling}.

It should be noted that mechanical resonators have also been
coupled to non-superconducting resonant circuits
\cite{brown_PRL_circuit_cooling, hao_APL_nearfield}. Sideband
cooling from room temperature to $45 \un{K}$ and $100 \un{K}$
respectively has been demonstrated in these systems. The first
method differs from the abovementioned superconducting cavities
in that it consists of a stripline resonator with lower
frequency ($100 \un{MHz}$) and lower quality factor ($234$). In
the second experiment, the resonator is an open-ended coaxial
cable with a resonance frequency of $11 \un{GHz}$ and a $Q$ of
80.

\subsection{Transmission modulation based detection methods}
\label{ssec:detectors:transmission} Mechanical motion can also
be detected by measuring the transmission of electrons or
photons through devices which embed a movable part. The motion
of the mechanical resonator modulates the electron or photon
transmission and the output signal contains spectral
information about the resonator dynamics. Below we will discuss
in detail two basic configurations of this detection scheme:
optical waveguides and electron tunneling devices, including
electron shuttles.
\begin{figure}[tb]
\centering
\includegraphics{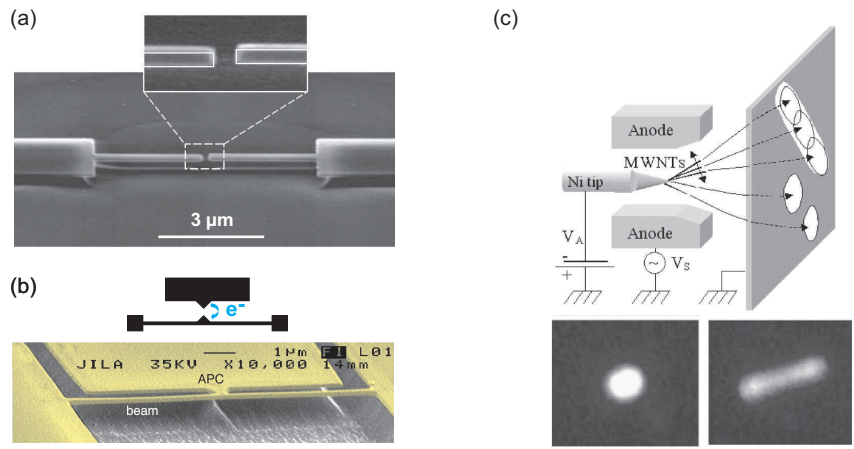}
\caption{(a) Readout using the light transmitted from one
waveguide (left) to another one (right). When one of the
cantilevers moves, the transmission of the light changes. This is
a non-interferometric optical method to detect mechanical motion. Reprinted by permission from Macmillan Publishers Ltd: Nature Nanotechnology 4, 377--382, copyright 2009.
(b) An
atomic-point-contact displacement detector. A tunnel contact
between the beam and a nearby electrode is made using
electromigration. When the Au beam is far from the electrode, the
tunneling probability for electrons (see inset) is low and the
resistance is high. Closer to the electrode, the transmission
increases and the resistance is lower. Reprinted figure with
permission from N. E. Flowers-Jacobs, D. R. Schmidt, K. W.
Lehnert, Intrinsic noise properties of atomic point contact
displacement detectors, Phys. Rev. Lett. 98 (9) (2007) 096804.
Copyright 2007 by the American Physical Society. (c) Setup to
measure multi-walled nanotube vibrations via field emission.
Electrons are emitted by applying a voltage $V_A = -500$ to $-900
\un{V}$ between the nanotube and a nearby ($3 \un{cm}$) screen.
The electrons hit the screen at a certain position and show up a
bright spot. The insets at the bottom show the spots for a
silicon-carbide resonator that is driven off-resonance (left) and
on resonance (right). In the latter case the spot is blurred due
to the motion of the resonator. Adapted from
Refs.~\cite{purcell_PRL_tuning_nanotube}
and~\cite{perisanu_APL_highQ_SiC}. Reprinted with permission from
S.~T. Purcell {\it et al.} Tuning of nanotube
  mechanical resonances by electric field pulling, Phys. Rev. Lett. 89~(27)
  (2002) 276103. Copyright 2002 by the American Physical Society and S.~Perisanu {\it et al.} Appl. Phys. Lett. 90~(4) (2007) 043113.
Copyright 2007, American Institute of Physics.
\label{fig:detectors:transmission}}
\end{figure}

\subsubsection{Optical waveguides}
As mentioned above, light can be guided on a chip using optical
waveguides. These are pieces of transparent material (e.g. Si,
SiO$_2$ or SiN) which are carved out of the underlying
substrate. Light can propagate through these waveguides and can
be directed to any place on the chip. Just outside the
waveguide the electromagnetic field is not zero and decays
exponentially with the distance from the waveguide. When two
waveguides are placed close together, light can be transmitted
from one to the other\footnote{A related, but different method
is demonstrated in Ref.~\cite{vlaminck_APL_evanescent}, where
photons are transferred from one waveguide to another, which
runs parallel to the former. The coupling between the
waveguides depends on the distance between them (i.e., on the
displacement) and photons that enter the second waveguide are
``lost'' and not detected at end of the first waveguide.}. The
transmission between the waveguides depends on the distance
between them and they can be employed as a position detector.
The method is also compatible with operation in water
\cite{nordstroem_APL_transmission}, which is important for
(bio)sensor applications, and it has also been used to make
optoelectronic switches
\cite{pruessner_JMM_optoelectronic_switch}. Li {\it et al.}
\cite{li_natnano_broadband_transmission} have used the
transmission-modulation principle to measure the mechanical
motion of cantilever structures made in a silicon-on-insulator
platform with high sensitivity. In particular, they placed the
suspended ends of two waveguides facing each other at a
distance of 200 nm as shown in Fig.
\ref{fig:detectors:transmission}a. The transmission of photons
through the gap depends on the distance and the exact position
of the cantilevers (misalignment). The thermal motion of the
cantilevers modulates the transmission and the transmitted
light is fed into a photodiode. The spectrum clearly reveals
multiple resonances of the cantilevers with a sensitivity of
$40 \un{fm/\rHz}$ at room temperature. An advantage of this
method over optical cavities is that the detection scheme
allows for transduction of nanomechanical motion over a wide
range of optical frequencies, instead of working only at
certain well-defined wavelengths.

\subsubsection{Electron tunneling}
When two metallic electrodes are placed close (up to a few
nanometer) to each other, electrons can tunnel through the gap
between them. With a voltage applied between both electrodes, a
net current flows that depends exponentially on their distance.
This exponential distance dependence can be used to measure
displacements of mechanical resonators. Flowers-Jacobs {\it et
al.} \cite{flowers_PRL_APC} have measured the displacement of
the doubly-clamped gold beam shown in Fig.
\ref{fig:detectors:transmission}b using an atomic point contact
(APC) made by electromigration
\cite{park_APL_electromigration}. Since the tunneling
resistance between the APC and the beam is high, a tank circuit
is used for impedance matching to $50 \un{\Omega}$
high-frequency amplifiers. The Brownian motion of the beam is
observed at a temperature of $250 \un{mK}$ with a shot-noise
limited imprecision of $2.3 \un{fm/\rHz}$. From the temperature
dependence of the thermal noise spectra, a backaction
temperature of $0.7 \un{K}$ is found, which is larger than
expected from the momentum carried by the tunneling electrons.
Kan {\it et al.} used a similar technique where an STM tip was
positioned above a MEMS resonator~\cite{kan_APL_tunneling}. The
current modulation due to the resonator motion was down-mixed
(Sec. \ref{sssec:detectors:mixing}) to overcome bandwidth
limitations. \added{Based on these methods an interesting new
type of detector is proposed
\cite{doiron_PRL_momentum_detector}. By incorporating the
movable tunnel junction into a loop threaded by a magnetic
flux, the APC can be used to measure the \emph{momentum}
instead of the position of the resonator. Also, other more
sophisticated schemes are envisioned for detecting entanglement
in the mechanical quantum oscillator
\cite{schmidt_PRL_entanglement_APC}.}

Another method is based on field emission of electrons from a
vibrating tip. Consider, for instance, a multi-walled carbon
nanotube \cite{purcell_PRL_tuning_nanotube} mounted under vacuum
in a field emission setup (see Fig.
\ref{fig:detectors:transmission}c). A large voltage is applied
between the carbon nanotube and an observation screen which lights
up at the position where electrons are impinging. Electrons are
accelerated from the tip of the nanotube to the screen by the
electric field. Motion of the nanotube cantilever results in a
small blurring of the spot at the screen. The thermal motion of
the nanotube already blurs the spot a bit, but when the vibration
amplitude is enlarged by applying a RF driving signal on nearby
electrodes the blurring of the spot becomes more pronounced on
resonance. The electric field also pulls on the nanotube, thereby
increasing the resonance frequency
\cite{purcell_PRL_tuning_nanotube}. The method has been used to
build a nanotube radio \cite{jensen_NL_CNT_radio}, a mass sensor
with atomic resolution \cite{jensen_natnano_masssensing}, and high
$Q$ silicon carbide nanowire resonators at room temperature
\cite{perisanu_APL_highQ_SiC}.

In more complicated circuits shuttling of electrons has been
considered \cite{erbe_APL_shuttle, erbe_PRL_shuttle,
koenig_natnano_shuttle}. These devices operate as follows:
there are two electrodes with a movable metallic island in
between. When the resonator is closer to one of the electrodes,
the transmission between the island and this electrode
increases, allowing for electrons to jump on the island. The
electrostatic force drives the island with the
negatively-charged electrons to the other electrode. This way,
they are transported to the other electrode by the resonator
(mechanical transport of electrons). In steady state this
shuttling occurs at the resonance frequency of the oscillating
island. Interestingly, the voltage applied between the two
electrodes can amplify this motion and lead to large amplitude
oscillations; an example of an electomechanical
instability\footnote{This shuttling instability can thus be
viewed as an example of backaction induced self-oscillations as
discussed in the previous Section. When pursuing the quantum
limit in these devices, people refer to the ability to
transport exactly one charge carrier per cycle from source to
drain; the zero-point motion does not play a crucial role in
achieving this.}. This subject has been extensively reviewed by
Gorelik {\it et al.} \cite{shekter_JCTN_shuttle_review} and we
refer to this work for further reading.

\subsection{Capacitive detectors}
In many nano-electromechanical devices, movable metallic parts
form capacitances with nearby metallic electrodes or ground
planes. Displacement of the mechanical structure inevitable leads
to a change in these capacitances which can be detected
electrically if they are large enough. Likewise, the motion of the
mechanical resonator can be actuated by applying voltages between
the different electrodes as we have discussed in Sec.
\ref{sssec:nems:nanotubes}. This capacitive actuation and read-out
of mechanical motion has been applied in several ways. Here, we
will discuss three popular methods.

\subsubsection{Single-electron transistors}
\label{sssec:detectors:set} 
\begin{figure}[tb]
\centering
\includegraphics{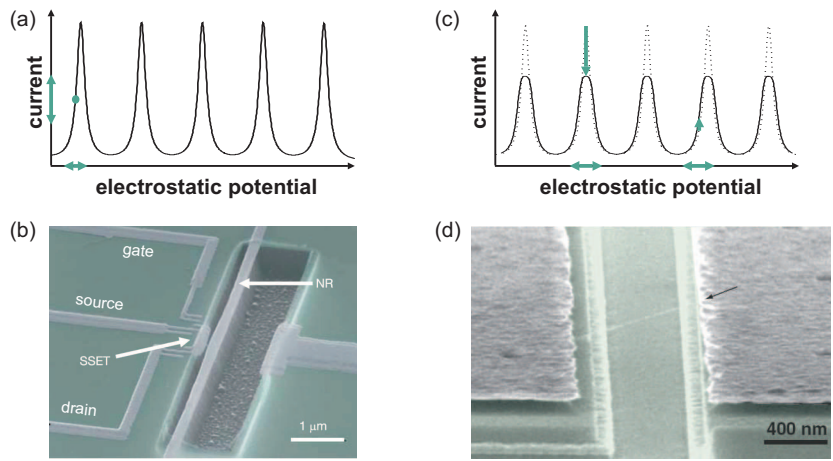}
\caption{(a) Illustration of the dependence of the current through
a SET on the electrostatic potential. The current is almost zero
in the Coulomb blockade regime and displays so-called Coulomb
peaks where the blockade is absent. The resonator motion modulates
the potential (horizontal arrow) and this leads to a proportional
change in the current. (b) The device used in the Schwab group.
The superconducting single-electron transistor (SSET) is made from
aluminum and is connected to a source and drain electrode. The
potential of the SET can be varied using a gate electrode and is
also influenced by the position of the beam resonator. From M.~D. LaHaye {\it et al.},
Science 304 (2004) 74--77. Reprinted with permission from AAAS. (c) Illustration of self-detection
via rectification. Motion of the resonator leads to a decrease in
the {\it average} current on top of the Coulomb peak, whereas deep in
the blockade region it leads to a (small) increase. (d) SEM
picture of a ultra-high quality factor nanotube resonator. From G.~A. Steele {\it et al.},
Science 325 (2009) 1103--1107. Reprinted with permission from AAAS.
\label{fig:detectors:SET}}
\end{figure}

One of the first mesoscopic devices
that was used for position detection \replaced{was}{is} the
single-electron transistor (SET) \cite{knobel_nature_set}. This
device consists of a metallic island that is connected by
tunnel contacts to a source and drain electrode. The total
capacitance of the island is so small that the energy required
to add one electron surpasses the thermal energy
\cite{grabert_single_charge_tunneling} and electrons can only
enter and leave the island one by one (sequential tunneling).
The resulting single-electron current through the island is
very sensitive to the electrostatic environment. Thus, if a
nearby resonator is capacitively coupled to the island, and it
has a different electrostatic potential, the device can be used
as a displacement sensor, as illustrated in Fig.
\ref{fig:detectors:SET}. The first realization of such a device
was made by Knobel and Cleland \cite{knobel_nature_set}, where
the SET measured the position of a $116 \un{MHz}$ beam
resonator with a sensitivity of $2 \un{fm/\rHz}$. This value
corresponds to a position resolution a factor of 100 above the
standard quantum limit, as is indicated in Table
\ref{tab:detectors:sensitivity}. The SET was operated as a
mixer, a technique that we will discuss in more detail in Sec.
\ref{sssec:detectors:mixing}. Subsequent work by LaHaye {\it et
al.}~\cite{lahaye_science} demonstrated, using a
superconducting radio-frequency SET, a sensitivity of four
times the standard quantum limit. The role of the RF SET is to
provide impedance matching for the high-frequency resonator
signal ($19.7 \un{MHz}$).

Using an improved version of this device (Fig.
\ref{fig:detectors:SET}b), Naik {\it et al.} \cite{naik_nature}
observed the backaction of the superconducting SET on the
mechanical resonator; a change in the capacitance leads to a
change in the average charge (``occupation'')
\cite{armour_PRB_shot_SET}, which in turn leads to a change in
the electrostatic force the resonator experiences. Depending on
the bias conditions (gate voltage) of the SET, shifts in the
resonance frequency and the damping rate were observed. Thus
the backaction force on the detector, characterized by the
function $\lambda_F$ that was introduced in Sec.
\ref{sssec:basic:lineardetectors}, is bias dependent for this
type of detector~\cite{blencowe_NJP_dynamics_SET,
clerk_NJP_qemsbath, blencowe_PR_QEMS, koerting_PRB_SSET}. The
effective resonator temperature has been measured for different
coupling strengths between the resonator and the SET. In the
experiment this is done by varying the voltage difference
between the resonator and the SET. For small couplings, the
resonator temperature equals that of the sample chip i.e.,
close to the base temperature of the dilution refrigerator. For
higher couplings, the resonator temperature is set by the
backaction temperature of the SET, and is raised to about $200
\un{mK}$.

It should also be possible to use a SET to sense the vibrations
of a suspended carbon nanotube, but the small capacitive
coupling between the nanotube and the SET island makes this a
challenging task \cite{huettel_NJP_QEMS}. It is more
advantageous to use the suspended carbon nanotube {\it itself}
as a self-detecting SET. Using current rectification and
frequency mixing, information about the driven motion of the
suspended nanotube has been obtained \cite{huettel_NL_highQ,
lassagne_science_coupled_cnt, steele_science_strong_coupling}.
In these experiments, a strong coupling between mechanical
motion and the charge on the nanotube has been observed. The
strong coupling is due to the electrostatic force generated by
individual electrons tunneling onto the nanotube. For the
devices used in these experiments, the change in equilibrium
position of the nanotube after adding a single electron easily
surpasses the zero-point motion. Typically, the single-electron
tunnel rate is much larger than the resonance frequency,
indicating that the backaction is determined by the average
number of electrons (``occupation'') on the nanotube. This
backaction leads to frequency shifts and changes in damping as
a function of gate voltage. To be more specific, the damping
increases dramatically with the amount of current flowing
through as the electron tunneling produces a large stochastic
backaction force.

In Refs. \cite{huettel_NL_highQ,
steele_science_strong_coupling} readout using current
rectification\footnote{The rectification and mixing measuring
method have in common that they both rely on the nonlinearity
of the current-voltage or current-gate-voltage characteristics
\cite{knobel_APL_SET_mixer}; the experimental implementation
is, however, different. It should also be noted that
rectification is more commonly used to detect displacement in
mechanical resonators. For example, in the experiments of Ref.~
\cite{jensen_NL_CNT_radio,jensen_natnano_masssensing} (Sec.
\ref{ssec:detectors:transmission}) the dc tunnel current
contains information about the vibrating carbon nanotubes since
a time-dependent displacement changes the
\replaced{source}{soured}-drain distance and thereby the
averaged current.} is employed. While the nanotube motion is
actuated by a RF signal on a nearby antenna, the detected
signal is at DC. The key to understand this is the notion that
nanotube motion effectively translates into an oscillating gate
voltage, which smears out the sharp features of the SET
current, as illustrated in Fig. \ref{fig:detectors:SET}c
\cite{chtchelkatchev_PRB_rectifying}. The technique is of
special interest as it constitutes a square-law detector (see
\appref{app:squared}). Moreover, it allows for the motion
detection with small currents, enabling the observation of
ultra high Q-factors, exceeding 100,000 at millikelvin
temperatures. The low dissipation enables the observation of
{\it single-electron tuning} and frequency tuning oscillations,
analogous to the Coulomb oscillations in the SET current.
Recently, this self-detecting rectification scheme has also
been employed for a thin aluminum beam \cite{pashkin_APL_SET}.

\subsubsection{Quantum-point contacts}
A quantum-point contact (QPC) is a narrow constriction in a
two-dimensional electron gas, whose conductance can be adjusted
using electrostatic potentials \cite{beenakker_overview}: Every
time another channel for electrons becomes available, the
conductance increases by one conductance quantum $2e^2/h$. In
practice these sharp steps are smoothed by temperature and when
the QPC is biased near such a step it is very sensitive to
changes in the electrostatic potential. If mechanical
vibrations modulate the electrostatic fields, the QPC can also
be used as a position detector. Cleland {\it et al.} have used
this principle to measure the vibrations of a beam made out of
a single-crystal GaAs heterostructure with a sensitivity of $3
\un{pm/\rHz}$ \cite{cleland_APL_QPC}. In this case the QPC is
part of the beam and frequency mixing (Sec.
\ref{sssec:detectors:mixing}) is used to detect the
displacement through the voltage across the QPC. The signal is
amplified through the piezoelectric effect in GaAs that will be
discussed in more detail in Sec.
\ref{ssec:detectors:piezoelectric}.

A similar sensitivity has been achieved in a completely
different setup by Poggio and coworkers
\cite{poggio_natphys_QPC}. Now a $5 \un{kHz}$ micromechanical
resonator is hanging above a QPC that is located on a different
substrate. Thermal noise specta have been recorded and the
authors have mapped out the transduction factor as a function
of cantilever position relative to the QPC. Furthermore, they
observe that the cantilever Q-factor is not affected by the QPC
source-drain current, indicating weak coupling and therefore
negligible backaction.

A QPC can also be used to probe vibrational modes of the host
crystal itself \cite{stettenheim_nature_QPC}. This substrate is
a truly three-dimensional resonator that consists of on the
order of $10^{20}$ atoms. Strictly speaking, the latter example
does not constitute of a capacitive detector as the
transduction is through the piezoelectric effect; the motion
directly influences the source-drain voltage and not the
conductance of the QPC.

\subsubsection{Frequency mixing}
\label{sssec:detectors:mixing} Frequency mixing has been
adopted in top-down solid-state devices as a versatile
technique to convert high-frequency motion into a low-frequency
signal \cite{cleland_APL_QPC, knobel_nature_set,
bargatin_APL_piezo_mixer}. It is most often used in combination
with capacitive detection techniques, although it is also used
in combination with electron tunneling (Sec.
\ref{ssec:detectors:transmission}) and piezoelectric resonators
(Sec. \ref{ssec:detectors:piezoelectric}). The basic principle
is as follows. Consider the generic relation between the input
(i.e., the displacement $u$) and the detector output $v$
introduced in Sec. \ref{sec:basic}: $v(\omega) =
\lambda_v(\omega) u(\omega)$. By modulating the transduction
$\lambda_v$ at a frequency $f_{LO}$ (often called the local
oscillator frequency), the displacement at frequency $f_R$ is
converted into a signal with frequencies $f_R + f_{LO}$ and
$f_R - f_{LO}$. The latter component at the difference
frequency can be chosen to be at a frequency far below the
resonator frequency: for example in the kHz range. The signal
at this frequency is not affected by the RC time of the
measurement setup and can therefore be measured
straightforwardly.

The technique has become of particular interest for detecting
vibrational motion of bottom-up devices. Sazonova {\it et al.}
\cite{sazonova_nature} were the first to apply frequency mixing
to suspended carbon nanotube resonators. They observed multiple
gate-tunable resonances with Q-factors on the order of 100 at
room temperature. Subsequently Witkamp {\it et al.}
\cite{witkamp_NL_bendingmode} identified the bending mode
vibrations of a carbon nanotube. Nowadays the technique has
been employed by many groups, not only restricted to carbon
nanotubes \cite{lassagne_NL_masssensing, chiu_NL_masssensing,
witkamp_APL_paddle, lassagne_science_coupled_cnt,
wang_science_nanotube_phase_transitions,
gouttenoire_SM_FM_nanotube}, but also applied to suspended
graphene sheets \cite{chen_natnano_graphene_mixing,
singh_NT_graphene_mixing}, free-hanging semiconducting
nanowires \cite{zhu_nanotech_ZnO_nanowire, fung_APL_nanowire}
and charge-density-wave sheets \cite{sengupta_PRB_CDW_NEMS}.
Furthermore, several variations to the original mixing scheme
have been implemented, including FM
\cite{gouttenoire_SM_FM_nanotube} and AM modulation
\cite{witkamp_APL_paddle}.

\begin{figure}[tb]
\centering
\includegraphics{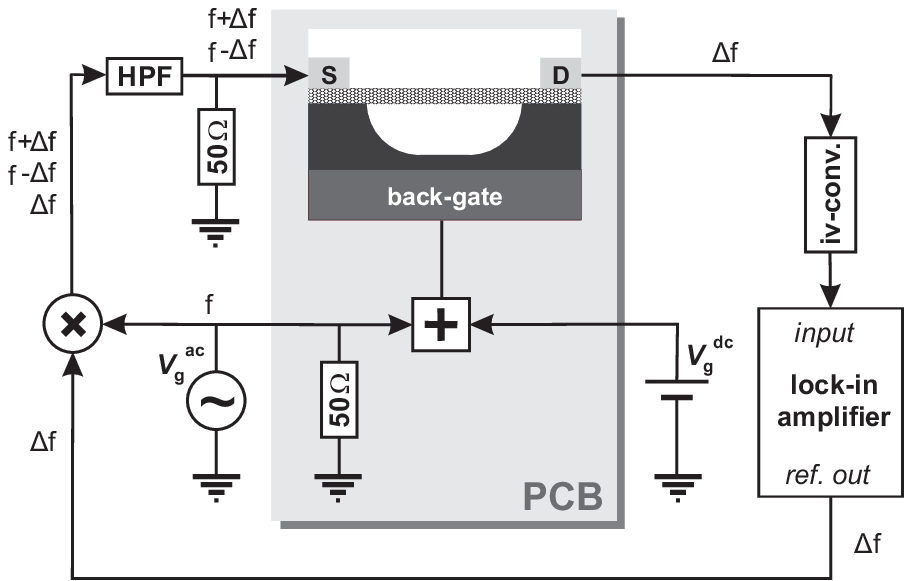}
\caption{Schematic overview of the measurement electronics used by
Witkamp {\it et al.} for position detection of a suspended
nanotube using frequency mixing \cite{witkamp_NL_bendingmode}. A
radio frequency (RF) generator applies an ac voltage to the
back-gate electrode to drive the suspended nanotube. A dc gate
voltage is added via a bias-T (indicated by the ``+''). The same
generator is used to generate the ac bias voltage by mixing its
output with the reference output of the lock-in amplifier. At the
source electrode the voltage has spectral components at $f+\Delta
f$ and $f-\Delta f$, whereas the gate voltage is oscillating at
frequency $f$. The nanotube mixes both signals, which results in
an output current at the drain electrode with spectral components
at $\Delta f$, $f+\Delta f$, $f-\Delta f$, $2f+\Delta f$ and
$2f-\Delta f$. The $\Delta f$ part of the current flowing through
the nanotube is converted into a voltage and is measured with a
lock-in amplifier. The printed circuit board (PCB) with the
sample, bias-T and $50 \unitt{\Omega}$ terminator is located
inside the vacuum chamber of a probe station.}
\label{fig:detectors:mixing}
\end{figure}

A more detailed understanding of this self-detecting method can
be obtained by considering, for example, the suspended carbon
nanotube of Fig. \ref{fig:nems:nanotube}. The conductance of a
semiconducting nanotube depends on the induced charge, which is
the product of the gate voltage and the gate capacitance, i.e.
$G = G(C_g V_g)$. As we have shown before, the gate capacitance
is position dependent and therefore the conductance varies in
time with the resonator frequency. In nanotube mixing
experiments, the current is measured, which is the product of
the conductance and the bias voltage: $I = V_b~\pderl{G}{u}
\cdot u$, from which we identify $\lambda_v =
V_b~\pderl{G}{u}$. The modulation of the transduction can thus
be done by applying an ac bias voltage: $V_b = V_{ac}
\cos(\omega_{LO} t)$, as explained in Fig.
\ref{fig:detectors:mixing}. In these experiments backaction
comes again from the electrons flowing through the nanotube
\cite{lassagne_science_coupled_cnt}.

\subsection{Piezoelectric resonators}
\label{ssec:detectors:piezoelectric} A different mechanism to
transduce a displacement into an electrical signal uses
piezoelectricity. In a piezoelectric material a stress induces
an electric polarization, or equivalently, an electric
displacement $\bm{\mathcal{D}}$ \cite{newnham_materials}. This
results in a voltage difference across the resonator that can
be used to infer the displacement. Piezoelectric resonators are
different from most other methods discussed in this Section as
no separate detectors are needed; they detect their own motion
instead. Moveover, they are also different from the other
self-detecting resonators (e.g. the suspended carbon nanotubes
and graphene resonators that were described above) since no
external signal needs to be applied in order to measure the
displacement (for example, in the nanotube experiments of
Sections \ref{sssec:detectors:set} and
\ref{sssec:detectors:mixing} one has to apply a source-drain
voltage to transduce the displacement). Still, from the
discussion in Sec. \ref{ssec:basic:sensitivity} it is clear
that backaction forces should also act on a piezoelectric
resonators when measuring the generated voltage. This happens
through the converse piezoelectric effect, where an electric
field generates a strain in the material. Referring back to the
definition of stress and strain in Sec. \ref{sec:nems}, the
direct and converse piezoelectric effect are
\cite{newnham_materials}:
\begin{equation}
\mathcal{D}_i = d_{ijk}\sigma_{jk}, \hspace{1cm} \strain_{ij} =
d_{ijk}\mathcal{E}_k.
\end{equation}
Here $\ten{d}$ is the third-rank tensor with the piezoelectric
coefficients. Similar to the properties of the elasticity
tensor, many of the elements \deleted{of} $d_{ijk}$ are zero,
or related to other elements, depending on the symmetries of
the crystal structure of the material. A necessary requirement
to have at least one nonzero element is that the material's
unit cell does not have an inversion center. For most
piezoelectric materials, including GaAs and AlN, the
piezoelectric coefficients are of the order of pC/N and they
can have both positive and negative values.

Piezoelectric displacement detection of small structures has
first been employed with AFM cantilevers
\cite{beck_APL_AFM_piezo, watanabe_RSI_piezo_AFM,
itoh_JVSTB_piezo_AFM}, before extending the technique to the
nanomechanical domain. Tang {\it et al.} used this detection
scheme\footnote{In this experiment the measured signal
contained both piezoelectric and piezoresistive components.} to
detect the motion of GaAs/AlGaAs beams
\cite{tang_APL_GaAs_piezo}. The beams are made asymmetric to
ensure that the piezoelectric signal is not nulled by the
opposite stresses at both sides of the neutral plane (Sec.
\ref{ssec:nems:eulerbernoulli}). Cleland {\it et al.} used
piezoelectricity to detect the motion of GaAs beams with an
integrated QPC in it. There, a deflection of the beam induces
in-plane stress and the resulting out-of-plain field
$\bm{\mathcal{D}}$ acts as an effective gate voltage. This then
changes the current through the QPC \cite{cleland_APL_QPC}. An
important consideration is that piezoelectric materials cannot
generate large currents. The resonator should therefore be
connected to a high-impedance load, such as the gate electrode
of the QPC \cite{cleland_APL_QPC}, to the gate of a SET
\cite{knobel_APL_piezo_SET}, or a high-impedance amplifier
\cite{tang_APL_GaAs_piezo}.

Mahboob and Yamaguchi used the two-dimensional electron gas in
a flexural beam resonator made of a GaAs/AlGaAs heterostructure
to measure the displacement and they demonstrated parametric
amplification by modulating the beam's resonance frequency
using the converse piezo-electric effect
\cite{mahboob_APL_piezoelectric}. The converse piezo-electric
effect was also used to tune the frequency of a doubly-clamped
zinc-oxide nanowire by Zhu and
coworkers~\cite{zhu_nanotech_ZnO_nanowire}. Okamoto {\it et
al.} found an enhancement of backaction effects due to
excitation of carriers in an piezoelectric GaAs resonator
\cite{okamoto_PRL_optomechanical}. Many of the abovementioned
effects were combined by Masmanidis and coworkers who used a
three-layer structure, consisting of p-doped, intrinsically
doped, and n-doped GaAs to demonstrate piezoelectric actuation,
frequency tuning, and nanomechanical bit operations
\cite{masmanidis_science_piezoelectric}. By electrically
shifting the carriers with respect to the neutral plane, the
strength of the actuation could be adjusted. Finally, by
capacitively coupling a bulk AlN resonator to a qubit,
O'Connell {\it et al.} obtained the high coupling rates
required to perform quantum operations on a mechanical
resonator \cite{oconnell_nature_quantum_piezo_resonator}. This
experiment was already mentioned in the Introduction and will
be discussed in more detail in Sec.
\ref{ssec:prospects:hybrid}.

\subsection{Flux-based position detectors}
\label{ssec:detectors:flux}
\subsubsection{SQUIDs}
Superconducting \added{quantum} interference devices (SQUIDs)
are well known for detecting small magnetic signals, such as
those generated by our brains. These devices consist of a
superconducting ring in which one or more Josephson junctions
are incorporated \cite{squidhandbook}. The voltage across the
loop depends on the amount of magnetic flux that threads the
loop, $\Phi_{mag}$. This flux-dependence\footnote{SQUIDs are
also used indirectly for position detection, where the SQUID is
used as low-noise voltage amplifier that amplifies the signal
generated by capacitive detection \cite{vinante_PRL_feedback}.
In this section, we only focus on direct position detection
with SQUIDs.} has been employed to sense the motion of a
cantilever with a small magnetic particle attached to it
\cite{usenko_APL_SQUID}.

\begin{figure}[tb]
\centering
\includegraphics{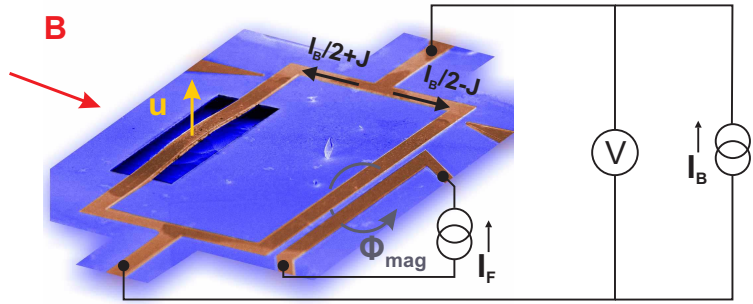}
\caption{Schematic overview of a dc SQUID position detector,
including the suspended beam resonator and measurement setup. The
SQUID loop is indicated in brown. The output voltage of the SQUID
depends on the bias current $I_B$ that is sent through the SQUID
and also on the amount of flux $\Phi_{mag}$ through the loop. A
magnetic field $B$ transduces a beam displacement $u$ into a
change in magnetic flux and subsequently in a change in the output
voltage $V$. The flux $\Phi_{mag}$ is fine-tuned with a stripline
current $I_F$. Backaction results from the circulating current in
the SQUID loop $J$. Reprinted figure with permission from M. Poot
{\it et al.}, Phys. Rev. Lett. 105 (2010) 207203. Copyright 2010
by the American Physical Society. \label{fig:detectors:squid}}
\end{figure}

It is, however, also possible to detect displacements by {\it
incorporating} a mechanical resonator in the SQUID loop
\cite{zhou_PRL_proposal, buks_RPB_rf_decoherence,
xue_NJP_controlable_coupling, blencowe_PRB_quantumanalysis,
buks_EPL_qndfock, wang_NJP_cooling, nation_PRB_squid_cavity,
zhang_PRA_quantum_feedback, xia_PRL_groundstate,
pugnetti_PRB_ratchet, pugnetti_EPL_resonant_coupling}: In the
presence of a constant magnetic field, a change in the
resonator displacement changes the loop area and thereby the
flux through the loop. This is illustrated in Fig.
\ref{fig:detectors:squid}. Recently, Etaki {\it et al.} have
used a dc SQUID as a sensitive detector of the position of an
integrated mechanical resonator. They detected the driven and
thermal motion of a 2-MHz buckled-beam resonator with
femtometre resolution at millikelvin temperatures
\cite{etaki_natphys_squid} and employed active feedback cooling
to cool the resonator to $20 \un{mK}$ \cite{poot_APL_feedback}.
In the present experiments, the sensitivity is limited by the
cryogenic amplifier and not yet by the, in principle quantum
limited SQUID itself.

Backaction has also been observed, leading to tunable shifts in
the resonance frequency and damping of the resonator
\cite{poot_PRL_backaction}. Different from the backaction in the
capacitive readout schemes, the backaction of the SQUID has an
inductive character as it is caused by the Lorentz force generated
by the current circulating in the loop of the SQUID. This current
also runs through the resonator (Fig. \ref{fig:detectors:squid})
and the magnetic field that couples displacement and flux,
generates a Lorentz force on the resonator.

\subsubsection{Magnetomotive}
A fast and relatively easy way to actuate and read-out
nanomechanical motion in clamped-clamped resonators is the
magnetomotive technique \cite{graywall_PRL_magnetomotive,
cleland_APL_magnetomotive, huang_nature_GHz,
shim_science_coupled_resonators}. Nowadays, this method is
mainly used to determine the vibrational frequencies and to
characterize dynamic properties such as the Q-factor and
nonlinear behavior. It is, however, of limited use for QEMS
experiments and we will therefore only briefly describe the
mechanism: An ac current $I_{ac}$ is sent through the
conducting (part of the) beam which is placed in a strong
static magnetic field. The ac current causes an ac Lorentz
force that drives the beam $\mu_n B \ell I_{ac}$. At the same
time, motion of the beam induces a time-varying voltage
(Faraday's law) at the driving frequency $V_{emf} = \mu_n B
\ell \replaced{\dot u}{v}$. By sweeping the driving frequency,
the resonance can be found by measuring the frequency response
of the voltage over the beam. The required large magnetic field
is typically generated using superconducting magnets in a
cryogenic environment, but recently room temperature
magnetomotive actuation and detection has been demonstrated
with a $2 \un{T}$ permanent magnet \cite{venstra_APL_arrays}.
An important feature of this technique is that observed
mechanical modes can be distinguished from electronic
resonances by varying the magnetic field. Since both the
driving force and the detector signal are proportional to $B$,
the mechanical signal scales as $B^2$
\cite{gaidarzhy_APL_multielement}.

\subsection{Level spectroscopy in suspended quantum dots}
\begin{figure}[tb]
\centering
\includegraphics{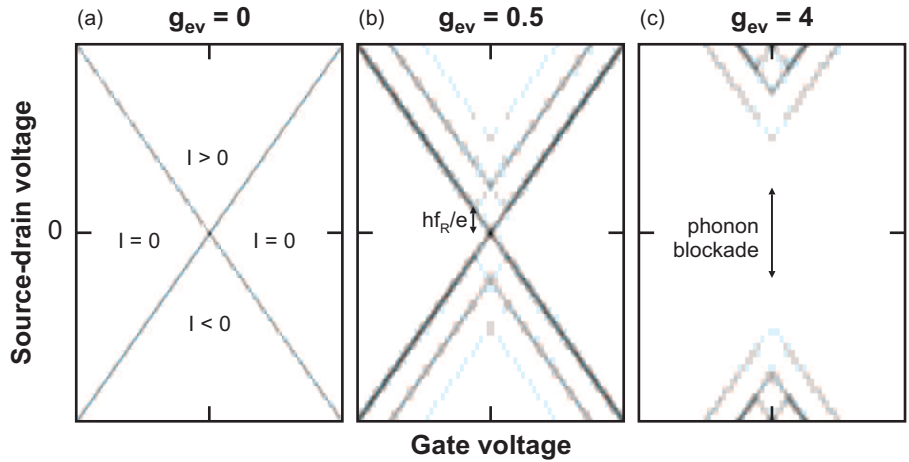}
\caption{Stability diagrams for zero (a), intermediate (b), and
large (c) electron-vibron coupling $g_{ev}$. A stability diagram
is a grayscale plot of the differential conductance plotted
against the gate and source-drain voltage. Coulomb peaks \replaced{(Fig.
\ref{fig:detectors:SET})}{as shown in Fig.
\ref{fig:detectors:SET}} appear here as dark lines, whereas regions
with low differential conductance d$I$/d$V$
are white. (a) shows a regular stability diagram
without signatures of the mechanical resonator. There are regions
where the current is blocked and regions where current flows.
These are separated by the high-conductance lines (dark). In (b) a
finite electron-vibron coupling is present and mechanical
excitations appear as lines parallel to the original lines of high
conductance (vibrational sidebands). The source-drain voltage where these lines
cross the Coulomb diamond edges, equals $hf_R/e$ as
indicated. For large $g_{ev}$ the current
is blocked at low source-drain voltage as shown in (c)
and vibrational excitations can only been seen at high bias voltage.
\label{fig:detectors:frankcondon}}
\end{figure}

At low temperatures, suspended quantum dots may reveal information
about their vibrational states: in the parameter regime dominated
by Coulomb blockade physics, the quantum mechanical level spectrum
of the confined electronic system can be characterized by
transport measurements. The levels show up as steps in the
current-voltage characteristics. A more accurate measurement
involves the recording of a so-called stability diagram in which
the differential conductance dI/dV plotted as function of gate
voltage $V_G$ and source-drain voltage $V$. Now the excitations
(due to electronic or vibrational degrees of freedom) show up as
lines, whose energy can directly be read out
\cite{thijssen_PSSB_overview} as illustrated in Fig.
\ref{fig:detectors:frankcondon}. In the case of vibrational states
these are also called vibrational side bands and when the
harmonics of a particular vibration are excited, the lines form a
spectrum with equidistant spacing. Thus, electron tunneling
through suspended quantum dots can excite vibrational modes and
these modes can then be detected as steps in the current-voltage
characteristics. Their observation involves, however, two
important considerations: the energy resolution and the
electron-vibron coupling. Concerning the energy resolution of
transport spectroscopy measurements one must have $\hbar \omega_R
> k_B T$ so that measurements are always performed in the quantum
limit of the mechanical mode.

Secondly, the electron-vibron coupling must be high enough; it
can be characterized by a dimensionless parameter $g_{ev} =
\halfl (\Delta u/u_0)^2$, where $\Delta u$ is the shift of the
resonator position\footnote{Actually, this is the shift in the
location of the minimum of the potential energy of the
resonator. The tunneling of the electron is assumed to take
place on a timescale that is much faster than the resonator can
react.} induced by adding one elementary charge, and $u_0$ is
the zero point motion of the mechanical oscillator
\cite{mccarthy_PRB_FC}. The parameter $g_{ev}$ determines the
step height in the current-voltage characteristic. We can
consider three regimes in describing the influence of
vibrational modes on transport: the weak electron-vibron
coupling regime with $g_{ev} \ll 1$, the intermediate regime
with $0.1 < g_{ev} < 1$ and the strong coupling ($g_{ev}  \gg
1$) limit. The boundaries of the intermediate regime
are\added{, however,} somewhat arbitrary. Figure
\ref{fig:detectors:frankcondon} shows calculated stability
diagrams for three different values of the electron-vibron
coupling. In the weak coupling regime (see panel (a)) only the
regular Coulomb step is present (no side bands are visible) and
consequently vibrational modes cannot be detected in a
transport experiment. Only for sufficiently large
electron-vibron coupling, one or multiple so-called
Franck-Condon steps can be observed \added{in the current,
which show up as lines in the stability diagram} (the
intermediate regime, panel (b)). This was first demonstrated in
molecular junctions exciting C$_{60}$ radial breathing
modes~\cite{park_nature_C60}, and later in suspended carbon
nanotubes probing the radial breathing mode
\cite{leroy_nature}, or the longitudinal, stretching
modes~\cite{sapmaz_PRL_longitudinal}. In the strong
electron-vibron coupling limit, the vibrational induced
excitations are only seen for larger bias voltages as the
height of the first steps is exponentially suppressed. In panel
(c) this is clearly visible: the grey lines representing the
vibrational excitations are only present for high source-drain
voltages. Importantly, the suppression holds for any gate
voltage and as a result the current at low bias is suppressed
in the whole gate range~\cite{koch_PRL_FC_blockade,
pistolesi_PRB_blockade}. Degeneracy points in the Coulomb
diamonds are no longer visible in the stability diagrams and
one speaks of phonon blockade of transport. Phonon blockade has
been observed in suspended quantum dots embedded in a
freestanding GaAs/AlGaAs membrane~\cite{weig_PRL_suspended_QD}
and subsequently in suspended carbon nanotube quantum
dots~\cite{leturcq_natphys_FC}.

Recently, is has been shown that inelastic electron tunneling
spectroscopy (IETS) can also be used to gain information about
vibrational modes in suspended quantum dots. H\"uttel {\it et
al.}~\cite{huettel_PRL_pumping} observed a harmonic excitation
spectrum connected to the longitudinal stretching modes of a
suspended carbon nanotube in the Coulomb-blockaded regime,
while temperature only allows the observation of a single
excitation. The non-equilibrium occupation of the modes is
explained by the pumping via electronic states, revealing a
subtle interplay between electronic and vibrational degrees of
freedom.

\section{Prospects} \label{sec:propects}
\begin{figure}[tb]
\centering
\includegraphics{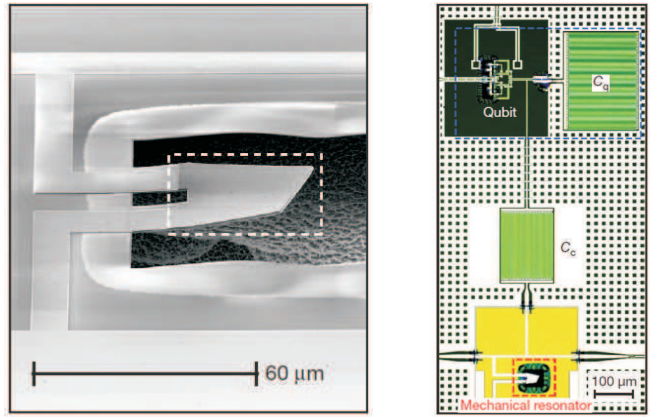}
\caption{A mechanical resonator coupled to a Josephson qubit. The
mechanical resonator (left) is cooled cryogenically to its ground
state and manipulation of its quantum state at the single phonon level has been demonstrated. The quantum properties of the resonator are probed by the
superconducting qubit which is positioned on the top side of the
right panel. Reprinted by permission from Macmillan Publishers Ltd: Nature
 464, 697--703, copyright 2010.
} \label{fig:prospects:qubit}
\end{figure}

So far, we have mainly concentrated on the limits of (linear)
displacement detection. As discussed before, there are other
ways of demonstrating quantum behavior. For example, square-law
detection would directly probe the energy eigenstates of the
mechanical resonator (Sec. \ref{sec:basic} and App.
\ref{app:squared}). Another approach is to couple a mechanical
quantum oscillator to another quantum system such as the
well-studied Josephson qubit \cite{lahaye_nature_qubit}: the
state of the mechanical resonator changes the state of the
qubit, which can then be \replaced{probed}{read off} to provide
information about the mechanical states. A general problem in
such schemes is that the coupling needs to be sufficiently
strong so that the exchange of quantum states occurs before
decoherence sets in. In April 2010, the first experimental
realization of a coupled quantum system involving a mechanical
resonator (see Fig.~\ref{fig:prospects:qubit}) has been
reported by the groups of A. Cleland and J.
Martinis~\cite{oconnell_nature_quantum_piezo_resonator}. They
demonstrated the superposition and coherent control of the
quantum states of a mechanical resonator. The key aspect of
their experiment is the use of a piezoelectric material for the
resonator which boosts the coupling; it would be extremely hard
to reach similar coupling strengths by using electrostatic
forces alone (see also Table \ref{tab:detectors:coupling}). We
will briefly come back to this ground breaking experiment in
the second part of this Section when discussing coupling
mechanical systems to other quantum systems from a more general
point of view (see Sec. \ref{ssec:prospects:hybrid}).

Also concerning the linear detection schemes many challenges
lie ahead. Strong coupling, ground-state preparation and
detection imprecision below the quantum limit have all been
reached now. These achievements set the stage for further
studies on controlling and detecting non-classical states of
mechanical motion; eventually one would like to have a
mechanical resonator in its ground state which is strongly
coupled to a single photon so that quantum state of single
phonons and photons can be exchanged. Large coupling strengths
and high Q-factors are necessary ingredients for reaching this
goal. It is not a priori clear which detector scheme will be
the best suited. The optical and superconducting cavities have
the advantage that the underlying concepts are well known and
have studied in detail; for detection schemes based on other
(mesoscopic) devices (e.g. superconducting SQUIDs, carbon-based
resonators) the understanding is clearly not at the same level.
For example, the backaction mechanisms in these cases are not
understood in all details as these coupled detector-resonator
systems cannot be mapped directly onto the Hamiltonian
describing cavity dynamics (Eq.~\ref{eq:basic:hamiltonian}).
More theoretical and experimental research is needed to
elucidate on the underlying mechanisms and physics. In the next
subsection, we will briefly discuss some of the issues and
challenges of linear detectors in more detail.

\subsection{NEMS as quantum-limited sensitive detectors}
It is clear that in the coming years more efficient read-out
and cooling techniques will be employed. For optical systems
challenges lie in increasing the coupling strength and to
construct optical setups that can be incorporated in dilution
refrigerators. In that respect, the recently developed on-chip
optic experiments are promising. In the gravitational wave
community the \replaced{next}{new} generation of LIGO will use
squeezed states to further improve the displacement sensitivity
\cite{ciolfi_LIGO_squeezed}. For the electronic systems,
reduction of detector noise is a major issue which can be
achieved by the implementation of quantum-limited amplifiers
(SQUIDs \cite{koch_APL_quantumnoise, muck_APL_nearQL}, point
contacts \cite{clerk_PRB_ql_linear,
clerk_PRB_mesoscopic_detectors, gurvitz_PRB_noninvasive} or
Josephson parametric amplifiers \cite{takahasi_communication,
teufel_natnano_beyond_SQL}). Furthermore, nonlinear (quantum)
effects (see e.g. Refs.~\cite{peano_PRB_quantum_duffing,
katz_PRL_quantum_duffing, woolley_NJP_quantum_coupled}) have so
far received less attention and this may become an interesting
research line.

For cooling experiments high signal-to-noise ratio's are
important as well as high Q-factors. Since Q-factors are
limited by material properties, mechanical oscillators of new
materials (preferably in a crystalline form) will have to be
fabricated and characterized. We should add here that the
understanding of the mechanisms controlling dissipation in
mechanical resonators is also still open. For example, the role
of tension in elevating the Q-factor is not completely
understood. The temperature dependence of the Q-factor is
another open-standing problem; in some cases a very strong
dependence is observed, while in other cases the dependence is
weak. What limits dissipation and what is the role of
microscopic defect such as two-level fluctuators~
\cite{ekinci_RSI_overview, imboden_APL_diamond,
unterreithmeier_PRL_damping}? Furthermore, for low-mass
resonators new effects may start to play a role such as the
influence of adsorbents on the resonator surface and nonlinear
damping terms associated with the induced tension introduced in
Sec.~\ref{sssec:nems:nanotubes}.
\begin{figure}[tb]
\centering
\includegraphics{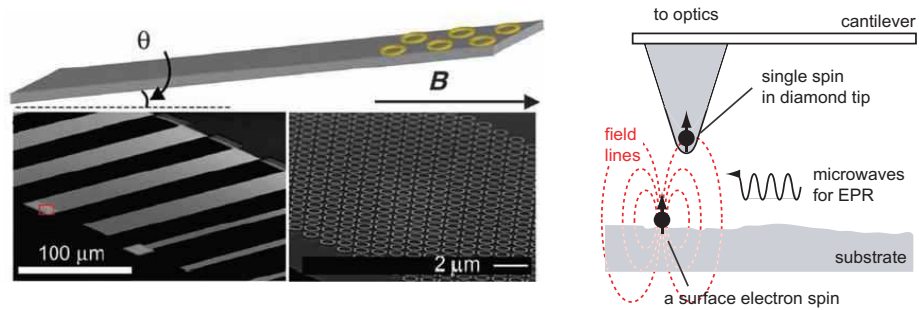}
\caption{Left: a cantilever resonator containing normal metal
rings to study persistent currents. In a magnetic field, the
currents produce a torque on the cantilever which shifts the
cantilever frequency. By measuring the resonance frequency, the
persistent currents could be studied with unprecedented
sensitivity. From A.~C. Bleszynski-Jayich {\it et al.}, Science
326 (2009) 272--275. Reprinted with permission from AAAS. Right:
Schematic representation of the diamond-based scanning spin
microscope proposed in Refs. \cite{chernobrod_JAP_scanning_spin,
berman_JoP_scanning_spin,degen_APL_scanning_diamond}. The single
spin of a NV defect in a diamond tip is a
sensitive magnetometer for the local magnetic field with nanoscale
spatial resolution. Magnetic fields near the surface shift the
electron spin resonance (EPR) frequency of the NV center, which
can for example be detected by exciting the EPR transition with a
microwave field and monitoring the change in photoluminescence of
the probe spin. Reprinted with permission from C.~L. Degen, Appl.
Phys. Lett.  92~(24) (2008) 243111. Copyright 2008, American
Institute of Physics.} \label{fig:prospects:sensitive_detectors}
\end{figure}

The push for refining detection schemes will undoubtedly lead to
the construction of better sensors, which eventually will be
quantum-limited. These will lead to new applications aimed at both
solving fundamental science questions as well as the development
of commercial products. An example of the former direction is the
recent detection of persistent currents in metal rings fabricated
on nanoscale cantilevers
~\cite{bleszynski-jayich_science_persistent_currents}, illustrated
in Fig. \ref{fig:prospects:sensitive_detectors}a. The precision is
one order of magnitude better than previous experiments on
persistent currents, which are mainly based on SQUIDs. It is also
expected that the newly developed optics techniques (i.e., on-chip
optics) will find their way in novel applications in controlling,
stopping and storing light \cite{chang_AIP_slowing_light,
safavi-naeini_NJP_phonon_photon}. Another promising example is the
detection of single spins using magnetic resonance force
microscopy analogous to the magnetic resonance imaging. Single
electron spins have already been
detected~\cite{rugar_nature_single_spin}; the next step is the
detection of nuclear spins which requires a thousand fold increase
in the sensitivity. An interesting recent proposal (Fig.
\ref{fig:prospects:sensitive_detectors}b) is the use of NV centers
so that operation at room temperature becomes
feasible~\cite{degen_APL_scanning_diamond}. If successful, this
would yield a revolution in imaging: Many chemical elements carry
a nuclear magnetic moment, so that a sensitive enough detector can
determine their identity and arrangement in more complex
molecules. Finally, NEMS may find an application in mass sensing
\cite{ekinci_APL_piezo_mass, yang_NL_masssensing,
chiu_NL_masssensing, jensen_natnano_masssensing,
lassagne_NL_masssensing, naik_natnano_masssensing,
gil_natnano_nanowire, eom_physrep_biological}: the state-of-the
art is that single gold atoms can be measured (see Sec.
\ref{sec:detectors}). The goal is to be able to detect masses with
resolution better than 1 Dalton (the mass of a hydrogen atom) so
that each element can be identified making it a mass spectrometer
\cite{ekinci_JAP_mass_sensing_limits}.

\subsection{Hybrid quantum {\it mechanical} systems}
\label{ssec:prospects:hybrid} 
\begin{figure}[tb]
\centering
\includegraphics{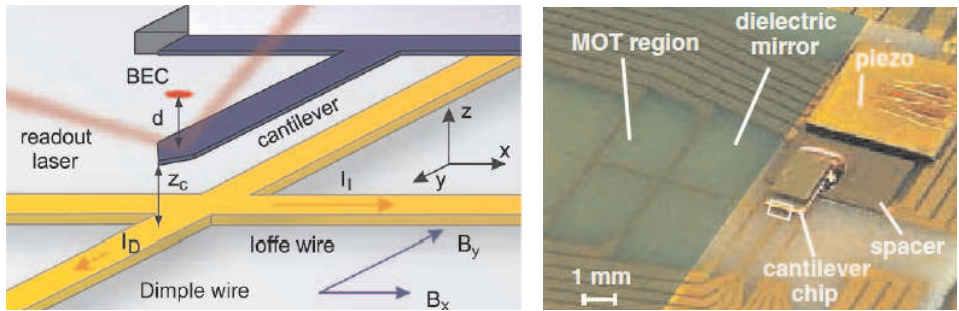}
\caption{Bose-Einstein condensate (BEC) on an atom-chip couples to
a cantilever. The cantilever motion modulates the potential of the
atoms, thereby coupling the cantilever position to atomic motion.
Left: schematic overview of the setup. The atoms of the BEC detect
the cantilever vibrations which can independently measured via the
readout laser. Right: photograph of the atom chip. Reprinted
figures with permission from D. Hunger {\it et al.}, Phys. Rev.
Lett. 104 (2010) 143002. Copyright 2010 by the American Physical
Society} \label{fig:prospects:atom_chip}
\end{figure}

In most of this review we have
treated optical systems with movable parts as stand-alone
systems. Interesting new quantum systems can be built when
coupling optical set-ups to other quantum systems such as an
ensemble of atoms in a Bose-Einstein condensate. The coupled
system would form a hybrid quantum system in which hybrid
strong coupling would enable the creation of atom-oscillator
entanglement and quantum state transfer. A theoretical proposal
appeared in 2007~\cite{treutlein_PRL_atom_chip_theory} and in
2008 the interaction between a Bose-Einstein condensate and the
optical field in cavity was studied
experimentally~\cite{brennecke_science_BEC_cavity}
demonstrating strong backaction dynamics. Subsequent
theoretical work showed that strong coupling between a
mechanical resonator and a single atom should be
feasible~\cite{hammerer_PRL_single_atom}. Strong coupling with
a single atom has indeed been observed in the blue-detuned
cavity regime: Experiments with a single, trapped Mg$^+$ ion
interacting with two laser beams showed the stimulated emission
of centre-of-mass phonons~\cite{vahala_natphys_phonon_laser}.
At high driving, coherent oscillating behavior is observed,
which can be viewed as the mechanical analogue to an optical
laser, i.e., a phonon laser. In another approach, the coupling
between vibrations of a micromechanical oscillator and the
motion of Bose-condensed atoms on a chip are mediated by
surface forces experienced by the atoms placed at one micron
from the mechanical structure~\cite{hunger_PRL_atom_chip}. As
illustrated in Fig. \ref{fig:prospects:atom_chip}, the
Bose-Einstein condensate probes the vibration of a cantilever.
By adjusting the magnetic trapping potential, the discrete
eigenmodes of the condensate can be tuned, which in turn tunes
the coupling to the cantilever. Systems of many resonators
coupled to each other via atoms in an optical lattice are
envisioned~\cite{geraci_PRA_atoms}.

A different approach to create entangled states of mechanical
motion by coupling a mechanical quantum oscillator to a solid
state device. In the experiment of the group of Cleland and
Martinis ~\cite{oconnell_nature_quantum_piezo_resonator}, a
Josephson qubit served as a two-level system probing the
mechanical states of a 3~ng FBAR resonator (c.f. Fig.
\ref{fig:prospects:qubit}). With a mechanical frequency of
6~GHz, a mechanical Q-factor of about 100 and a coupling of
62~MHz, there was just enough time to perform a few quantum
operations on the mechanical resonator before the quantum state
decohered.

If it would be possible to increase the Q-factor while
maintaining a strong coupling, quantum mechanical manipulation
of long-lived phonon states comes into reach. For example, with
a Q-factor of $10^5$ and a frequency of 1~GHz, the mechanical
state would survive for $100~\un{\mu s}$ before it
significantly decoheres. This time is typically much longer
than the coherence time of a superconducting qubit, indicating
that quantum information can be stored in the mechanical motion
and transferred back when needed. The long phonon life times
should also be contrasted to the short-lived states in optical
cavities: with a frequency of several hundreds of THz and a
state-of-the-art optical $Q$ of $10^6$, the photon lifetime is
a few ns. Also in this case the mechanical states can be used
as storage or delay unit for quantum information if efficient
exchange of photons and phonons can be achieved (i.e., strong
coupling). From a more general point of view and as a clear
direction for the future, one can envision the construction of
hybrid systems involving quantum mechanical oscillators that
exploit and combine the strengths of the individual quantum
systems. The mechanical quantum resonators would then serve as
reservoirs of long-lived states with the advantage that they
can be well coupled to a variety of other quantum systems.

As a final note, let us return to the crossover between quantum
and classical systems as discussed in the Introduction. The
quantum experiments of
Refs.~\cite{oconnell_nature_quantum_piezo_resonator,teufel_arXiv_groundstate}
were performed on mechanical oscillators with a mass of 3~ng
and 48~pg respectively. This means that a mechanical object
consisting of about $10^{12}$ atoms can still behave quantum
mechanically. The crossover has thus not been reached in
experiments and continues therefore to be a subject of future
studies. Again this aspect shows that the field of mechanical
systems in the quantum regime is still largely unexplored: It
will still rapidly develop in the years to come bringing many
exciting new experiments and discoveries.

\section{Acknowledgements}
We thank Andreas H\"uttel and Daniel Schmid for their
suggestions about the manuscript. We are indebted to Samir
Etaki, Benoit Witkamp, Yaroslav Blanter, Francois Konschelle,
Miles Blencowe, Jack Harris, and Hong Tang for the discussions
on a wide range of the topics covered in this Report, and for
their suggested improvements. This work was supported by FOM,
NWO (VICI grant), NanoNed, and the EU FP7 STREP projects QNEMS
and RODIN.

\appendix
\renewcommand{\thesection}{\Alph{section}}

\section{Complex Green's function and displacement}
\label{app:complex} One is often interested in knowing the
amplitude $A(t)$ and phase $\varphi(t)$ of the displacement
when it is written as $u(t) = A(t) \cos(\omega_R t +
\varphi(t))$. Since this is only one equation for two
functions, the amplitude and phase are not uniquely defined.
Consider, for example, the solution $A(t) = u(t)/\cos(\omega_R
t)$ and $\varphi(t) = 0$. \replaced{Already}{Even} for a pure
sine wave \added{$u(t) = sin(\omega_R t)$} this results in a
rapidly varying amplitude. The usual notion of the amplitude
and phase are that when the signal is close to sinusoidal, the
amplitude and phase are that of the sine wave. This can be
realized using the complex displacement, which is defined as
$u_c(t) \equiv A_c \exp(i \omega_R t)$ with the requirement
that $\real[u_c(t)] = u(t)$. A convenient way of implementing
this, is using the complex extension of the resonator Green's
function, cf. Eq. \ref{eq:basic:green}:
\begin{equation}
u_c(t) = h_c(t)\conv F(t)/k_R, \hspace{1cm} h_c(t) = -i
e^{i\omega_R t} \cdot e^{-\frac{\omega_R t}{2Q}} \Theta(t),
\end{equation}
so that $\real[h_c] = h_{HO}$. The amplitude and phase of the
resonator displacement are in this case given by the modulus
and argument of the complex amplitude: $A(t) = |A_c|$ and
$\varphi(t) = \angle A_c$ respectively.

Another notion that is often used in the literature is that of
quadratures. Now the displacement is written as $u(t) = U(t)
\cos(\omega_R t) + P(t) \sin(\omega_R t)$. Here $U$ is the
in-phase component and $P$ is the out-of-phase component
(sometimes called quadrature). The quadrature representation is
related to the amplitude-and-phase representation by: $U =
A\cos\varphi$, $P = A\sin\varphi$, and by $A^2 = U^2 + P^2$,
$\tan \varphi = P/U$. The quadratures are also readily
calculated using the complex Green's function: $U = \real[u_c]$
and $P = \imag[u_c]$. Finally, we note that if the displacement
is in the in-phase quadrature, i.e. $u(t) = A\cos(\omega_R t)$,
then the velocity $\tderl{u}{t} \approx -\omega_R
A\sin(\omega_R t)$ is in the out-of-phase quadrature.

\section{Optimal filtering of $v(t)$}
\label{app:optimal} 
\begin{figure}[tb]
\centering
\includegraphics{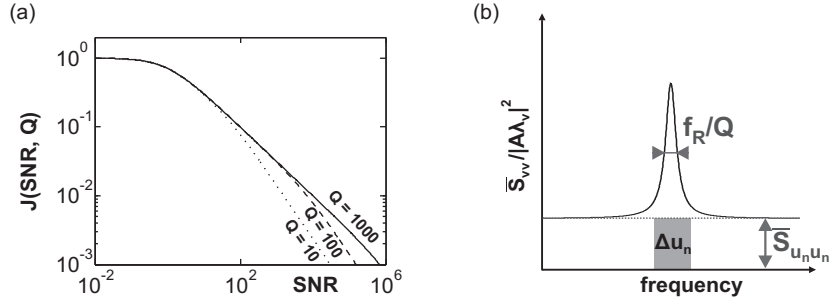}
\caption{Plots of the function $J(\text{SNR}, Q)$ for \replaced{three}{two}
different values of $Q$. The right\deleted{-hand} side \added{of the plot where $\text{SNR} \gg 1$} corresponds to the
situation where the resonator peak is well above the detector
noise level and the resolution is good. On the other hand, on the
left side \added{(i.e. for $\text{SNR} \ll 1$)} the mechanical signal is buried in the imprecision noise
and the error is as large as the resonator signal \added{giving $J = 1$}. (b) \replaced{An}{The}
alternative definition of $\Delta u_n^2$ (cf. Eq.~\ref{eq:basic:resolution_equiv}) is the area of the gray \replaced{rectangle}{area}
indicated in the noise spectrum. \label{fig:app:spectrum}}
\end{figure}

In the presence of both position
(imprecision) and force noise, one wants to reconstruct the
resonator motion in the absence of the detector, $u_i(t) =
h_R(t) \conv F_n(t)$, as good as possible from the measured
time trace $v(t)$ of the detector output. This is done by
finding the estimator $\hat u = g(t) \conv v(t)$ that minimizes
the resolution squared: $\Delta u^2 = \expect[(u_i - \hat
u)^2]$. Using the autocorrelation functions and converting
these into noise PSDs using the Wiener-Khinchin theorem
\cite{shanmugan_random, clerk_overview} the resolution is
written as:
\begin{equation}
\Delta u^2 = R_{u_i u_i}(0) - 2 R_{u_i \hat u}(0) + R_{\hat u \hat
u}(0) = \frac{1}{2\pi}\int_{-\infty}^\infty \left[S_{u_i
u_i}(\omega) - 2 G(\omega) S_{u_i v}(\omega) + |G(\omega)|^2
S_{vv} \right] \intd \omega. \label{eq:basic:resolution}
\end{equation}
Minimizing this w.r.t. $G$, yields the optimal filter
$G_{\text{opt}} = S_{v u_i}/S_{vv}$ \cite{shanmugan_random},
where:
\begin{eqnarray}
S_{v u_i}  = S_{u_i v}^* & = & A H_R (H_R')^* \lambda_v^*S_{F_n
F_n},
\\
S_{vv}  = S_{vv}^* & = & A^2 |H_R'|^2 |\lambda_v|^2\left(S_{F_n
F_n} + A^2 S_{\Phi_{det,n} \Phi_{det,n}} \right) \nonumber
\\
& + & S_{v_n v_n} + 2 A^2 \real \left[ \lambda_v (H_R')^*
S_{\Phi_{det,n} v_n} \right],
\end{eqnarray}
so that the squared resolution is given by:
\begin{equation} \Delta u^2 =
\frac{1}{2\pi}\int_0^\infty \left[\overline S_{u_i u_i}(\omega) -
|G_{\text{opt}}(\omega)|^2 \overline S_{vv} \right] \intd \omega.
\label{eq:basic:resolution_opt}
\end{equation}
Depending on the properties of the detector and the coupling
$A$, two important limits can be distinguished:
\begin{itemize}
\item The detector exerts backaction force noise, but the
    displacement noise is negligible: $\overline S_{v_n
    v_n} = 0$. In this case the integral in Eq.
    \ref{eq:basic:resolution_opt} is easily solved and one
    finds:
\begin{equation}
\Delta u_{BA}^2 = \langle u_i^2 \rangle \cdot \left( 1+
\frac{\overline S_{F_n F_n}}{A^2 \overline S_{\Phi_{det,n}
\Phi_{det,n}}}\right)^{-1} = \langle u_i^2 \rangle \cdot \left( 1+
\frac{\overline S_{F_n F_n}}{\overline S_{F_{BA,n}
F_{BA,n}}}\right)^{-1}. \label{eq:basic:resolution_ba}
\end{equation}
The resolution thus increases (gets worse) with increasing
$A$ as the influence of the detector\added{, i.e. the
force noise $F_{BA,n}(t)$,} on the resonator motion grows.
For small values of $A$ this goes as $\Delta u_{BA} \propto
A$.
\item The detector adds displacement imprecision noise
    whereas  the backaction noise is very small, i.e.
    $\overline S_{\Phi_{det,n} \Phi_{det,n}} = 0$. In this
    case Eq. \ref{eq:basic:resolution_opt} reduces to:
\begin{equation}
\Delta u_n^2 = \langle u_i^2 \rangle \cdot J\left(\text{SNR},
Q\right), \text{~with~} \text{SNR} = \frac{Q^2}{k_R^2}
\frac{\overline S_{F_n F_n}}{\overline S_{u_n u_n} },
\label{eq:basic:resolution_imp}
\end{equation}
under the assumption that the detector noise PSD referred
to the detector input $\overline S_{u_n u_n} = \overline
S_{v_n v_n}/A^2|\lambda_v|^2$ is white. $J(\text{SNR}, Q)$
is a function (see the plots in Fig.
\ref{fig:app:spectrum}a) that depends on the quality factor
and the signal-to-noise ratio, SNR. $J$ tends to 1 when the
signal-to-noise ratio is well below unity. In that case,
the signal contains so much noise that it hardly contains
information about the displacement and using the signal
$v(t)$ is not going to give much more information then just
assuming that the resonator is at $u = 0$. The average
error that is made in the latter case is $\langle u_i^2
\rangle^{1/2}$.

When $\text{SNR} \rightarrow \infty$ the function $J(\text{SNR},
Q)$ goes to zero as $Q/(2^{1/2}~\text{SNR}^{3/4})$. The resolution
improves with increasing coupling as the resonator signal is
amplified more and more with respect to the noise floor $\overline
S_{v_n v_n}$.

For practical purposes it is convenient to use a slightly
different definition of the resolution
\cite{blencowe_PR_QEMS, lahaye_science,
etaki_natphys_squid} that does not involve the
signal-to-noise ratio dependent estimator:
\begin{equation}
\Delta u_n^2 \equiv \overline S_{u_n u_n} \frac{\pi}{2}
\frac{f_R}{Q} = \frac{\langle u_i^2 \rangle}{\text{SNR}},
\label{eq:basic:resolution_equiv}
\end{equation}
which is readily extracted from the measured noise spectra
as indicated in Fig. \ref{fig:app:spectrum}b. This
definition is based on the fact that one can measure the
position during a time $\sim Q/f_R$ before the resonator
has forgotten its initial amplitude and phase (see also
Fig. \ref{fig:basic:oscillator_noise}). Note again, that
this is the imprecision noise of the detector which does
not take the effect of backaction noise into account.
\end{itemize}

\section{Square-law detection}
\label{app:squared} In a $u$-squared detector the output of the
detector depends quadratically on the displacement. In analogy
with Eq.~\ref{eq:basic:convout} the detector output is now
$v(t) = A \lambda_v^{(2)}(t) \conv u(t)^2 + v_n^{(2)}(t)$,
where the superscript indicates that this is the response of
the square-law detector. A resonator oscillating with frequency
$f_R$, results in a detector output with frequency components
at $2f_R$ and at dc. As with every detector, the output also
contains imprecision noise, $v_n^{(2)}$. Again, by increasing
the coupling strength $A$ the signal-to-noise ratio can be
improved. In Sec. \ref{ssec:basic:sensitivity} it was shown
that for a linear detector this leads to an increased
backaction on the resonator, so it is interesting to see if the
square-law detector has backaction. The backaction of a
detector can vanish when the commutator of the quantity that is
measured ($\hat u^2 = u_0^2 (\hat a^\dagger + \hat a)^2$) at
different times is zero \cite{braginsky_science_QND}. In the
Heisenberg representation \cite{griffiths_qm} we have $[\hat
u^2 (t_1), \hat u^2 (t_2)] = 4i u_0^2 \sin(\omega_R \{t_2-
t_1\})\cdot(\hat u (t_1)\hat u (t_2)+\hat u (t_2)\hat u (t_1))
\neq 0$ for the square detector. Quantum mechanics thus
requires that the detector exerts a backaction force noise on
the resonator to comply with the uncertainty principle. Also
the quantum limit on the resolution for a square detector can
in principle be derived using the formalism outlined in
Sec.~\ref{ssec:basic:sensitivity}.

In many implementations, however, the detector has a narrow
bandwidth and its output does not contain the signal at
frequency $2f_R$, but only the dc component (see e.g.
Ref.~\cite{huettel_NL_highQ}). The operator corresponding to
the $u^2$-detection can be written as $\hat u^2 = u_0 ^2(\hat
a^2 + \hat a ^{\dagger 2} + \hat a \hat a^{\dagger} + \hat a
^{\dagger} \hat a)$. The first two terms oscillate at twice the
resonator frequency and when they are discarded, one finds that
this detector detects $\hat a \hat a^{\dagger} + \hat a
^{\dagger} \hat a = 2\hat n + 1$. In other words, it detects
the number of quanta or the energy in the resonator
\cite{jacobs_EPL_energy_measurement}. Using the fact that
$\tderl{\hat n}{t} = 0$, one finds that $[n(t_1), n(t_2)] = 0$
and that this detector can be backaction evading
\cite{braginsky_science_QND}. The measurement of the resonator
energy will change the phase of resonator, but leaves the
energy unaffected \cite{jacobs_EPL_energy_measurement}.





\end{document}